\documentclass[12pt,a4paper]{article}

\usepackage{algorithm}
\usepackage{algpseudocode}
\usepackage{geometry}
\usepackage{enumerate}
\usepackage[utf8]{inputenc}
\usepackage[T1]{fontenc}
\RequirePackage{amsmath,amssymb}
\RequirePackage{mathrsfs}
\usepackage{amsfonts}
\RequirePackage{amsfonts,bm}
\RequirePackage{dsfont}
\RequirePackage{braket}
\usepackage{subfigure}
\RequirePackage{tabularx}
\usepackage{multirow}
\RequirePackage{nicefrac}
\RequirePackage{graphicx}
\usepackage{tikz-feynman}
\usepackage{multicol}
\RequirePackage{booktabs}
\RequirePackage{colortbl}
\RequirePackage{xcolor}
\RequirePackage[symbol]{footmisc}
\usepackage{mathtools}
\usepackage{comment}
\usepackage{longtable}
\usepackage{tikz,tikz-feynman}
\usepackage{subcaption}
\usepackage{caption}
\usepackage{bbold}
\usepackage{blkarray}
\usepackage{float}
\usepackage{rotating}
\usepackage{hhline}
\usepackage{tikz}
\usepackage{multirow}
\usepackage[section]{placeins}

\usepackage{booktabs, tabu}

\usepackage[hypcap]{caption}

\usepackage[compress]{cite}

\newenvironment{aline}
    {\begin{equation}
    \begin{aligned}
    }
    { 
    \end{aligned}
    \end{equation}
    \ignorespacesafterend
    }
    
\def\AEF{A.E. Faraggi}

\def\EJP#1#2#3{{\it Eur.\ Phys.\ Jour.}\/ {\textbf C#1} (#2) #3}

\def\JHEP#1#2#3{{\it JHEP}\/ {\textbf #1} (#2) #3}
\def\NPB#1#2#3{{\it Nucl.\ Phys.}\/ {\textbf B#1} (#2) #3}
\def\PLB#1#2#3{{\it Phys.\ Lett.}\/ {\textbf B#1} (#2) #3}
\def\PRD#1#2#3{{\it Phys.\ Rev.}\/ {\textbf D#1} (#2) #3}

\def\beq{\begin{equation}}
\def\eeq{\end{equation}}
\def\beqn{\begin{eqnarray}}
\def\eeqn{\end{eqnarray}}

\usepackage{empheq}
\usepackage[most]{tcolorbox}
\usepackage{blkarray}
\newtcbox{\mymath}[1][]{%
    nobeforeafter, math upper, tcbox raise base,
    enhanced, colframe=blue!30!black,
    colback=blue!30, boxrule=1pt,
    #1}

\usepackage{datetime}
\newdateformat{monthyeardate}{\monthname[\THEMONTH] \THEYEAR}

\newcommand{\CC}[2]{C{#1\atopwithdelims[]#2}}
\newcommand{\GG}[2]{{#1\atopwithdelims[]#2}}

\newcommand{\ba}{\begin{eqnarray}}
\newcommand{\ea}{\end{eqnarray}}

\newcommand{\vth}{\vartheta} 
\newcommand{\vthb}{\bar{\vartheta}} 
\newcommand{\smb}[2]{\hspace{-0.02cm}\left[\begin{smallmatrix}#1\\#2\end{smallmatrix} \right]}
\newcommand{\smbar}[4]{\hspace{-0.02cm}\left[\begin{smallmatrix}#1\\#2\end{smallmatrix} \middle| \begin{smallmatrix}#3\\#4\end{smallmatrix}  \right]}

\newcommand{\one}{\mathds{1}} 
\newcommand{\Sv}{\bm{S}} 

\newcommand\T[1]{\bm{T_#1}}
\newcommand\bv[1]{\bm{b_#1}}

\numberwithin{equation}{section}

\begin{document}

% \samepage{
% \setcounter{page}{1}
% \rightline{LTH ---------------}
% \rightline{------------ 2023}

% \vfill
% \begin{center}
%   {\Large \bf{
%      Moduli Stabilisation}}

% \vspace{1cm}
% \vfill

% % {\large Alonzo R. Diaz Avalos $^{1}$\footnote{E-mail address: a.diaz-avalos@liverpool.ac.uk} and
% % Alon E. Faraggi$^{1,2}$\footnote{E-mail address: alon.faraggi@liverpool.ac.uk} }

% % \vspace{1cm}

% % {\it $^{1}$ Dept.\ of Mathematical Sciences, University of Liverpool, Liverpool
% % L69 7ZL, UK\\}
% % \vspace{.08in}

% % {\it $^{2}$ CERN, Theoretical Physics Department, CH--1211 Geneva 23, Switzerland\\}

% \vspace{.025in}
% \end{center}

% \vfill
% \begin{abstract}
% \noindent
% Moduli Stabilisation ...

% \end{abstract}

% \smallskip}

\begin{titlepage}
\samepage{
\setcounter{page}{1}
%\rightline{LTH-1332}
\rightline{\monthyeardate\today}

\vfill
\begin{center}
  {\Large \bf{
  A Perturbatively Stable Non-Supersymmetric \\ \bigskip String Model   with AdS Vacuum}}

\vspace{1cm}
\vfill

{\large 
Ignatios Antoniadis$^{1,2,3}$
 \footnote{E-mail address: antoniad@lpthe.jussieu.fr},
\\ \medskip
Alonzo R. Diaz Avalos$^{3}$
\footnote{E-mail address: a.diaz-avalos@liverpool.ac.uk}
 and 
Alon E. Faraggi$^{3}$
\footnote{E-mail address: alon.faraggi@liverpool.ac.uk}

\vspace{1cm}

{\it $^{1}$ High Energy Physics Research Unit, Faculty of Science, Chulalongkorn University, Bangkok 1030, Thailand\\}
\vspace{.08in}

{\it $^{2}$ Laboratoire de Physique Th\'eorique et Hautes Energies - LPTHE Sorbonne Universit\'e, CNRS, 4 Place Jussieu, 75005 Paris, France\\}
\vspace{.08in}

{\it $^{3}$ Dept.\ of Mathematical Sciences, University of Liverpool, Liverpool
L69 7ZL, UK\\}

\vspace{.025in}

}
\end{center}

\vfill

\begin{abstract}
\noindent
We present a construction of a perturbatively stable non-supersymmetric type II closed string model in four dimensions. It is based on a freely acting Scherk-Schwarz $\mathbb{Z}_2$-deformation of a supersymmetric construction which is recovered in appropriate decompactification limits. The model exhibits also the so-called misaligned supersymmetry with alternating signs for the number difference between bosons and fermions at successive mass levels. The tree-level spectrum is tachyon free for any value of the radii and moduli. At one loop level, the scalar potential has a non-supersymmetric minimum at the self-dual (free fermionic) point with negative energy, around which all tree-level massless scalars acquire positive masses. The model is thus non-supersymmetric and perturbatively stable.   
\end{abstract}
\vfill
\smallskip}

\end{titlepage}
\newpage

% \tableofcontents

\section{Introduction}
Low energy supersymmetry has not been observed and experimental bounds on new physics beyond the Standard Model are in the multi-TeV region. On the other hand, supersymmetry is a consistent ingredient of string theory, although its breaking scale may be at very high energies, perhaps not far from the string scale. It is therefore important to study properties of non-supersymmetric string theories and their phenomenological implications. One of the main interesting open problems is the question of moduli stabilisation related to the stability of non-supersymmetric strings. In contrast to the supersymmetric case, such an analysis cannot be restricted to tree-level since in the absence of supersymmetry, several tools like holomorphicity and non-renormalisation theorems loose their  validity. In particular, tree-level moduli and massless scalars may acquire tachyonic masses at one loop, leading to instability. For this reason, in this work, we address this problem within a simple framework of type II strings offering calculability. \\

\noindent
We first define a four-dimensional (4d) $\mathcal{N}=2$ supersymmetric model within the free-fermionic formulation~\cite{Antoniadis:1986rn, Kawai:1986ah, Antoniadis:1987wp} that corresponds to a particular point of the moduli space of a symmetric $\mathbb{Z}_2 \times \mathbb{Z}_2$ orbifold with
enhanced symmetry \cite{Athanasopoulos:2016aws}. We then turn on the $T^2\times T^2\times T^2$ torus moduli of the orbifold and break supersymmetry by a freely acting $\mathbb{Z}_2$-deformation which contains appropriate shifts of the compactification lattice momenta, in a way to avoid the appearance of tachyons for any value of the moduli. It turns out that $\mathcal{N}=1$ supersymmetry is restored from the (Left) L- or (Right) R-movers in appropriate decompactification limits of the three $T^2$s. Finally, we compute the tree-level spectrum and show that it is free of tachyons. 
%Actually, supersymmetry is broken mainly in the untwisted sector of the orbifold; all $\mathbb{Z}_2$-twisted sectors with NS-NS (Neveu-Schwarz) scalars are $\mathcal{N}=1$ supersymmetric. 
Moreover, the spectrum exhibits the property of misaligned supersymmetry: the difference between bosons and fermions has an alternating sign at successive mass levels.
The model is therefore stable at tree-level, but not supersymmetric.\\

\noindent
We then compute the one loop potential for the toroidal moduli and show that it has a global minimum at the free-fermionic point with a negative energy. Actually, the model has a $T$-duality symmetry which is a subgroup of $SO(2,2;\mathbb{Z})/[SO(2)\times SO(2)]$ for each $T^2$ and the free-fermionic point corresponds to the self-dual point under this symmetry.
%$\left\{SO(2,2;\mathbb{Z})/[SO(2)\times SO(2)]\right\}^3$, where the power 3 is due to the three $T^2$s.
Next, we proceed by computing the one-loop corrections to all tree-level massless bosons around the new vacuum. For each $T^2$, there are two emergent $U(1)$s and three types of untwisted scalars:
\begin{enumerate}[(i)]
\item the fluctuations of the two complex moduli $T$ and $U$ around their self-dual points, for each $T^2$; 
\item two emergent complex scalars accompanying the emergent abelian gauge bosons for each $T^2$ (corresponding to their internal components in six dimensions) that we call in the following `single charged' (although neutral), since they arise from the L or R part of the $T^2$ lattice which is factorised at the fermionic point; 
\item eight emergent real scalars that we call `double charged', coming from lattice combinations of different torii.
\end{enumerate}
Despite the terminology, all scalars are neutral under the emergent $U(1)$s.
Finally, there are twisted scalars arising from the three $\mathbb{Z}_2$-twisted orbifold sectors, forming chiral multiplets of $\mathcal{N}=1$ supersymmetry for L- or R-movers, which is locally preserved at the corresponding fixed points.
It turns out that all one-loop corrections provide positive (non-tachyonic) masses to the above scalars. The model is therefore perturbatively stable at weak string coupling.
Moreover, this shows that the free-fermionic self-dual point corresponds to a minimum of the scalar potential, as we show also explicitly by computing its full moduli-dependence. Actually this minimum is metastable, as there are also unbounded directions. In these directions, there is therefore a maximum which we also compute numerically and find that it possesses a negative energy which is however within the Breitenlohner-Freedman stability bound in anti-de Sitter (AdS) spacetime~\cite{Breitenlohner:1982jf}, making it locally stable.\\

\noindent
Of course, since the one loop vacuum energy is non-vanishing and negative, the graviton two-point function at zero momentum is negative as if there is generated an apparent tachyonic graviton mass. This is obvious from the expansion of the square root of the determinant of the metric around flat space. Actually, in the Einstein frame, this corresponds to a negative dilaton potential vanishing at weak coupling. Thus, one has to stabilise the dilaton in order to draw a definite conclusion on the vacuum of the theory. This is discussed in the last part of our work. \\

\noindent
The outline of our paper is the following. In Section~2, we present the construction of the model and compute its one loop partition function (subsection~2.1); we extract the massless spectrum (subsection~2.2 for the supersymmetric model and subsection~2.3 for the deformed non-supersymmetric) and show the absence of tachyons over the whole moduli space (subsection~2.4); we compute the one loop scalar potential, derive the duality symmetries and show that it possesses a global minimum at the self-dual (free fermionic) point with negative energy (subsection~2.5). In Section~3, we compute the one loop masses around this minimum for all bosonic tree-level massless fields: the graviton (subsection~3.1), the geometric moduli (subsection~3.2), the `single charged' vectors and scalars (subsection~3.3), the `double charged' scalars (subsection~3.4), as well as the twisted scalars (subsection~3.5). In Section~4, we discuss the dilaton stabilisation and the perturbatively stable vacuum of the model. Finally, Section~5 contains our concluding remarks. There are also four appendices containing the definitions of Theta functions (Appendix A), the list of the massless states (Appendix B for the supersymmetric and Appendix C for the non-supersymmetric) and the one-loop correlators used in our computations (Appendix D). 

\section{String Vacua Analysis}

\setlength{\parindent}{0pt}

\subsection{Partition Function}
In this model we analyze a four dimensional (4d) symmetric $\mathbb{Z}_2 \times \mathbb{Z}_2$ orbifold in Type II string theory \cite{Gregori:1999ny},
described by the following basis vectors of boundary conditions in the free fermionic formulation
\begin{align}
\label{basis}
{\one}&=\{\psi^\mu,\
\chi^{1,\dots,6},y^{1,\dots,6}, \omega^{1,\dots,6} \; | \; \bar{\psi}^\mu, \bar{\chi}^{1,\dots,6}, \bar{y}^{1,\dots,6}, \bar{\omega}^{1,\dots,6} \},\nonumber\\
\Sv&=\{{\psi^\mu},\chi^{1,\dots,6}\},\nonumber\\
\bar{\Sv}&=\{ \; | \; \bar{\psi}^\mu, \bar{\chi} ^{1,\dots,6}\},\nonumber\\
\bv{1}&=\{\psi^\mu, \chi^{12} ,y^{34},y^{56}\; | \; \bar{\psi}^\mu, \bar{\chi}^{12}, \bar{y}^{34},
\bar{y}^{56} \},\\
\bv{2}&=\{\psi^\mu, \chi^{34} ,y^{12},y^{56}\; | \; \bar{\psi}^\mu, \bar{\chi}^{34}, \bar{y}^{12},
\bar{y}^{56} \},\nonumber\\
\T{1}&=\{y^{12},w^{12}\; | \; \bar{y}^{12},\bar{w}^{12}\}.\nonumber \\
\T{2}&=\{y^{34},w^{34}\; | \; \bar{y}^{34},\bar{w}^{34}\}.\nonumber 
\end{align}
where the index $\mu$ refers to 4d space-time, while the remaining fermions describe the internal degrees of freedom corresponding to the compactified (super)space.
Each vector denotes the periodic fermions, while the remaining ones are antiperiodic. Those on the left side of the vertical line are the Left-movers, while on the right side with a bar are Right-movers.
Supersymmetry is explicitly broken from $\mathcal{N} = \left(4,4\right) \rightarrow \left(1,1\right)$ by the $\mathbb{Z}_2 \times \mathbb{Z}_2$ action given by the $\bv{1}$, $\bv{2}$ vectors.  
The definition of the model requires to specify also the so-called
GSO projections matrix $C$, imposing modular invariance, which completely specifies the spectrum at the free-fermionic point by acting as projectors on the different sectors. With the following choice
\begin{equation}
\label{GSO_0}
\small
\CC{\bm{v_i}}{\bm{v_j}}= 
\begin{blockarray}{cccccccc}
& \one & \Sv & \bar{\Sv} & \bv{1} & \bv{2} & \T{1} & \T{2}\\
\begin{block}{c(rrrrrrr)}
\one &\;\; 1&1& 1& 1&1&1&1 \;\; \\ 
\Sv &\;\; 1&-1& 1& -1&-1 & -1 & -1\;\; \\ 
\bar{\Sv} &\;\; 1&1&-1&-1&-1 &-1 & -1 \;\; \\ 
\bv{1} &\;\; 1&1&1&1&1 &1 &1 \;\; \\ 
\bv{2} &\;\; 1&1&1&1& 1&1 &1 \;\; \\
\T{1} &\;\; 1&-1&-1&1& 1 &1 &1 \;\; \\
\T{2} &\;\; 1&-1&-1&1& 1 &1 &1 \; \; \\
\end{block}
\end{blockarray}
\end{equation}
the partition function at the free fermionic point can be written as follows
\begin{aline}\label{PF_FF}
    Z=\,&\frac{1}{\eta^{12}\; \bar{\eta}^{12}} \, \frac{1}{2^{7}} \sum_{indices}  (-1)^{a+b+k+l}\\[0.2cm]
    &\times \vth\smb{a}{b}_{\psi^\mu} \vth\smb{a+h_2}{b+g_2}_{\chi^{12}} \vth\smb{a+h_1}{b+g_1}_{\chi^{34}} \vth\smb{a-h_1-h_2}{b-g_1-g_2}_{\chi^{56}}\\[0.2cm]
    &\times \vthb\smb{k}{l}_{\bar{\psi}^\mu} \vthb\smb{k+h_2}{l+g_2}_{\bar{\chi}^{12}} \vthb\smb{k+h_1}{l+g_1}_{\bar{\chi}^{34}} \vthb\smb{k-h_1-h_2}{l-g_1-g_2}_{\bar{\chi}^{56}}\\[0.2cm]
    &\times \left|\vth\smb{\zeta_1}{\delta_1}_{\omega^{12}} \vth\smb{\zeta_1+h_2}{\delta_1+g_2}_{y^{12}}\right|^2 
    \left|\vth\smb{\zeta_2}{\delta_2}_{\omega^{34}} \vth\smb{\zeta_2+h_1}{\delta_2+g_1}_{y^{34}} \right|^2
    \left|\vth\smb{\zeta_3}{\delta_3}_{\omega^{56}} \vth\smb{\zeta_3+h_1+h_2}{\delta_3+g_1+g_2}_{y^{56}}\right|^2\\
\end{aline}
with $\psi^\mu$ the four-dimensional complexified fermions, $\chi^{IJ} = \chi^I + i \chi^J$ the complexified internal fermions with $\chi^I$ the real components, and $y^{IJ}, \omega^{IJ}$ the complex fermionised internal bosons $X^{IJ} = X^I + i X^J$ such that $\partial X^I \sim y^I \omega^I$. The Dedekind $\eta$-function and the Jacobi $\vartheta$-functions depending on the complex modulus $\tau$ of the world-sheet torus are defined in Appendix~A. Here, $h_i, g_i$ are the $\mathbb{Z}_2 \times \mathbb{Z}_2$ orbifold twists, while $a, b$ and $k, l$ are the spin structures along the two non-contractible cycles of the world-sheet torus for the L- and R-movers, respectively. \\

The fermions associated to the internal bosons can be rewritten in a lattice form by performing a double Poisson resummation, such that the moduli fields can be reinserted back into the partition function. Denoting by $T^{(i)} = T^{(i)} _1 + i T^{(i)} _2$, $U^{(i)} = U^{(i)} _1 + i U^{(i)} _2$ the real and imaginary parts of the moduli associated to the $i$-th torus, the partition function for our choice of basis vectors can be written as \cite{R2, R4, No_Scale, Pert, No_Scale_2, Al_1, Al_2}

\begin{aline}\label{PF}
    Z=\,&\frac{1}{\eta^{12} \; \bar{\eta}^{12}} \, \frac{1}{2^{7}} \sum_{indices}  (-1)^{a+b+k+l} \\[0.2cm]
    &\times \vth\smb{a}{b}_{\psi^\mu} \vth\smb{a+h_2}{b+g_2}_{\chi^{12}} \vth\smb{a+h_1}{b+g_1}_{\chi^{34}} \vth\smb{a-h_1-h_2}{b-g_1-g_2}_{\chi^{56}}\\[0.2cm]
    &\times \vthb\smb{k}{l}_{\bar{\psi}^\mu} \vthb\smb{k+h_2}{l+g_2}_{\bar{\chi}^{12}} \vthb\smb{k+h_1}{l+g_1}_{\bar{\chi}^{34}} \vthb\smb{k-h_1-h_2}{l-g_1-g_2}_{\bar{\chi}^{56}}\\[0.3cm]
    &\times \Gamma^{(1)}_{2,2}\smbar{H_1}{G_1}{h_2}{g_2}(T^{(1)},U^{(1)}) 
    \; \Gamma^{(2)}_{2,2}\smbar{H_2}{G_2}{h_1}{g_1}(T^{(2)},U^{(2)}) 
    \; \Gamma^{(3)}_{2,2}\smbar{H_3}{G_3}{h_1+h_2}{g_1+g_2}(T^{(3)},U^{(3)})
\end{aline}
where the twists $H_i,G_i$ correspond to the boundary conditions $\zeta_i,\delta_i$ of~\eqref{PF_FF} and 
the fermionised partition function corresponds to the particular case with the moduli at the free fermionic point $T^{\left(i\right)*} = 1+i$, $U^{\left(i\right)*} = 1+i$ (see below, section~2.5). The $i$-th lattice has the following form
\begin{equation}
    \label{torus_1}
    \Gamma^{(i)}_{2,2}\smbar{H_i}{G_i}{h_j}{g_j}(T^{(i)},U^{(i)}) = 
    \begin{cases}
      \; \Gamma^{(i)}_{2,2}\GG{H_i}{G_i}(T^{(i)},U^{(i)}) & \text{if $h_j = g_j = 0$}\\
      & \\
      \; \frac{1}{2} \sum_{\zeta_i, \delta_i = 0} ^1 (-1)^{\zeta_i G_i + \delta_i H_i + H_i G_i} \left|\vth\smb{\zeta_i}{\delta_i} \vth\smb{\zeta_i+h_j}{\delta_i+g_j}\right|^2  & \text{otherwise}\\ 
    \end{cases}       
\end{equation}
with the shifted lattice defined as
\begin{equation}
    \label{torus_2}
    \Gamma^{(i)}_{2,2}\GG{H_i}{G_i}(T^{(i)},U^{(i)}) = \sum_{m_1, m_2, n_1, n_2} (-1)^{G_i m_2 } q^{\frac{1}{2} p_L ^2} \bar{q}^{\frac{1}{2} p_R ^2}
\end{equation}
where $q=e^{2i\pi\tau}$ and the Left and Right momenta $p_L$, $p_R$ defined with a momentum shift along the second circle of the world-sheet torus as follows
% \begin{equation}
%     P_L = \frac{m_2  - U^{(i)}m_1 + T^{(i)} \left( n_1  + U^{(i)} \left( n_2+ \frac{H_i}{2} \right) \right)}{\sqrt{T^{(i)}_2 U^{(i)}_2}}
%     \label{momenta_left}
% \end{equation}
% \begin{equation}
%     P_R = \frac{m_2  - U^{(i)}m_1 + \bar{T}^{(i)} \left( n_1  + U^{(i)} \left( n_2 + \frac{H_i}{2} \right) \right)}{\sqrt{T^{(i)}_2 U^{(i)}_2}}
%     \label{momenta_right}
% \end{equation}
\begin{equation}
    p_L = \frac{G^{-1}}{\sqrt{2}} \left( -m + Bn+Gn \right)
    \label{momenta_left}
\end{equation}
\begin{equation}
    p_R = \frac{G^{-1}}{\sqrt{2}} \left( -m + Bn-Gn \right)
    \label{momenta_right}
\end{equation}
\begin{equation}
    m= \begin{pmatrix}
        m_1 \\
        m_2
    \end{pmatrix}, \qquad
    n= \begin{pmatrix}
        n_1 \\
        n_2 + \frac{H_i}{2}
    \end{pmatrix}
\end{equation}
and background fields $G_{IJ}$ and $B_{IJ}$ 
\begin{equation}
    G = \frac{T_2}{U_2}
    \begin{pmatrix}
        1 & U_1\\
        U_1 & \left|U\right|^2
    \end{pmatrix}
    , \qquad 
    B = T_1
    \begin{pmatrix}
        0 & -1\\
        1 & 0
    \end{pmatrix}
\end{equation}
where the symmetrized product $\cdot$ is defined as $p\cdot k = p^I G_{IJ} k^J$ such that $p_L^2 = p_L^I G_{IJ} p_L ^J$, $p_R^2 = p_R^I G_{IJ} p_R ^J$. 
%In the following we will be denoting by $p_{i,L}$, $p_{i,R}$ the momenta associated to the $\left(i\right)$-th torus. \\

Given the phases $\CC{\Sv}{\T{i}}=\CC{\bar{\Sv}}{\T{i}}=-1$, two gravitinos survive the $\mathbb{Z}_2 \times \mathbb{Z}_2$ orbifold. Then supersymmetry is broken spontaneously by introducing the Scherk-Schwarz (SS) phase deformation into the partition function
\begin{equation}
    \left(-1\right)^{a+b+k+l} \longrightarrow \left(-1\right)^{a+b+k+l+ a\left(G_1 + G_3\right) + b\left(H_1+H_3\right) + H_1 G_1 +H_3 G_3 + kG_2 + l H_2 +H_2 G_2} 
\end{equation}
preserving modular invariance, such that the Left  and Right gravitinos acquire a tree-level mass squared shift. \\

Supersymmetry is then restored at the boundary of the moduli space $T_2^{(i)} \sim R_1 ^{(i)}R_2 ^{(i)} \rightarrow \infty$, where $R_{1,2}^{(i)}$ are the two radii of the $i$-th torus in the factorised case of two circles.
Performing a double Poisson resummation, in this limit the lattice becomes
\begin{align}
    \Gamma^{(i)}_{2,2}\GG{H_i}{G_i}(T^{(i)},U^{(i)}) & = \frac{T^{(i)} _2}{\tau_2} \sum_{m_1, m_2, n_1, n_2} e^{ - 2 \pi i T^{(i)} \det{A} - \frac{\pi T_2 ^{(i)}}{\tau_2 U_2 ^{(i)}} \left| \left( 1,U^{(i)} \right) \; A \; 
    \left(\begin{smallmatrix} \tau\\1  \end{smallmatrix} \right)
    \right|^2  }  \rightarrow \frac{T_2 ^{(i)}}{\tau_2} 
\end{align}
with
\begin{equation}
    A = 
    \begin{pmatrix}
        n_1 & m_1 \\
        n_2 + \frac{H_i}{2} \; & \;m_2 +\frac{G_i}{2}
    \end{pmatrix}
\end{equation}
where in this limit only the $m_1 = n_1 = m_2 = n_2 = H_i = G_i = 0$ term gives a non-vanishing result.

\subsection{Massless Spectrum of the supersymmetric model}
We describe first for clarity the $\mathcal{N}=2$ supersymmetric model defined in \eqref{basis} - \eqref{PF_FF}. At a generic point in the $(T^2)^3$ lattices, the massless spectrum, besides the gravity multiplet and the dilaton hypermultiplet, consists of three vector multiplets and three hypermultiplets from the untwisted sector and 48 $(16\times 3)$ hypermultiplets from the twisted orbifold sectors. This is precisely the spectrum of the $\mathbb{Z}_2\times\mathbb{Z}_2$ orbifold with discrete torsion~\cite{Vafa:1994rv}. It is easy to see that the role of the discrete torsion is played by the sign of the projection of $\mathbf{T_2}$ on $\mathbf{b_1}$, corresponding to the coefficient $\CC{\bm{b_1}}{\bm{T_2}}$. Note that there are no extra massless states at the fermionic (self-dual) point.
%At the self-dual point, there is an extra massless vector or hypermultiplet (depending if it comes from the $T$ or $U$ lattice) while at the fermionic point both of them become massless.

\subsection{Massless Spectrum of the deformed model}
The spectrum of our deformed (non-supersymmetric) model can be read from the basis vectors and the phases matrix in the partition function. See Appendix \ref{appendix:spectrum} for the full spectrum sector-by-sector. In particular the tree-level massless untwisted states from the NS-NS (Neveu-Schwarz) sector $\textbf{NS}$ comprise of the graviton, the complex axio-dilaton scalar and the three complex structure moduli and three K\"ahler moduli. The would-be supersymmetric partners instead acquire a tree-level non-zero mass shift after supersymmetry breaking. In particular, the Left and Right gravitinos from the $\bm{S}$ and $\bm{\bar{S}}$ sector, as well as the scalars and vectors from the $\bm{S}+\bm{\bar{S}}$ sector have masses
\begin{equation}
    m^2 _{\bm{S}} = \frac{1}{T_2 ^{\left(1\right)} U_2 ^{\left(1\right)}} + \frac{1}{T_2 ^{\left(3\right)} U_2 ^{\left(3\right)}}, \qquad m^2 _{\bm{\bar{S}}} = \frac{1}{T_2 ^{\left(2\right)} U_2 ^{\left(2\right)}}, \qquad m^2 _{\bm{S}+\bm{\bar{S}}} = \frac{1}{T_2 ^{\left(1\right)} U_2 ^{\left(1\right)}} + \frac{1}{T_2 ^{\left(2\right)} U_2 ^{\left(2\right)}} + \frac{1}{T_2 ^{\left(3\right)} U_2 ^{\left(3\right)}}
\label{gravitino-mass}
\end{equation}
such that at $T_2 ^{\left(1\right)}, T_2 ^{\left(3\right)} \rightarrow \infty $ Left supersymmetry is restored and at $T_2 ^{\left(2\right)} \rightarrow \infty $ Right supersymmetry is restored. \\

Other untwisted massless bosonic states come from the $\bm{T_1}$, $\bm{T_2}$, $\bm{T_3} = \bm{\one} + \bm{S} + \bm{\Sv} + \bm{T_1} + \bm{T_2} $ sectors
\newcolumntype{C}[1]{>{\centering\let\newline\\\arraybackslash\hspace{0pt}}m{#1}}
% \begin{center}
% \begin{tabular}{C{4cm} C{4cm} C{4cm}}
% $e^{\pm i k_L ^1 \cdot X^1} \bar{\psi}^\mu$, & $\psi^\mu e^{\pm i k_R ^2 \cdot \bar{X}^2}$, & $e^{\pm i k_L ^3 \cdot X^3} \bar{\psi}^\mu$ \\
% $e^{\pm i k_L ^1 \cdot X^1} \bar{\chi}^I$,  $I = 1,2$& $\chi^I e^{\pm i k_R ^2 \cdot \bar{X}^2}$,  $I = 3,4$, & $e^{\pm i k_L ^3 \cdot X^3} \bar{\chi}^I$,  $I = 5,6$ 
% \end{tabular}
% \end{center} \bm{X^{12}}
\begin{aline}\label{UTmassless}
\frac{1}{\sqrt{2}} \; \left( e^{i k_L ^1 \cdot \bm{X^{12}}}+e^{-i k_L ^1 \cdot \bm{X^{12}}} \right) \bar{\psi}^\mu, \qquad & \qquad \frac{1}{\sqrt{2}} \; \left( e^{i k_L ^1 \cdot \bm{X^{12}}} - e^{-i k_L ^1 \cdot \bm{X^{12}}} \right) \bar{\chi}^I, \quad I = 1,2\\
\frac{1}{\sqrt{2}} \; \psi^\mu \left( e^{i k_R ^2 \cdot \bm{\bar{X}^{34}}} + e^{-i k_R ^2 \cdot \bm{\bar{X}^{34}}}\right) , \qquad & \qquad  \frac{1}{\sqrt{2}} \; \chi^I \left( e^{i k_R ^2 \cdot \bm{\bar{X}^{34}}} - e^{-i k_R ^2 \cdot \bm{\bar{X}^{34}}} \right), \quad I = 3,4, \\
 \frac{1}{\sqrt{2}} \; \left( e^{i k_L ^3 \cdot \bm{X^{56}}} +  e^{-i k_L ^3 \cdot \bm{X^{56}}} \right) \bar{\psi}^\mu, \qquad & \qquad  \frac{1}{\sqrt{2}} \; \left( e^{i k_L ^3 \cdot \bm{X^{56}}} - e^{-i k_L ^3 \cdot \bm{X^{56}}} \right) \bar{\chi}^I, \quad I = 5,6
\end{aline}
given by the following momenta combinations $k^i$ at the free fermionic point, with $i$ referring to the $i$-th torus
\begin{center}
\begin{tabular}{lccccll} 
    $k_L^1, \; k_L^3 = \begin{pmatrix} 0 \\ \frac{1}{\sqrt2}\end{pmatrix}$, & $k_R^1, \; k_R^3 = \begin{pmatrix} 0 \\ 0 \end{pmatrix}$, && & &
    $m = \begin{pmatrix} 0 \\ -1 \end{pmatrix}   $,&$ 
     n = \begin{pmatrix} 0 \\ 0 \end{pmatrix}$, \\
    $k_L^1, \; k_L^3 = \begin{pmatrix} \sqrt{2} \\ - \frac{1}{\sqrt2}\end{pmatrix}$, & $k_R^1, \; k_R^3 = \begin{pmatrix} 0 \\ 0 \end{pmatrix}$ && & &
    $m = \begin{pmatrix} -1 \\ -1 \end{pmatrix}   $,&$
     n = \begin{pmatrix} 1 \\ -1 \end{pmatrix}$, \\
     $k_L^1, \; k_L^3 = \begin{pmatrix} 0 \\ - \frac{1}{\sqrt2}\end{pmatrix}$, & $k_R^1, \; k_R^3 = \begin{pmatrix} 0 \\ 0 \end{pmatrix}$ && & &
    $m = \begin{pmatrix} 0 \\ 1 \end{pmatrix}   $,&$
     n = \begin{pmatrix} 0 \\ -1 \end{pmatrix}$, \\
     $k_L^1, \; k_L^3 = \begin{pmatrix} - \sqrt{2} \\ \frac{1}{\sqrt2}\end{pmatrix}$, & $k_R^1, \; k_R^3 = \begin{pmatrix} 0 \\ 0 \end{pmatrix}$ && & &
    $m = \begin{pmatrix} 1 \\ 1 \end{pmatrix}   $,&$
     n = \begin{pmatrix} -1 \\ 0 \end{pmatrix}$, \\
\end{tabular}
\end{center}
and
\begin{center}
\begin{tabular}{lccccll} 
    $k_L ^2 = \begin{pmatrix} 0 \\ 0 \end{pmatrix}$, & $k_R^2 = \begin{pmatrix} 0 \\ \frac{1}{\sqrt2}\end{pmatrix}$, & & & &
    $m = \begin{pmatrix} -1 \\ -1 \end{pmatrix}   $,&$ 
     n = \begin{pmatrix} 0 \\ -1 \end{pmatrix}$, \\
    $k_L ^2 = \begin{pmatrix} 0 \\ 0 \end{pmatrix}$ & $k_R^2 = \begin{pmatrix} \sqrt{2} \\ - \frac{1}{\sqrt2}\end{pmatrix}$, & & & &
    $m = \begin{pmatrix} 0 \\ 1 \end{pmatrix}   $,&$
     n = \begin{pmatrix} -1 \\ 0 \end{pmatrix}$, \\
    $k_L ^2 = \begin{pmatrix} 0 \\ 0 \end{pmatrix}$ & $k_R^2 = \begin{pmatrix} 0 \\ - \frac{1}{\sqrt2}\end{pmatrix}$, & & & &
    $m = \begin{pmatrix} 1 \\ 1 \end{pmatrix}   $,&$
     n = \begin{pmatrix} 0 \\ 0 \end{pmatrix}$, \\
    $k_L ^2 = \begin{pmatrix} 0 \\ 0 \end{pmatrix}$ & $k_R^2 = \begin{pmatrix} - \sqrt{2} \\ \frac{1}{\sqrt2}\end{pmatrix}$, & & & &
    $m = \begin{pmatrix} 0 \\ -1 \end{pmatrix}   $,&$
     n = \begin{pmatrix} 1 \\ -1 \end{pmatrix}$, \\
\end{tabular}
\end{center}

which comprise of 6 `charged' vector bosons and 6 complex `charged' scalars, with momenta $k$ in the corresponding lattice given above such that $\left( k_L ^1\right)^2 = \left( k_R ^2\right)^2 = \left( k_L ^3\right)^2 =1$, $\left( k_R ^1\right)^2 = \left( k_L ^2\right)^2 = \left( k_R ^3\right)^2 =0$.
In particular, the vectors are gauge bosons at the symmetry enhancement point along any direction in the moduli space, massive at tree-level at any point in the moduli space except at the free-fermionic point. 
% , as can be seen for example from the their masses in Fig. \ref{fig:T_Mass} as a function of $T_2$. 
% \begin{figure}[H]
% \centering
% \includegraphics[width=0.4\linewidth]{T_Mass.pdf}
% \caption{Mass for the states from the  $\T{i}$ sector. }
% \label{fig:T_Mass}
% \end{figure}
% with the same behaviour for the scalars. 
\\
Finally, in the untwisted spectrum there are the following four scalars that become massless at the fermionic point, with internal momenta only from the $\bm{T_1}+\bm{T_2}$ and $\bm{T_2}+\bm{T_3}$ sectors
% \begin{equation}
%     e^{i k_L ^1 \cdot X^1} e^{i k_R ^2 \cdot \bar{X}^2}, \qquad e^{i k_L ^3 \cdot X^3} e^{i k_R ^2 \cdot \bar{X}^2} 
% \end{equation}
\begin{aline}
    &\frac{1}{2} \left( e^{i k_L ^1 \cdot \bm{X^{12}}} + e^{-i k_L ^1 \cdot \bm{X^{12}}} \right)  \left( e^{i k_R ^2 \cdot \bm{\bar{X}^{34}}} + e^{-i k_R ^2 \cdot \bm{\bar{X}^{34}}} \right), \\
&\frac{1}{2} \left( e^{i k_L ^3 \cdot \bm{X^{56}}} + e^{-i k_L ^3 \cdot \bm{X^{56}}} \right)  \left( e^{i k_R ^2 \cdot \bm{\bar{X}^{34}}} + e^{-i k_R ^2 \cdot \bm{\bar{X}^{34}}} \right)
\end{aline}
with
\begin{aline}
    & k_L ^1, k_R ^2, k_L ^3 = \pm\begin{pmatrix} \sqrt{2} \\ - \frac{1}{\sqrt2}\end{pmatrix}, \pm\begin{pmatrix} 0 \\ \frac{1}{\sqrt2}\end{pmatrix} \\
    & k_R^1, k_L ^2, k_R ^3 = \begin{pmatrix} 0 \\ 0 \end{pmatrix}
\end{aline}
and combinations of winding and momenta $m$, $n$ as before for two different tori,
such that $\left( k_L ^1\right)^2 = \left( k_R ^2\right)^2 = \left( k_L ^3\right)^2 =1$, $\left( k_R ^1\right)^2 = \left( k_L ^2\right)^2 = \left( k_R ^3\right)^2 =0$, as well as the four states with  $L \leftrightarrow R$ reversed
\begin{aline}
& \frac{1}{2} \left( e^{i k_L ^2 \cdot \bm{X^{34}}} + e^{-i k_L ^2 \cdot \bm{X^{34}}} \right)  \left( e^{i k_R ^1 \cdot \bm{\bar{X}^{12}} } + e^{-i k_R ^1 \cdot \bm{\bar{X}^{12}}} \right), 
\\
& \frac{1}{2} \left( e^{i k_L ^2 \cdot \bm{X^{34}}} + e^{-i k_L ^2 \cdot \bm{X^{34}}} \right)  \left( e^{i k_R ^3 \cdot \bm{\bar{X}^{56}}} + e^{-i k_R ^3 \cdot \bm{\bar{X}^{56}}} \right)
\end{aline}
with
\begin{aline}
    & k_R ^1, k_L ^2, k_R ^3 = \pm\begin{pmatrix} \sqrt{2} \\ - \frac{1}{\sqrt2}\end{pmatrix}, \pm\begin{pmatrix} 0 \\ \frac{1}{\sqrt2}\end{pmatrix} \\
    & k_L^1, k_R ^2, k_L ^3 = \begin{pmatrix} 0 \\ 0 \end{pmatrix}
\end{aline}
by exchanging $m$, $n$ of the two tori coordinates, such that $\left( k_L ^1\right)^2 = \left( k_R ^2\right)^2 = \left( k_L ^3\right)^2 =1$, $\left( k_R ^1\right)^2 = \left( k_L ^2\right)^2 = \left( k_R ^3\right)^2 =0$. \\

In the twisted spectrum, after SS supersymmetry breaking, states from the $\bm{b_i}$ sectors will acquire a tree-level mass, as can be seen from Appendix \ref{appendix:spectrum}, while will remain massless the states from $\bm{b_2} + \bm{\bar{S}}$ with the would-be supersymmetric partner $\bm{b_2} + \bm{\bar{S} + \bm{S}}$, and $\bm{b_1} + \bm{S}$, $\bm{b_3} + \bm{S}$ with the would-be supersymmetric partners $\bm{b_1} + \bm{S} + \bm{\bar{S}} $, $\bm{b_3} + \bm{S} + \bm{\bar{S}} $, as well as combinations with additional $\bm{T_i}$ sectors. Note that the twisted massless scalars and the fermionic counterparts give the same and opposite in sign tree-level contribution to the partition function, but supersymmetry is still broken as we will show when computing the one-loop mass correction. \\

\begin{figure}[H]
    \centering
    \subfigure[]{\includegraphics[width=0.7\linewidth]{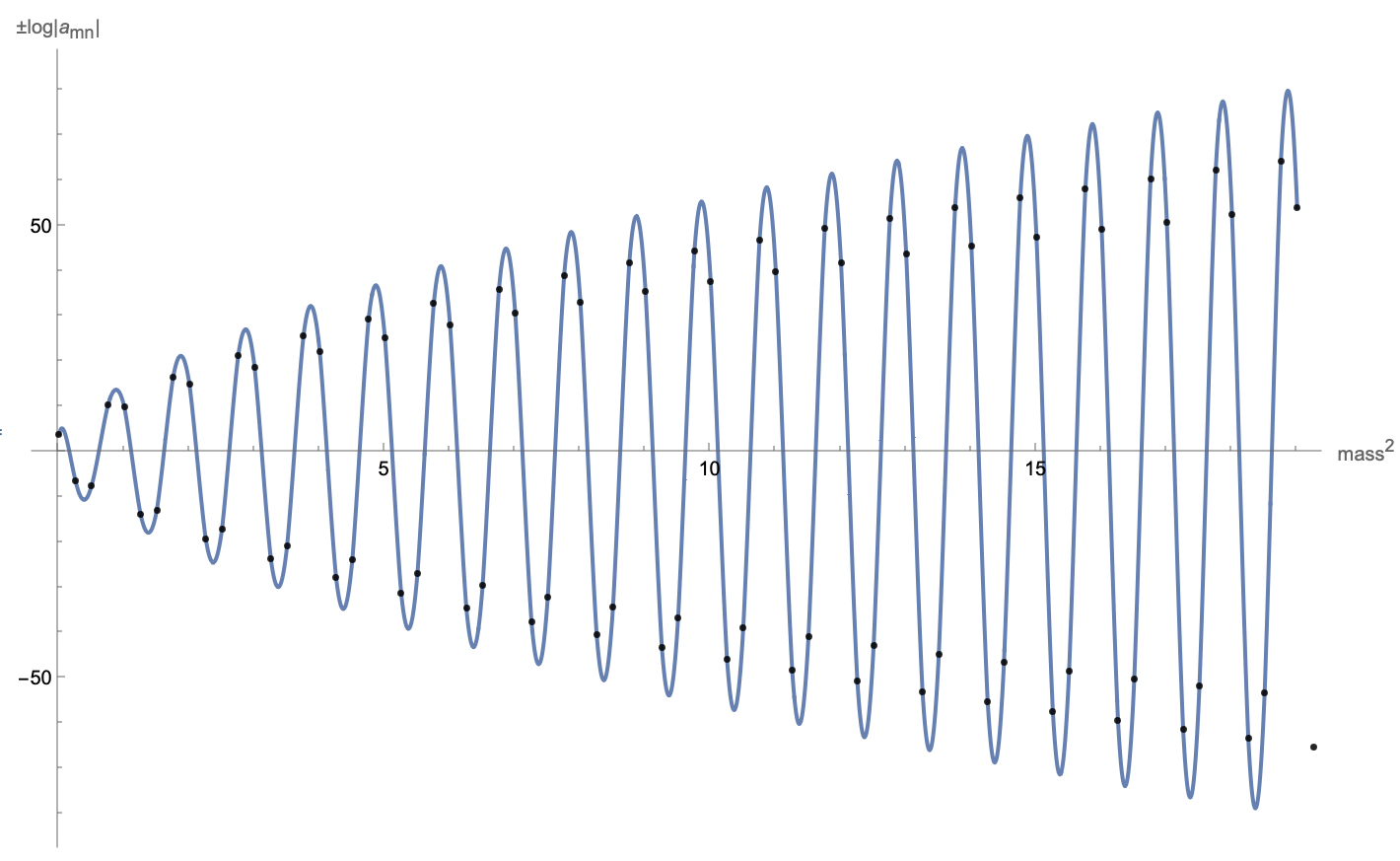}} 
    \captionsetup{justification=centering}
    \caption{Signed logarithm of the net number of bosons minus fermions at each mass level.}
    \label{fig:Misaligned}
\end{figure}

Although supersymmetry is broken, by performing a $q$-expansion of the partition function
\begin{equation}
    Z = \sum_{m,n} a_{mn} \; q^m \bar{q}^n
\end{equation}
with $a_{nm}$ the net number of bosons minus fermions at each mass level, it can be noted in Figure \ref{fig:Misaligned} the oscillatory behavior of misaligned supersymmetry \cite{Misaligned}.

\subsection{Tachyon Analysis}

We now discuss possible tree-level tachyons. At the free-fermionic point they have been projected out from the spectrum by the GSO phases, but other tachyonic states may appear as soon as we move away from that point. The mass of a state, using the momenta as in $\eqref{momenta_left}$, $\eqref{momenta_right}$ is the following
\begin{equation}
    M_L ^2 = \frac{1}{2} p_L^2 + N_L - \frac{1}{2}
\end{equation}
\begin{equation}
    M_R ^2 = \frac{1}{2} p_R^2 + N_R - \frac{1}{2}
\end{equation}
It is then clear that tree-level tachyons might appear when there are no oscillator, $N_L = N_R = 0$, and when there is a moduli dependence. In particular, these states can come from the  $\bm{T_i}$ and $\bm{T_i} + \bm{T_j}$ sectors, dependent on the moduli, and correspond to massless states at the free-fermionic point. Any other combination of basis vectors will produce massive states. \\

%The mass of a state, using the momenta as in $\eqref{momenta_left}$, $\eqref{momenta_right}$ is the following
%\begin{equation}
%    M_L ^2 = \frac{1}{2} p_L^2 + N_L - \frac{1}{2}
%\end{equation}
%\begin{equation}
%    M_R ^2 = \frac{1}{2} p_R^2 + N_R - \frac{1}{2}
%\end{equation}\\

%To check the presence of tachyons we set $N_L = N_R =0$. 
%Expanding the masses for the $\bm{T_i}$ sector, $H_i=1$, using the definition of the momenta in $\eqref{momenta_left}$, $\eqref{momenta_right}$ for the $\left(i\right)$-th torus
%\begin{equation}
%    M_L ^2 = -\frac{m_{i,2}}{4} - \frac{m_{i,1} n_{i,1}}{2} - \frac{m_{i,2} n_{i,2}}{2} + \dotsc + N_L - \frac{1}{2}
%\end{equation}
%\begin{equation}
%    M_R ^2 = \frac{m_{i,2}}{4} + \frac{m_{i,1} n_{i,1}}{2} + \frac{m_{i,2} n_{i,2}}{2} + \dotsc + N_R - \frac{1}{2}
%\end{equation}
%where the dots are common terms for the Left and Right masses, which comprehend the moduli dependent terms.
Inspecting the phase in the partition function
\begin{aline}
    & (-)^{\displaystyle{a+b+k+l + aG_1 + b H_1 + H_1 G_1 + aG_3 + b H_3 + H_3 G_3 + kG_2 + l H_2 + H_2 G_2}}
\end{aline}
and considering for instance the $H_1=1$ sector, $a=k=h_1=h_2=H_2=H_3=0$, the phase above becomes
\begin{aline}
   (-)^{\displaystyle{l + G_1}}
\end{aline}
%One has to consider also the additional phase in the lattice definition $\eqref{torus_2}$ of the $\left(i\right)$-th torus, $(-1)^{G_i m_{i,2}}$, such that the total phase is
%\begin{aline}
%     (-)^{\displaystyle{l + G_1 \left(1 +m_{1,2} \right) + G_2 m_{2,2} + G_3 m_{3,2}}}
%\end{aline}
%The summation over $G_1$ imposes $m_{1,2}$ to be odd, so that by looking at the formulae for the Left and Right masses there is no level matching for tachyonic states. 
%In addition, 
By summing over $l$ we have
\begin{equation}
    \sum_l \left(-1\right)^l \vthb^4 \smb{0}{l}  =  16 \sqrt{\bar{q}} + \dots
\end{equation}
which cancels the tachyonic contribution $\bar{\eta}^{-12} \sim \sqrt{\bar{q}} + \dots$  in the partition function, forcing the presence of a right oscillator, independently of the point in the moduli space. Thus the lowest level-matched states are the massless states $e^{i k_L ^1 \cdot X^1} \bar{\chi}^1$,  $e^{i k_L ^1 \cdot X^1} \bar{\psi}^\mu$ as showed before, which appear at the free-fermionic point and are massive at any other point in the moduli space, as can be checked by studying the mass formula as a function of the moduli. The same argument applies for the states from the $\bm{T_2}$ and $\bm{T_3}$ sectors as well as the $\bm{T_i}+\bm{T_j}$ sectors. \\

%forces $N_R = \frac{1}{2}$ producing only massless level-matched states at the fermionic point at most with an oscillator on the Right side to balance the mismatch, giving the states  $e^{i k_L ^1 \cdot X^1} \bar{\chi}^1$,  $e^{i k_L ^1 \cdot X^1} \bar{\psi}^\mu$ as showed before. The same argument applies to the states from the $\bm{T_2}$ and $\bm{T_3}$ sectors.\\

%For the $\bm{T_i}+\bm{T_j}$ sector, for $i=1$, $j=3$ massless states are projected out upon summation over $b+l$. 
%For the $\bm{T_1}+\bm{T_2}$ sector, with $H_1=H_2=1$ and with the constraint of $m_{1,2}=m_{2,2}=$ odd, the minimum value the mass can acquire without any oscillator is precisely $\frac{1}{2}$ for any value of the moduli, such that no tachyonic state appears, while producing the $e^{i k_L ^1 \cdot X^1} e^{i k_R ^2 \cdot \bar{X}^2}$ massless scalars at the fermionic point. The same argument applies for the $\bm{T_2}+\bm{T_3}$ sector. 

\subsection{One-loop potential}

The one-loop potential as a function of the $(T^2)^3$ moduli fields is defined as follows
\begin{equation}
    V \left(T^{\left(i\right)},U^{\left(i\right)}\right) = - \frac{1}{2 \left( 2 \pi \right)^4} \int_\mathcal{F} \frac{d^2 \tau}{\tau_2 ^3} Z
\end{equation}

\begin{figure}[H]
    \centering
    \subfigure[]{\includegraphics[width=0.49\linewidth]{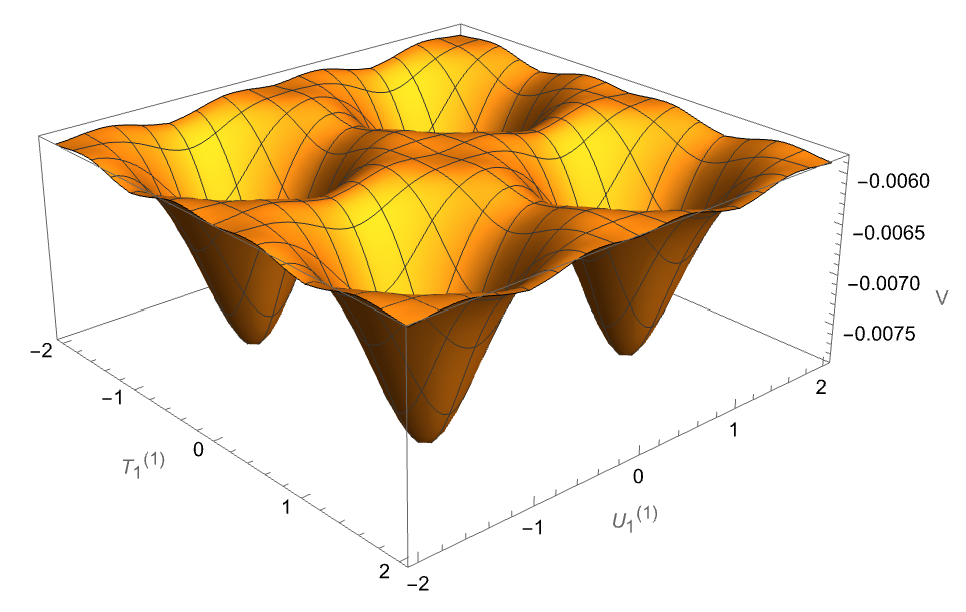}}
    \subfigure[]{\includegraphics[width=0.49\linewidth]{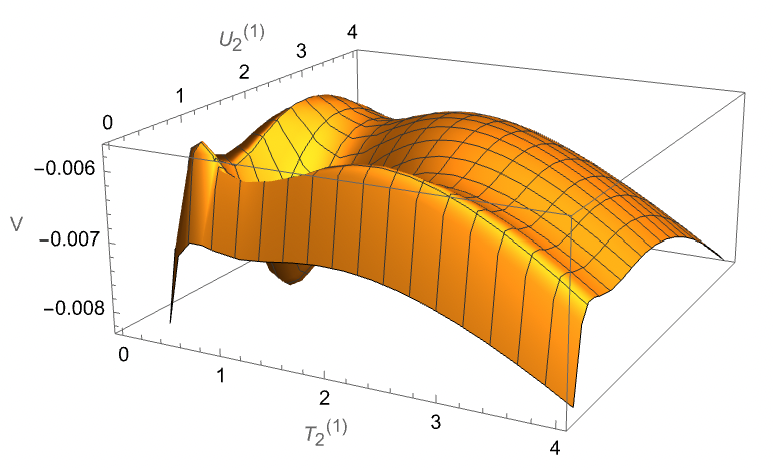}}
    \captionsetup{justification=centering}
    \caption{One-loop potential as a function of $T_1 ^{\left(1\right)}$, $U_1 ^{\left(1\right)}$ (left) and $T_2 ^{\left(1\right)}$, $U_2 ^{\left(1\right)}$ (right), with other moduli fixed at FF point.}
    \label{fig:Model_1_1}
\end{figure}

\begin{figure}[H]
    \centering
    \subfigure[]{\includegraphics[width=0.49\linewidth]{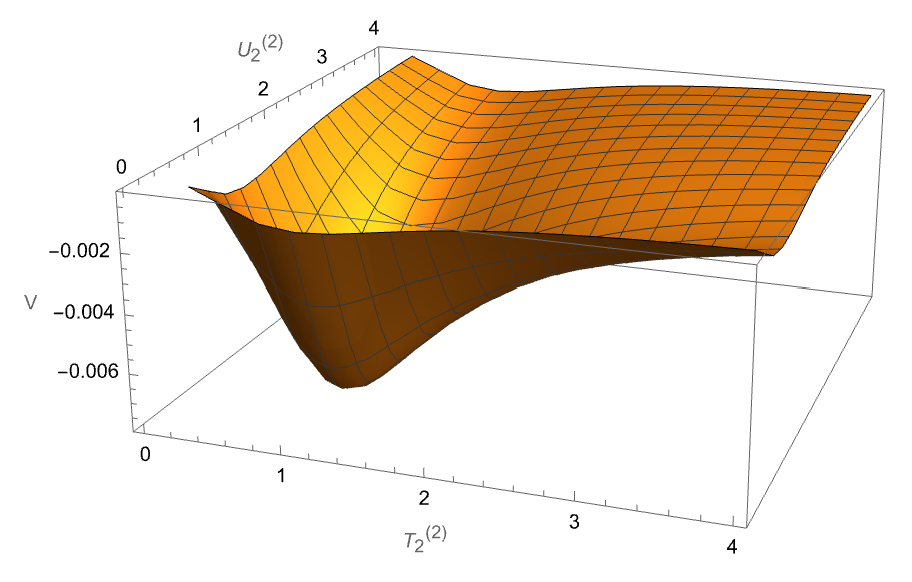}} 
    \subfigure[]{\includegraphics[width=0.49\linewidth]{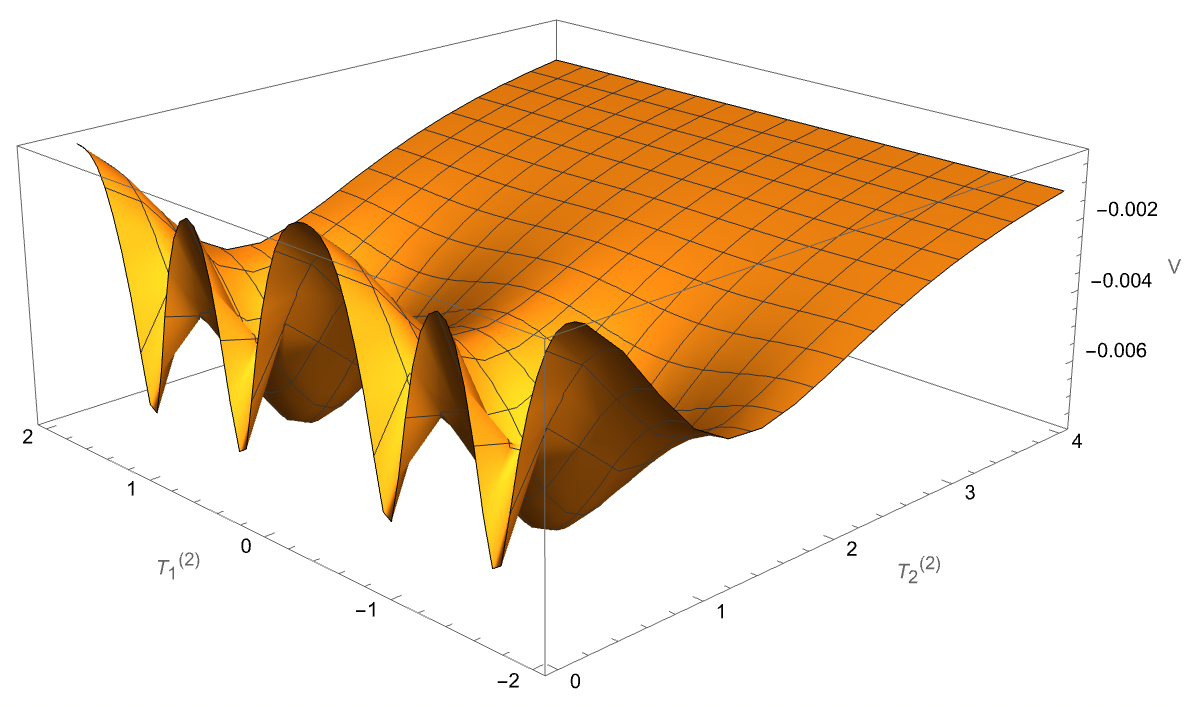}} 
    \captionsetup{justification=centering}
    \caption{One-loop potential as a function of $T_2 ^{\left(2\right)}$, $U_2 ^{\left(2\right)}$ (left) and $T_1 ^{\left(2\right)}$, $T_2 ^{\left(2\right)}$ (right), with other moduli fixed at FF point.}                                                                                                                                                  
    \label{fig:Model_1_2}
\end{figure}

\noindent 
The potential as a function of the 12 real moduli fields corresponding to $T_i, U_i$ is checked to have a minimum at the free-fermionic point, with the Hessian positive definite
\begin{equation}
    Hessian = 
    \begin{pmatrix}
        0.015581...*\mathbb{1}_4 &  \mathbb{0}_4 & \mathbb{0}_4 \\
        \mathbb{0}_4 & 0.027943...*\mathbb{1}_4 & \mathbb{0}_4 \\
        \mathbb{0}_4 & \mathbb{0}_4 & 0.015581...*\mathbb{1}_4
    \end{pmatrix}
\end{equation}
Note that the vacuum is metastable. As examples, in Figures \ref{fig:Model_1_1}, \ref{fig:Model_1_2} we display plots of the potential as a function of two moduli fields (the real or the imaginary parts in two different inequivalent tori in Fig.~\ref{fig:Model_1_1}, or two different moduli of the same torus) with the other moduli fixed at the free fermionic (FF) point. Moreover, in Fig.~\ref{fig:Single-modulus}, the potential is plotted as a function of the imaginary part of the $T$-modulus of the first torus, $T_2^{(1)}$. 

The multiplicity of minima along the real part or any modulus is due to the invariance of the potential under a shift by 2 (see below discussion of $T$-dualities). In the decompactification limit of the second torus $T_2^{(2)}\to\infty$, or of the first and third torus simultaneously $T_2^{(1)},T_2^{(3)}\to\infty$ the potential vanishes asymptotically since supersymmetry is recovered, as discussed in Section~2.1 and, thus, in these directions the potential has only one extremum which is a global minimum at the fermionic self-dual point with negative energy, see Fig.~\ref{fig:Model_1_2}. On the other hand, the decompactification limit of the first or third torus, keeping the size of the other fixed, is non supersymmetric with more bosons than fermions leading to an unbounded direction towards minus infinity, see Fig.~\ref{fig:Model_1_1}.and \ref{fig:Single-modulus}. As a result, in these directions, the potential develops a maximum corresponding to a point where the modulus mass becomes tachyonic at one loop, see Fig.~\ref{fig:Single-modulus}. The description of the maximum is presented in Section~3.2. It turns out that the maximum is stable because its curvature is inside the Breitenlohner-Freedman (BF) bound in anti-de Sitter (AdS) space~\cite{Breitenlohner:1982jf}.\\

After the orbifold action the T-duality group $SL \left(2, \mathbb{Z} \right)$ is broken into a subgroup $\Gamma$. For the particular choice of internal shifts, the generator of the T-duality transformation is given by~\cite{Florakis:2016ani}
\begin{equation}\label{GammaInv}
    \begin{pmatrix}
        1 & -2 \\
        1 & -1
    \end{pmatrix}
    \in \Gamma,  \qquad 
    T^{(i)} \rightarrow \frac{T^{(i)}-2}{T^{(i)}-1}
\end{equation}
%\begin{align}
%    T^{(i)}_1 \rightarrow T^{(i)}_1+2, \qquad \qquad T^{(i)}_2 \rightarrow \frac{1}{T^{(i)}_2} \\
%    U^{(i)}_1 \rightarrow U^{(i)}_1+2, \qquad \qquad U^{(i)}_2 \rightarrow \frac{1}{U^{(i)}_2}
%\end{align}
together with an $SL\left(2;\mathbf{Z}\right)$ translation
\begin{align}\label{GammaTra}
    T^{(i)} \rightarrow T^{(i)} +2 \qquad \qquad U^{(i)} \rightarrow U^{(i)}+2
\end{align}
as well as the mirror symmetry transformation
\begin{equation}
    T^{(i)} \leftrightarrow U^{(i)}, \qquad m_{i,1} \leftrightarrow - n_{i,1}
\end{equation}
The group generated by \eqref{GammaInv} and \eqref{GammaTra} is 
\begin{equation}
\Gamma^1(2)=\left\{ 
\begin{pmatrix}
        a & b \\
        c & d
    \end{pmatrix}
\in SL(2;\mathbb{Z})\, \text{with}\, a,d=1\,\text{mod}\, 1\,\text{and}\, b=0\,\text{mod}\,2
\right\}
\end{equation}
where $ad-bc=1$ and the group elements act on the modulus, say $T$, via fractional linear transformations $T\to (aT+b)/(cT+d)$.\\

In particular the radial coordinate of the modulus transforms as
\begin{align}
    1 + i T^{(i)} _2 \rightarrow  1 + \frac{i}{T^{(i)} _2}
\end{align}
With the one-loop Cosmological constant $\Lambda$ defined as the one-loop potential at its minimum, $T^{\left(i\right)*} = 1+i$, $U^{\left(i\right)*} = 1+i$, we can perform a $q$, $\bar{q}$ expansion and integrating term-by-term, we get the following numerical result for the vacuum energy
\begin{equation*}
    \Lambda = -0.007889715628901839 \text{ } M_s^4
\end{equation*}
where we included the string mass dimension $M_s$. For the evaluation of the masses we will include this mass term after the computation of the 2-point function.

%\noindent
%The masses are always positive, being the double derivative at the minimum of the potential. 
%Something we can note is that the real and imaginary parts of the moduli are always the same and that the masses of the $U^{(i)}$ moduli are $m^2 _{U_i } = m^2 _{T_i }$ for $T^{(i)}, U^{(i)}$ in the same torus. \\

% \noindent
% It is interesting to study the case of a different choice for our lattice. This time we choose the shift to act on the momenta of the first circle together with a winding shift along the second circle. \\
% The shifted lattice has now this form
% \begin{equation}
%     \Gamma^{(i)}_{2,2}\GG{H_i}{G_i}(T^{(i)},U^{(i)}) = \sum_{m_1, m_2, n_1, n_2} (-1)^{G_i \left( m_1+n_2\right)} q^{\frac{1}{4} \left| P_L\right|^2} \bar{q}^{\frac{1}{4} \left| P_R\right|^2}
% \end{equation}
% with the left and right momenta as
% \begin{equation}
%     P_L = \frac{m_2 + \frac{H_i}{2} - U^{(i)}m_1 + T^{(i)} \left( n_1 + \frac{H_i}{2} + U^{(i)} n_2 \right)}{\sqrt{T^{(i)}_2 U^{(i)}_2}}
% \end{equation}
% \begin{equation}
%     P_R = \frac{m_2 + \frac{H_i}{2} - U^{(i)}m_1 + \bar{T}^{(i)} \left( n_1 + \frac{H_i}{2} + U^{(i)} n_2 \right)}{\sqrt{T^{(i)}_2 U^{(i)}_2}}
% \end{equation}

% \noindent
% For this model the minimum is at the free fermionic point $T^* = i$, $U^* = \frac{1+i}{2}$, see Figures \ref{fig:Model_2_T1},\ref{fig:Model_2_T2},\ref{fig:Model_2_U1},\ref{fig:Model_2_U2}, and we get the following numerical results
% \begin{equation*}
%     \Lambda = -0.007889715628901839`
% \end{equation*}

\section{One-loop Masses}
In this section we compute one-loop masses of tree-level massless states, which comprise of the graviton, axio-dilaton and antisymmetric tensor, the moduli fields, the single charged scalars and vector bosons, the double charged scalars and the twisted states. In Appendix \ref{appendix:correlators} we give the different bosonic and fermionic correlators needed. 

\subsection{Graviton}

The vertex operator for the graviton and axio-dilaton at 0-momentum in the $-1$ picture is given by
\begin{equation}
  h_{\mu\nu}\,  \psi^\mu \; \bar{\psi}^\nu
\end{equation}
where $h_{\mu\nu}$ is the symmetric polarisation tensor. In order to move into the $0$ picture we use the supercurrent $T_F \sim \psi^\mu \partial X^\mu$ with the following tree-level operator product expansion (OPE)
\begin{equation}
    \langle  \psi^\mu \left(z \right) \psi^\nu \left(w \right) \rangle \sim \delta^{\mu \nu} \frac{1}{z-w} 
\end{equation}
such that the vertex in the $0$ picture is given by
\begin{equation}
  h_{\mu\nu}\,  \partial X^\mu \; \bar{\partial}X^\nu 
\end{equation}
The corresponding 2-point function, normalised such that the most singular term $z^{-2} \bar{z}^{-2}$ has coefficient $1$ with an additional $\pi^{-2}$ factor, is the following \cite{Graviton}
\begin{aline}
    m^2 = \int & \frac{d^2 z}{\pi^2} \;  
    h_{\mu \nu} h_{\alpha \beta}  \langle  \partial X^\mu \partial X^\alpha  \rangle \langle \bar{\partial} X^\nu \bar{\partial} \bar{X}^\beta  \rangle +
    h_{\mu \nu} h_{\alpha \beta}  \langle  \partial X^\mu \bar{\partial} X^\beta  \rangle  \langle \bar{\partial} X^\nu  \partial X^\alpha \rangle\\
    & =  h_{\mu \nu} h_{\alpha \beta} 
    \int \frac{d^2 z}{\pi^2} \; \left\langle \eta^{\mu \alpha} \eta^{\nu \beta}\left( \partial^2 \log \vth_1 +\frac{\pi}{\tau_2} \right) \left( \bar{\partial}^2 \log \vthb_1 +\frac{\pi}{\tau_2} \right)  + \eta^{\mu \beta} \eta^{\nu \alpha} \frac{\pi^2}{\tau_2 ^2} \right\rangle \\
    & = h_{\mu \nu} h^{\nu \mu} \left\langle\frac{1}{\tau_2} \right\rangle = \Lambda \text{ } M_s ^{-2} g_s ^2 = {\Lambda\over M_p^2}
\end{aline}
where $g_s$ is the string coupling, $M_p$ is the reduced Planck mass, and $h_{\mu \nu} h^{\nu \mu} = 1$ for the symmetric polarisation tensor, comprising the graviton and dilaton, and $-1$ for the antisymmetric tensor. This apparent graviton mass is just the consequence of the one loop cosmological constant, when expanding the square-root of the determinant of the metric to second order around flat space.

% \begin{center}
% \begin{tikzpicture}[baseline=(b)]
% \begin{feynman}
% \diagram [ baseline=(b), horizontal=a to b, layered layout] {
% a  -- [photon] b -- [half left, looseness=1.5, edge] c -- [half left, looseness=1.5, edge] b, c  -- [photon] d
% }; 
% \end{feynman}
% \end{tikzpicture} 
% \quad + \quad
% \begin{tikzpicture}[baseline=(b)]
% \begin{feynman}
% \diagram [ baseline=(b), horizontal=a to b, layered layout] {
% a  -- [photon] b -- [out=45,in=135,min distance=2.5cm] b  -- [photon] c
% }; 
% \end{feynman}
% \end{tikzpicture} 
% \quad + \quad
% \begin{tikzpicture}[baseline=(b)]
% \begin{feynman}
% \diagram [small, baseline=(c), horizontal=b to a, layered layout] {
% a -- [photon] k -- [photon] b -- [photon] d -- [photon] j, b -- [boson] c -- c ,
% {[same layer] b, c},
% }; 
% \path (c)--++(90:0.5) coordinate (A);
% \draw (A) circle(0.5);
% \end{feynman}
% \end{tikzpicture}%
% \end{center}

\subsection{Moduli fields}

The moduli remain massless at tree-level even after supersymmetry breaking, being untouched by the SS deformation, but will acquire in general a one-loop positive or negative contribution. \\

\begin{figure}[H]
    \centering
    \includegraphics[width=0.6\linewidth]{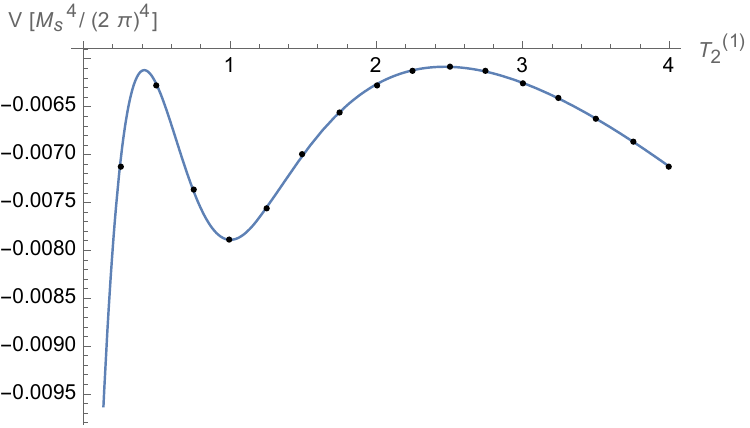}
    \captionsetup{justification=centering}
    \caption{One-loop potential as a function of $T_2 ^{\left( 1 \right)}$, with other moduli fixed at FF point.}
    \label{fig:Single-modulus}
\end{figure}

By inspecting the tree-level mass formula for the gravitino \eqref{gravitino-mass}, as $T_2 ^{\left( 2 \right)} \longrightarrow \infty$  Right supersymmetry is recovered, with the one-loop potential approaching zero from below. 
For the Left supersymmetry instead, the gravitino becomes massless when both $T_2 ^{\left( 1 \right)}$, $T_2 ^{\left( 3 \right)} \rightarrow \infty$. By fixing all moduli at the free-fermionic point and varying only $T_2 ^{\left( 1 \right)}$, or equivalently $T_2 ^{\left( 3 \right)}$, the potential is unbounded from below that follows a maximum as shown in Figure \ref{fig:Single-modulus}.

The vertex operator for the moduli fields at 0-momentum at the FF point in the $-1$ picture is 
\begin{equation}
    v _{IJ}\left( \phi\right) \; \chi^I \; \bar{\chi}^J
\end{equation}
Using the supercurrent $T_F \sim \chi^I \partial X^I$ with the following OPE
\begin{equation}
    \langle \chi^I \left(z \right)  \chi^J \left(w \right) \rangle \sim \delta^{IJ} \frac{1}{z-w}
\end{equation}
such that in the $0$ picture
\begin{equation}
    v _{IJ} \left( \phi\right) \; \partial X^I \; \bar{\partial}X^J
\end{equation}
with $I,J$ in the same torus, $\phi =T^{\left(i\right)}$ or $\phi =U^{\left(i\right)}$ and the matrix $v$ defined as follows \cite{R2}
\begin{equation}
    v_{IJ}\left(\phi\right) = \partial_\phi \left( G_{IJ} + B_{IJ} \right)
\end{equation}
with $\partial_{\phi} = \frac{1}{2} \partial_{\text{ Re$\left(\phi\right)$ }} - \frac{i}{2} \partial_{\text{ Im$\left(\phi\right)$ }}$, such that
\begin{equation}
    v_{IJ}\left(T\right) = -\frac{i}{2U_2}
    \begin{pmatrix}
        1 & U\\
        \bar{U} & \left|U\right|^2
    \end{pmatrix}
    , \qquad 
    v_{IJ}\left(U\right) = \frac{i T_2}{2U_2 ^2}
    \begin{pmatrix}
        1 & \bar{U}\\
        \bar{U} & \bar{U}^2
    \end{pmatrix}
\end{equation}
computed at the free-fermionic point. 
The 2-point function, performing the integration, is then given by
\begin{aline}
    v_{IJ}& \left(\phi\right) v^* _{MN}\left(\phi\right) \int \frac{d^2 z}{\pi^2} \; \langle \partial X^I \partial X^M \rangle \; \langle \bar{\partial} X^J \bar{\partial} X^N \rangle \\
    & = v_{IJ} \left(\phi\right) v^* _{MN} \left(\phi\right) \!\!\int \frac{d^2 z}{\pi^2} \!  \left\langle \left(G^{IM} \partial^2 \log   \vth_1 \left(z\right) + 4 \pi^2 p_L ^I p_L ^M \right) \!\left( 
    G^{JN} \bar{\partial}^2 \log   \vthb_1 \left(\bar{z}\right) +  4 \pi^2 p _R ^J p _R ^N  \right) \right\rangle  \\
    & = v_{IJ} \left(\phi\right) v^* _{MN}\left(\phi\right) \left\langle \tau_2 \left(-G^{IM}\frac{1}{\tau_2}+ 4 \pi p_L ^I p_L ^M \right) \left( - G^{JN}\frac{ 1 }{\tau_2 } +  4 \pi p _R ^J p _R ^N  \right)  \right\rangle
\end{aline}
where we have used \eqref{integral_1}, \eqref{integral_2} and normalised such that the most singular term in the worldsheet coordinate $z^{-2}\bar{z}^{-2}$ has coefficient $1$ with an additional $\pi^{-2}$ factor. It is understood in the 2-point function the sum over lattice momenta. The expression 
at the free-fermionic point can be rewritten as 
\begin{aline}
    & = \left\langle  \tau_2 \left(-\frac{1}{\tau_2}+ 2 \pi p_L ^2 \right) \left( - \frac{ 1 }{\tau_2 } +  2 \pi p_R ^2 \right) \right\rangle 
\end{aline}
with $p^2 = p \cdot p = p^I G_{IJ} p^J$, or alternatively as a double derivative of the potential with respect to the complex modulus
\begin{aline}
    & = \left\langle  \frac{4 }{\tau_2} \partial _{\phi} \partial_{\phi^*} \; \Gamma^{(i)}_{2,2}\GG{H_i}{G_i}(T^{(i)},U^{(i)})  + \frac{1}{\tau_2} \right\rangle
\end{aline}

%\textcolor{red}{kinetik term, there is a vertex with graviton (no dilaton) and the graviton goes into the vacuum, and have a propagator 1/k squares which cancel two momenta of the vertex, and on the other side there is usual tadpole which is a CC }

The last term proportional to the one-loop cosmological constant corresponds to the graviton tadpole attached to the tree-level moduli 2-point function using a vertex from its kinetic term. It provides a reducible contribution which at zero momentum becomes a contact term when the two momenta from the vertex cancel the graviton propagator and has thus to be removed from the evaluation of the one-loop masses. \\

The masses for the complex structure moduli $U^{(i)}$ and K\"ahler class moduli $T^{(i)}$ in the same $i$-th torus are equal, as well as the masses for the real and imaginary parts of the moduli. More precisely, the masses of the complex moduli in the three different tori $T^{\left(i\right)}$ at the minimum are 
\begin{aline}
    m^2 _{T^{\left(1\right)}} = m^2 _{T^{\left(3\right)}} & = 0.03131648778596434 \text{ }M_s ^{2} g_s ^2\\
%   0.02342677215706334 \\
    m^2 _{T^{\left(2\right)}} & = 0.05588420962623252 \text{ }M_s ^{2} g_s ^2\\
%   0.04799449399733152 \\
%    m^2 _{T^{\left(3\right)}} & = 0.03131648778596434 \text{ }M_s ^{2} g_s ^2\\
%   0.02342677215706334 \\
\end{aline}
%with
%\begin{equation}
%    m^2 _{T_1 ^{\left(i\right)}} = m^2 _{T_2 ^{\left(i\right)}} = \frac{1}{2} m^2 _{T ^{\left(i\right)}}, \qquad i =1,2,3
%\end{equation}
and
\begin{equation}
    m^2 _{U^{\left(i\right)}} = m^2 _{T^{\left(i\right)}} , \qquad i =1,2,3
\end{equation}

As mentioned before, away from the minimum, the $T_2 ^{\left(1\right)}$ and $T_2 ^{\left(3\right)}$ moduli develop a tachyonic instability at one-loop at the maxima points of the potential, each maximum point related to each other by T-duality. We examine this point as in Figure \ref{fig:Single-modulus}, by varying $T_2 ^{\left(1\right)}$ and fixing all other moduli at the free-fermionic point. The one-loop mass can be computed numerically as the double derivative of the potential with respect to the modulus at the maximum and gives the following result for the imaginary parts of the moduli
\begin{aline}
    m^2 _{T_2 ^{\left(1\right)}} = m^2 _{T_2 ^{\left(3\right)}} =  -0.0013716106156093262 \text{ }M_s ^{2} g_s ^2
\end{aline}
which however does not bring instabilities as this value is sitting inside the BF bound~\cite{Breitenlohner:1982jf} in AdS space
\begin{equation}
    m^2 _{T_2 ^{\left(1\right)}} \geq \frac{3}{4} {\Lambda \over M_p ^2}
\end{equation}

\subsection{Single charged scalars and vectors}

By defining $\boldsymbol{X^{IJ}} = \begin{pmatrix} X^I \\ X^J \end{pmatrix} $ and $\boldsymbol{\chi^{IJ}} = \begin{pmatrix} \chi^I \\ \chi^J \end{pmatrix} $, the vertex operators in the $-1$ picture for the gauge bosons and scalars in the different tori are the following
\begin{aline}
e^{i k_L ^1 \cdot \boldsymbol{X^{12}}} \; \bar{\psi}^\mu, \qquad & \qquad e^{i k_L ^1 \cdot \boldsymbol{X^{12}}} \;\bar{\chi}^I, \quad I = 1,2\\ 
\psi^\mu \; e^{i k_R ^2 \cdot \boldsymbol{X^{34}}}, \qquad & \qquad \chi^I \; e^{i k_R ^2 \cdot \boldsymbol{X^{34}}}, \quad I = 3,4\\
e^{i k_L ^3 \cdot \boldsymbol{X^{56}}} \; \bar{\psi}^\mu, \qquad & \qquad e^{i k_L ^3 \cdot \boldsymbol{X^{56}}}
\; \bar{\chi}^I, \quad I = 5,6
\end{aline}
with $k_L ^{1,3}, k_R ^2$ given below \eqref{UTmassless},
from which  we can construct the orbifold invariant combinations
\begin{aline}
\frac{1}{\sqrt{2}} \; \left( e^{i k_L ^1 \cdot \boldsymbol{X^{12}}}+e^{-i k_L ^1 \cdot \boldsymbol{X^{12}}} \right) \bar{\psi}^\mu, \qquad & \qquad \frac{1}{\sqrt{2}} \; \left( e^{i k_L ^1 \cdot \boldsymbol{X^{12}} } - e^{-i k_L ^1 \cdot \boldsymbol{X^{12}} } \right) \bar{\chi}^I, \quad I = 1,2\\
\frac{1}{\sqrt{2}} \; \psi^\mu \left( e^{i k_R ^2 \cdot 
\boldsymbol{\bar{X}^{34}}
} + e^{-i k_R ^2 \cdot \boldsymbol{\bar{X}^{34}} }\right), \qquad & \qquad \frac{1}{\sqrt{2}} \; \chi^I \left( e^{i k_R ^2 \cdot \boldsymbol{\bar{X}^{34}}} - e^{-i k_R ^2 \cdot \boldsymbol{\bar{X}^{34}}} \right), \quad I = 3,4\\
\frac{1}{\sqrt{2}} \; \left( e^{i k_L ^3 \cdot \boldsymbol{X^{56}}} +  e^{-i k_L ^3 \cdot \boldsymbol{X^{56}}} \right) \bar{\psi}^\mu, \qquad & \qquad \frac{1}{\sqrt{2}} \; \left( e^{i k_L ^3 \cdot \boldsymbol{X^{56}} } - e^{-i k_L ^3 \cdot 
\boldsymbol{X^{56}} } \right) \bar{\chi}^I, \quad I = 5,6
\end{aline}

and in the $0$ picture
\begin{aline}
\frac{k_L ^1}{\sqrt{2}} \cdot \boldsymbol{\chi^{12}}
\left( e^{i k_L ^1 \cdot \boldsymbol{X^{12}}
}-e^{-i k_L ^1 \cdot \boldsymbol{X^{12}}
} \right) \bar{\partial}X^\mu, \quad & \quad \frac{k_L ^1}{\sqrt{2}}  \cdot 
\boldsymbol{\chi^{12}} \left( e^{i k_L ^1 \cdot \boldsymbol{X^{12}}} + e^{-i k_L ^1 \cdot 
\boldsymbol{X^{12}} } \right) \bar{\partial}X^I, \quad I = 1,2\\
 \partial X^\mu \frac{k_R ^2}{\sqrt{2}} \cdot \boldsymbol{\bar{\chi}^{34}} \left( e^{i k_R ^2 \cdot \boldsymbol{\bar{X}^{34}} } - e^{-i k_R ^2 \cdot 
\boldsymbol{\bar{X}^{34}} }\right), \quad & \quad \partial X^I \frac{k_R ^2}{\sqrt{2}} \cdot \boldsymbol{\bar{\chi}^{34}} \left( e^{i k_R ^2 \cdot \boldsymbol{\bar{X}^{34}} } + e^{-i k_R ^2 \cdot \boldsymbol{\bar{X}^{34}}
} \right), \quad I = 3,4 \\
\frac{k_L ^3 }{\sqrt{2}} \cdot \boldsymbol{\chi^{56}} \left( e^{i k_L ^3 \cdot \boldsymbol{X^{56}}
} - e^{-i k_L ^3 \cdot \boldsymbol{X^{56}}
} \right) \bar{\partial}X^\mu, \quad & \quad \frac{k_L ^3}{\sqrt{2}}  \cdot \boldsymbol{\chi^{56}} \left( e^{i k_L ^3 \cdot \boldsymbol{X^{56}}
} + e^{-i k_L ^3 \cdot \boldsymbol{X^{56}}
} \right) \bar{\partial}X^I, \quad I = 5,6 
\end{aline}

The 2-point function for the first vector is 
\begin{aline}\label{2ptVector}
    - \frac{1}{2} \int & \frac{d^2 z}{\pi^2}  \; \left( k_L ^1 \right)^2 \left( \langle \; e^{i k_L ^1 \cdot \boldsymbol{X^{12}} 
    } e^{- i k_L ^1 \cdot \boldsymbol{X^{12}}  
    } \rangle + \langle \; e^{-i k_L ^1 \cdot \boldsymbol{X^{12}} 
    } e^{i k_L ^1 \cdot \boldsymbol{X^{12}} 
    } \rangle \right) \langle \chi^{I} \chi^{J} \rangle \; \langle \bar{\partial} X ^\mu \bar{\partial} X^\nu \rangle \; \eta_{\mu \nu} \; G_{IJ} \\
    & = - \int \frac{d^2 z}{\pi^2} \; \left\langle \left(\frac{\vth_1^{'2} \left(0 \right)}{\vth_1 ^2 \left(z \right)} \; e^{2 \pi i k_L ^1 \cdot p_L ^1 z } \; \frac{\vth\smb{a}{b}_{\chi^{12}}\left(z \right)}{\vth\smb{a}{b}_{\chi^{12}}\left(0 \right)} \right) \left( \bar{\partial}_{\bar{z}} ^2 \log \bar{\vth_1} \left(\bar{z}\right) + \frac{\pi}{\tau_2} \right) \right\rangle
\end{aline}
with $\left( k_L ^1 \right)^2 = 1$ and with proper normalization. The integrand is double-periodic over the torus with a double pole at $z, \bar{z} = 0$. It can then be written as an integral in terms of the Weierstrass function \cite{Weierstrass_Int}
\begin{aline}
    = - \int \frac{d^2 z}{\pi^2} \; 
    \left\langle \left( \wp - \frac{1}{3} \frac{\vth_1 ^{'''}}{\vth_1 ^{'}} + \frac{1}{2}\left(  2 \pi i k_L ^1 \cdot p_L ^1\right)^2 + \frac{1}{2} \frac{\vth\smb{a}{b}^{''} _{\chi^{12}}\left(0 \right)}{\vth\smb{a}{b}_{\chi^{12}}\left(0 \right)}  \right)
    \left( \frac{1}{3}\frac{\bar{\vth}_1^{'''}}{\bar{\vth}_1^{'}} - \bar{\wp} + \frac{\pi}{\tau_2}
    \right) \right\rangle = 0
\end{aline}
The vanishing of the integral can also be understood by the fact that the last parenthesis in \eqref{2ptVector} corresponding to the non-compact correlator $\langle \bar{\partial} X \bar{\partial} X \rangle$ is a total derivative with respect to $\bar z$ of the correlator $\langle X \bar{\partial} X \rangle$, while the holomorphic term in the first parenthesis has no simple pole, so that the integral reduces to a boundary contribution of a double periodic function which is zero.
We thus find that the one-loop masses for the gauge bosons are vanishing. \\

Similarly we can compute the 2-point function for the scalars
\begin{aline}
    - & \frac{1}{2} \int \frac{d^2 z}{\pi^2}  \; \left( k_L ^1 \right)^2 \left( \langle \; e^{i k_L ^1 \cdot \boldsymbol{X^{12}} 
    } e^{- i k_L ^1 \cdot \boldsymbol{X^{12}} } \rangle + \langle \; e^{-i k_L ^1 \cdot \boldsymbol{X^{12}} } e^{i k_L ^1 \cdot \boldsymbol{X^{12}} } \rangle \right) \langle \chi^{I} \chi^{J} \rangle \; \langle \bar{\partial} X ^\alpha \bar{\partial} X^\beta \rangle \;G_{IJ} \; G_{\alpha \beta} \\
    & = - \int \frac{d^2 z}{\pi^2} \; \left\langle \frac{\vth_1^{'2} \left(0 \right)}{\vth_1 ^2 \left(z \right)} \; e^{2 \pi i k_L ^1 \cdot p_L ^1 \; z } \; \frac{\vth\smb{a}{b}_{\chi^{12}}\left(z \right)}{\vth\smb{a}{b}_{\chi^{12}}\left(0 \right)} \left( \bar{\partial}_{\bar{z}} ^2 \log \bar{\vth_1} \left(\bar{z}\right) + \frac{1}{2} \left( 2 \pi p_R^1\right)^2 \right) \right\rangle \\
    & = - \int \frac{d^2 z}{\pi^2} \; \left\langle \left( \wp - \frac{1}{3} \frac{\vth_1 ^{'''}}{\vth_1 ^{'}} + \frac{1}{2}\left(  2 \pi i k_L ^1 \cdot p_L ^1\right)^2 + \frac{1}{2} \frac{\vth\smb{a}{b}^{''} _{\chi^{12}}\left(0 \right)}{\vth\smb{a}{b}_{\chi^{12}}\left(0 \right)}  \right) \left( \frac{1}{3}\frac{\bar{\vth}_1^{'''}}{\bar{\vth}_1^{'}} - \bar{\wp}  + \frac{1}{2} \left( 2 \pi p_R^1\right)^2 \right) \right\rangle \\
    & = \left\langle \tau_2 \left( - 2 \pi \left(k_L ^1 \cdot p_L ^1\right)^2 + \frac{1}{2\pi} \frac{\vth\smb{a}{b}^{''} _{\chi^{12}}\left(0 \right)}{\vth\smb{a}{b}_{\chi^{12}}\left(0 \right)} + \frac{1}{\tau_2} \right) \left( - 2 \pi \left(p_R^1\right)^2 + \frac{1}{\tau_2} \right)  \right\rangle
\end{aline}
Note that the twist indices corresponding to the plane with internal momentum must be set to zero, $h_2 = g_2 = 0$, and only the even theta functions for the fermion correlator contribute. 
The $\langle \frac{1}{\tau_2}  \rangle = \Lambda$ term corresponds to the tadpole diagram and is thus removed for the evaluation of the one-loop mass.
The numerical results of the one-loop masses for the `single charged' scalars in the $i$-th torus are the following
\begin{aline}
    m^2 _{T^{\left(1\right)}} = m^2 _{T^{\left(3\right)}} & = 0.03131648775627817 \text{ }M_s ^{2} g_s ^2\\
    m^2 _{T^{\left(2\right)}} & = 0.05588420948778768 \text{ }M_s ^{2} g_s ^2 %\\
  %  m^2 _{T^{\left(3\right)}} & = 0.03131648775627817 \text{ }M_s ^{2} g_s ^2 
\end{aline}

Equivalently, the same result can be obtained by fermionizing the bosons 
\begin{equation}
    e^{iX^I} \sim \omega^I + i y^I
\end{equation}
such that the orbifold invariant scalar vertex operators in the $0$ picture are
\begin{aline}
& \chi^{I} \omega^{I}
\;  \bar{y}^{J} \bar{\omega}^{J} , \quad I, J = 1,2\\
& y^{I} \omega^{I} \; \bar{\chi}^{J}
\bar{\omega}^J, \quad I, J = 3,4 \\
& \chi^{I} \omega^{I}
\;  \bar{y}^{J} \bar{\omega}^{J} , \quad I, J = 5,6\\
\end{aline}
The two point function for the first scalar is then given by
\begin{aline}
    \int \frac{d^2 z}{\pi^2} \left( 
    \frac{\vth \smb{a+h_2}{b+g_2}_{\chi^{12}} \left(z \right) }{ \vth \smb{a+h_2}{b+g_2}_{\chi^{12}} \left(0 \right) } \;
    \frac{\vth \smb{\zeta_1}{\delta_1}_{\omega^{12}} \left(z \right) }{ \vth \smb{\zeta_1}{\delta_1}_{\omega^{12}} \left(0 \right)  }
    \right)  
    \left( 
    \frac{\vthb \smb{\zeta_1+h_2}{\delta_1+g_2}_{\bar{y}^{12}} \left(z \right) }{ \vthb \smb{\zeta_1+h_2}{\delta_1+g_2}_{\bar{y}^{12}} \left(0 \right)  } \;
    \frac{\vthb \smb{\zeta_1}{\delta_1}_{\bar{\omega}^{12}} \left(z \right) }{ \vthb \smb{\zeta_1}{\delta_1}_{\bar{\omega}^{12}} \left(0 \right)  }
    \right)
\end{aline}
It can be checked that only the $h_2 = g_2 = 0$ term gives a non vanishing contribution upon summation over the indices. The anti-holomorphic side then corresponds to a squared fermion correlator which can be simplified by using the identity for the squared of the Szeg\"o kernel \eqref{square_Szeko} and performing the $z$-integration we get
\begin{aline}
    \left\langle 
    \tau_2 \left( 
    \frac{1}{2 \pi} \frac{\vth\smb{a}{b}^{''} _{\chi^{12}}\left(0 \right)}{\vth\smb{a}{b}_{\chi^{12}}\left(0 \right)} 
    +
    \frac{1}{2 \pi} \frac{\vth\smb{\zeta_1}{\delta_1}^{''} _{y^{12}}\left(0 \right)}{\vth\smb{\zeta_1}{\zeta_1}_{y^{12}}\left(0 \right)} 
    +
    \frac{1}{\tau_2}
    \right)  
    \left( 
    \frac{1}{2 \pi} \frac{\vthb\smb{\zeta_1}{\delta_1}^{''} _{\bar{y}^{12}}\left(0 \right)}{\vthb\smb{\zeta_1}{\zeta_1}_{\bar{y}^{12}}\left(0 \right)} + \frac{1}{\tau_2}
    \right) \right\rangle
\end{aline}
with 
\begin{equation}
    \frac{1}{2 \pi} \frac{\vthb\smb{\zeta_1}{\delta_1}^{''} _{y^{12}}\left(0 \right)}{\vthb\smb{\zeta_1}{\zeta_1}_{y^{12}}\left(0 \right)} = - 2 \pi \left( p_R ^1 \right)^2
\end{equation}
Also, by performing multiple Poisson resummations and by using the fact that $\left(k_L ^1 \right)^2 = 1$
\begin{aline}
    & \sum_{m_i, n_i} \left(-1\right)^{G_1 m_2} q^{\frac{1}{2} \left(p_L ^1\right) ^2} \bar{q}^{\frac{1}{2} \left( p^1 _R \right)^2} \; e^{2 \sqrt{2} \pi i z k^1 _L \cdot p^1 _L} \\
    & = \sum_{\zeta_1, \delta_1=0,1}\frac{1}{2} \left(-1\right)^{\zeta_1 G_1 + \delta_1 H_1 + H_1 G_1} \sum_{m_i, n_i} \prod_{j=1,2} 
    q^{\frac{1}{2} \left( n_j - \frac{\zeta_1}{2} \right)^2 } e^{2 \pi i \left( z - \frac{\delta_1}{2} \right) \left( n_j - \frac{\zeta_1}{2} \right) }
    \bar{q}^{\frac{1}{2} \left( m_j - \frac{\zeta_1}{2} \right)^2 } e^{i \pi \delta_1 \left( m_j - \frac{\zeta_1}{2} \right) } \\
    & = \sum_{\zeta_1, \delta_1=0,1}\frac{1}{2} \left(-1\right)^{\zeta_1 G_1 + \delta_1 H_1 + H_1 G_1}  
    \vth^2 \smb{\zeta_1}{\delta_1} \left(z \right) \vthb^2 \smb{\zeta_1}{\delta_1} \left(0 \right)
\end{aline}
such that
\begin{equation}
    \partial_z ^2 \; \frac{1}{4 \pi} \frac{\vth\smb{\zeta_1}{\delta_1}^2 _{y^{12}}\left(z \right)}{\vth\smb{\zeta_1}{\delta_1}^2_{y^{12}}\left(0 \right)}|_{z=0}   =
    \frac{1}{2 \pi} \frac{\vth\smb{\zeta_1}{\delta_1}^{''} _{y^{12}}\left(0 \right)}{\vth\smb{\zeta_1}{\delta_1}_{y^{12}}\left(0 \right)}   = - 2 \pi \left( k_l^1 \cdot p_L ^1 \right)^2
\end{equation}
getting the same formula for the one-loop scalar mass.

\subsection{Double charged scalars}
The vertex operators for `double charged' scalars at zero-momentum, in the $-1$ picture, are the following
\begin{aline}
e^{i k_L ^1 \cdot \boldsymbol{X^{12}}} 
e^{i k_R ^2 \cdot \boldsymbol{\bar{X}^{34}}}, \qquad
e^{i k_L ^3 \cdot \boldsymbol{X^{56}}} 
e^{i k_R ^2 \cdot \boldsymbol{\bar{X}^{34}}},  \\
e^{i k_L ^2 \cdot \boldsymbol{X^{34}}} 
e^{i k_R ^1 \cdot \boldsymbol{\bar{X}^{12}}}, \qquad
e^{i k_L ^2 \cdot \boldsymbol{X^{34}}} 
e^{i k_R ^3 \cdot \boldsymbol{\bar{X}^{56}}},
\end{aline}
with the orbifold invariant combinations
\begin{aline}
\frac{1}{2} \left( e^{i k_L ^1 \cdot \boldsymbol{X^{12}}} + e^{-i k_L ^1 \cdot \boldsymbol{X^{12}}} \right)  \left( e^{i k_R ^2 \cdot \boldsymbol{\bar{X}^{34}}} + e^{-i k_R ^2 \cdot \boldsymbol{\bar{X}^{34}}} \right), \quad
\frac{1}{2} \left( e^{i k_L ^3 \cdot \boldsymbol{X^{56}}} + e^{-i k_L ^3 \cdot \boldsymbol{X^{56}}} \right)  \left( e^{i k_R ^2 \cdot\boldsymbol{\bar{X}^{34}}} + e^{-i k_R ^2 \cdot \boldsymbol{\bar{X}^{34}}} \right) \\
\frac{1}{2} \left( e^{i k_L ^2 \cdot \boldsymbol{X^{34}}} + e^{-i k_L ^2 \cdot \boldsymbol{X^{34}}} \right)  \left( e^{i k_R ^1 \cdot \boldsymbol{\bar{X}^{12}}} + e^{-i k_R ^1 \cdot \boldsymbol{\bar{X}^{12}}} \right), \quad
\frac{1}{2} \left( e^{i k_L ^2 \cdot \boldsymbol{X^{34}}} + e^{-i k_L ^2 \cdot \boldsymbol{X^{34}}} \right)  \left( e^{i k_R ^3 \cdot \boldsymbol{\bar{X}^{56}}} + e^{-i k_R ^3 \cdot \boldsymbol{\bar{X}^{56}}} \right)
\end{aline}
We can then write the vertex operators in the $0$ picture

\begin{aline}
\frac{1}{2} \;  k_L ^1 \cdot \boldsymbol{\chi^{12}} \left( e^{i k_L ^1 \cdot \boldsymbol{X^{12}}} - e^{-i k_L ^1 \cdot \boldsymbol{X^{12}}} \right)  \; k_R ^2 \cdot \boldsymbol{\bar{\chi}^{34}} \left( e^{i k_R ^2 \cdot \boldsymbol{\bar{X}^{34}}} - e^{-i k_R ^2 \cdot \boldsymbol{\bar{X}^{34}}} \right), \\
\frac{1}{2} \;  k_L ^3 \cdot \boldsymbol{\chi^{56}} \left( e^{i k_L ^3 \cdot \boldsymbol{X^{56}}} - e^{-i k_L ^3 \cdot \boldsymbol{X^{56}}} \right)  \; k_R ^2 \cdot \boldsymbol{\bar{\chi}^{34}} \left( e^{i k_R ^2 \cdot \boldsymbol{\bar{X}^{34}}} - e^{-i k_R ^2 \cdot \boldsymbol{\bar{X}^{34}}} \right),
\label{double_charged_1_2}
\end{aline}
with the $L \leftrightarrow R$ exchange states
\begin{aline}
\frac{1}{2} \;  k_L ^2 \cdot \boldsymbol{\chi^{34}} \left( e^{i k_L ^2 \cdot \boldsymbol{X^{34}}} - e^{-i k_L ^2 \cdot \boldsymbol{X^{34}}} \right)  \; k_R ^1 \cdot \boldsymbol{\bar{\chi}^{12}} \left( e^{i k_R ^1 \cdot \boldsymbol{\bar{X}^{12}}} - e^{-i k_R ^1 \cdot \boldsymbol{\bar{X}^{12}}} \right), \\
\frac{1}{2} \;  k_L ^2 \cdot \boldsymbol{\chi^{34}} \left( e^{i k_L ^2 \cdot \boldsymbol{X^{34}}} - e^{-i k_L ^2 \cdot \boldsymbol{X^{34}}} \right)  \; k_R ^3 \cdot \boldsymbol{\bar{\chi}^{56}} \left( e^{i k_R ^3 \cdot \boldsymbol{\bar{X}^{56}}} - e^{-i k_R ^3 \cdot \boldsymbol{\bar{X}^{56}}} \right) \textcolor{white}{,}
\label{double_charged_2_1}
\end{aline}

For the first scalar, the 2-point function with proper normalisation is the following
\begin{aline}
    \frac{1}{4} \int & \frac{d^2 z}{\pi^2}  \; \left( k_L ^1 \right)^2 \left( k_R ^2 \right)^2  
    \left( \langle  e^{i k_L ^1 \cdot \boldsymbol{X^{12}}} e^{- i k_L ^1 \cdot \boldsymbol{X^{12}}} \rangle + \langle  e^{-i k_L ^1 \cdot \boldsymbol{X^{12}}} e^{i k_L ^1 \cdot \boldsymbol{X^{12}}} \rangle \right) \\
    & \times
    \left(\langle  e^{i k_R ^2 \cdot \boldsymbol{\bar{X}^{34}}} e^{- i k_R ^2 \cdot \boldsymbol{\bar{X}^{34}}} \rangle + \langle  e^{-i k_R ^2 \cdot \boldsymbol{\bar{X}^{34}}} e^{i k_R ^2 \cdot \boldsymbol{\bar{X}^{34}}} \rangle \right)\; 
    \langle \chi^{I} \chi^{J} \rangle \; 
    \langle \bar{\chi}^{M} \bar{\chi}^{N} \rangle \; G_{IJ} G_{MN}\\
    & = \int \frac{d^2 z}{\pi^2} \; 
    \left\langle \left(\frac{\vth_1^{'2} \left(0 \right)}{\vth_1 ^2 \left(z \right)} \; e^{2 \pi i k_L ^1 \cdot p_L ^1 \; z } \; \frac{\vth\smb{a}{b}_{\chi^{12}}\left(z \right)}{\vth\smb{a}{b}_{\chi^{12}}\left(0 \right)}
    \right)
    \left( \frac{\vthb_1^{'2} \left(0 \right)}{\vthb_1 ^2 \left(\bar{z} \right)} \; e^{2 \pi i k_R ^2 \cdot p_R ^2 \; \bar{z} } \; \frac{\vthb\smb{k}{l}_{\bar{\chi}^{34}}\left(\bar{z} \right)}{\vthb\smb{k}{l}_{\bar{\chi}^{34}}\left(0 \right)} \right) \right\rangle
\end{aline}
with $\left( k_L ^1 \right)^2 = \left( k_R ^2 \right)^2 = 1$, $I,J=1,2$ and $M,N = 3,4$, all twist indices set to zero, $h_i = g_i = 0$, with the only contribution coming from the even theta functions for the fermion correlator. Written in terms of the Weierstrass function and performing the $z$-integral it becomes
% \begin{aline}
%     \int & d^2 z  \; \left( k_L ^1 \right)^2 \left( k_R ^2 \right)^2  
%     \langle  e^{i k_L ^1 \cdot X^1} e^{- i k_L ^1 \cdot X^1} \rangle \; 
%     \langle  e^{i k_R ^2 \cdot \bar{X}^2} e^{- i k_R ^2 \cdot \bar{X}^2} \rangle \; 
%     \langle \chi^{12} \chi^{12} \rangle \; 
%     \langle \bar{\chi}^{34} \bar{\chi}^{34} \rangle \\
%     & = \int d^2 z \; 
%     \langle \left( \wp - \frac{1}{3} \frac{\vth_1 ^{'''}}{\vth_1 ^{'}} + \frac{1}{2}\left(  2 \pi i k_L ^1 \cdot p_L ^1\right)^2 + \frac{1}{2} \frac{\vth\smb{a}{b}^{''} _{\chi^{12}}\left(0 \right)}{\vth\smb{a}{b}_{\chi^{12}}\left(0 \right)}  \right) \\
%     & \times 
%     \left( \bar{\wp} - \frac{1}{3} \frac{\vthb_1 ^{'''}}{\vthb_1 ^{'}} + \frac{1}{2}\left(  2 \pi i k_R ^2 \cdot p_R ^2\right)^2 + \frac{1}{2} \frac{\vthb\smb{k}{l}^{''} _{\bar{\chi}^{34}}\left(0 \right)}{\vthb\smb{a}{b}_{\bar{\chi}^{34}}\left(0 \right)}  \right) \rangle \\
%     & = \langle \left( \frac{1}{2}\left(  2 \pi i k_L ^1 \cdot p_L ^1\right)^2 + \frac{1}{2} \frac{\vth\smb{a}{b}^{''} _{\chi^{12}}\left(0 \right)}{\vth\smb{a}{b}_{\chi^{12}}\left(0 \right)} + \frac{\pi}{\tau_2} \right) \left( \frac{1}{2}\left(  2 \pi i k_R ^2 \cdot p_R ^2\right)^2 + \frac{1}{2} \frac{\vthb\smb{k}{l}^{''} _{\bar{\chi}^{34}}\left(0 \right)}{\vthb\smb{a}{b}_{\bar{\chi}^{34}}\left(0 \right)} + \frac{\pi}{\tau_2} \right) \rangle
% \end{aline}
\begin{aline}
    = & \int \frac{d^2 z}{\pi^2} \; 
    \left\langle \left( \wp - \frac{1}{3} \frac{\vth_1 ^{'''}}{\vth_1 ^{'}} + \frac{1}{2}\left(  2 \pi i k_L ^1 \cdot p_L ^1\right)^2 + \frac{1}{2} \frac{\vth\smb{a}{b}^{''} _{\chi^{12}}\left(0 \right)}{\vth\smb{a}{b}_{\chi^{12}}\left(0 \right)}  \right)\right. \\
    & \left.\times 
    \left( \bar{\wp} - \frac{1}{3} \frac{\vthb_1 ^{'''}}{\vthb_1 ^{'}} + \frac{1}{2}\left(  2 \pi i k_R ^2 \cdot p_R ^2\right)^2 + \frac{1}{2} \frac{\vthb\smb{k}{l}^{''} _{\bar{\chi}^{34}}\left(0 \right)}{\vthb\smb{a}{b}_{\bar{\chi}^{34}}\left(0 \right)}  \right) \right\rangle \\
    & = \left\langle \tau_2 \left( -  2 \pi \left( k_L ^1 \cdot p_L ^1\right)^2 + \frac{1}{2 \pi} \frac{\vth\smb{a}{b}^{''} _{\chi^{12}}\left(0 \right)}{\vth\smb{a}{b}_{\chi^{12}}\left(0 \right)} + \frac{1}{\tau_2} \right) \left( - 2 \pi \left( k_R ^2 \cdot p_R ^2\right)^2 + \frac{1}{2 \pi} \frac{\vthb\smb{k}{l}^{''} _{\bar{\chi}^{34}}\left(0 \right)}{\vthb\smb{a}{b}_{\bar{\chi}^{34}}\left(0 \right)} + \frac{1}{\tau_2} \right) \right\rangle
\end{aline}
\\
Also in this case the $\Lambda$ term corresponds to a tadpole contribution to the 2-point function, and is thus removed for the evaluation of the one-loop masses. \\

\begin{figure}[H]
    \centering
    \includegraphics[width=1.\linewidth]{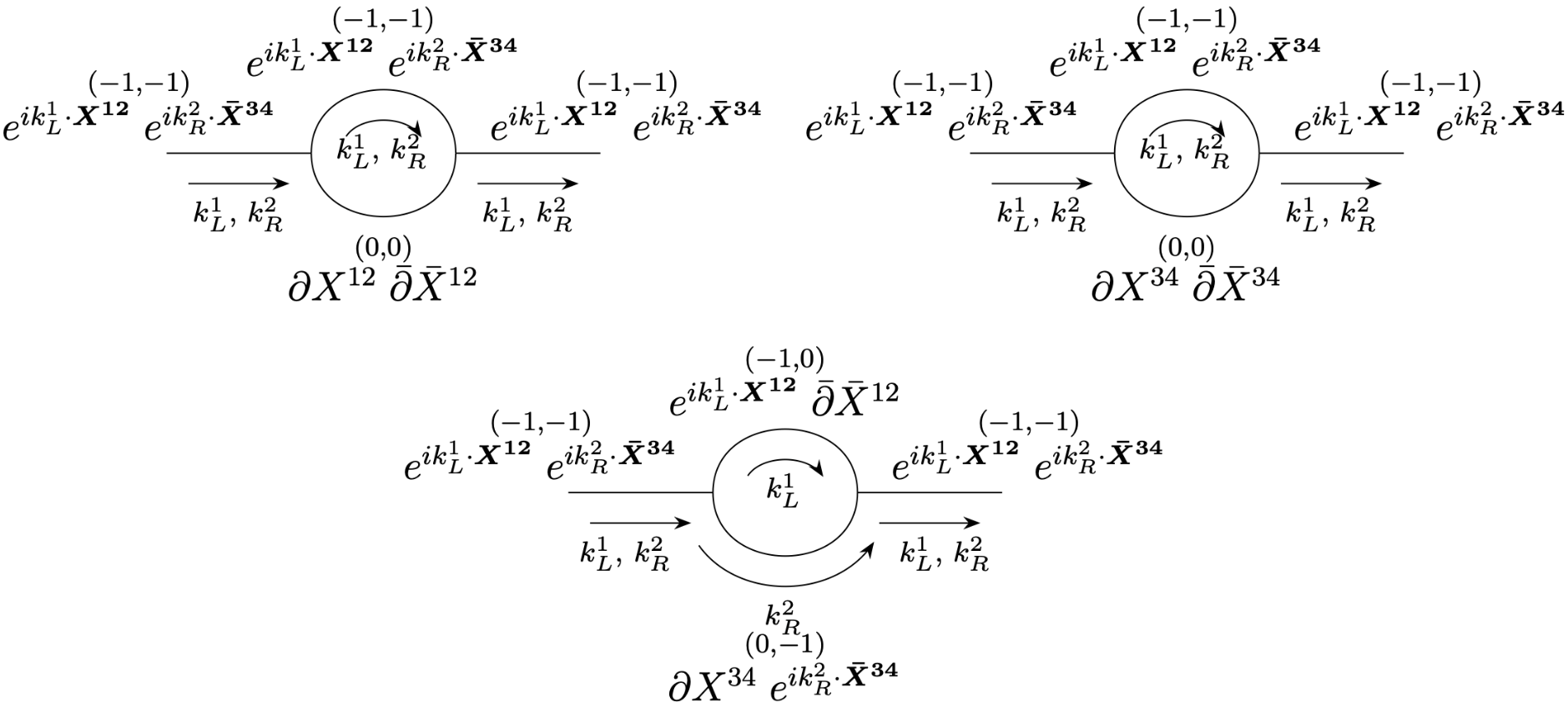}
    \captionsetup{justification=centering}
    \caption{IR divergent One loop Feynman diagrams for $\eqref{double_charged_1_2}$}
    \label{fig:Feynman_1}
\end{figure}

The integrand above is periodic and modular covariant but logarithmically divergent in the infrared (IR) region $\tau_2 \rightarrow \infty$ due to massless states in the loop, whose contribution do not cancel upon summation in the indices. We can show this by including the partition function for the amplitude
\begin{aline}
    A & = \frac{1}{2^{7}} \sum_{indices}  (-1)^{\tilde{\Phi} }  
    \sum_{p^1 _{L}, p^1 _{R} \in \Gamma^1}
    \sum_{p^2 _{L}, p^2 _{R} \in \Gamma^2}
    \sum_{p^3 _{L}, p^3 _{R} \in \Gamma^3} \;  \int d^2 \tau \, \frac{1}{\tau_2 ^2 \; \eta^{12}\bar{\eta}^{12}} \; \vth\smb{a}{b}^4 \vthb\smb{k}{l}^4
     \\[0.3cm]
    & \times \tau_2 \left( -2 \pi \left( k^1 _{L} \cdot p^1 _{L} \right)^2  + \frac{1}{2 \pi} \frac{\vth\smb{a}{b}''}{\vth\smb{a}{b}} + \frac{1}{\tau_2} \right) \left( -2 \pi \left( k^2 _{R} \cdot p^2 _{R} \right)^2  + \frac{1}{2 \pi} \frac{\vthb\smb{k}{l}''}{\vthb\smb{k}{l}} + \frac{1}{\tau_2} \right)  \\[0.3cm]
    &\times 
    q^{\frac{1}{2}\left(p^1_{L}\right) ^2} \bar{q}^{\frac{1}{2}\left(p^1_{R}\right) ^2} \left( -1 \right)^{G_1 m_{1,2}} \;
    q^{\frac{1}{2}\left(p^2_{L}\right) ^2} \bar{q}^{\frac{1}{2}\left(p^2_{R}\right) ^2} \left( -1 \right)^{G_2 m_{2,2}} \;
    q^{\frac{1}{2}\left(p^3_{L}\right) ^2} \bar{q}^{\frac{1}{2}\left(p^3_{R}\right) ^2} \left( -1 \right)^{G_3 m_{3,2}}
\end{aline}
with $\tilde{\Phi} = a + b+ k + l + a\left( G_1+G_3 \right) + b \left(H_1 +H_3 \right) + H_1 G_1 +H_3 G_3 + k G_2 +l H_2 +H_2 G_2$. Using the following expansions for the massless contribution
%The divergent contribution composes of two terms, one from the $\bm{NS}$ sector, with $\left(p^1 _{L}\right)^2 = \left(p^1 _{R}\right) ^2 = \left(p^2 _{L} \right)^2 = \left(p^2 _{R} \right) ^2 = 0$, $m_{1,2} = m_{2,2} = 0$, and an other from the $\bm{T_1}+\bm{T_2}$ sector, with $\left(p^1 _{L}\right)^2 = \left(p^2_{R}\right) ^2 = 1$, $\left(p^2 _{L}\right)^2 = \left(p^1 _{R}\right) ^2 = 0$, $k^1 _{L} \cdot p^1 _{L} = k^2 _{R} \cdot p^2 _{R}  = \pm1$, $m_{1,2} = m_{2,2} = \pm 1$
\begin{aline}
    & \vth\smb{0}{b}^4 = 1 + \dots \qquad \quad \frac{\vth\smb{0}{b}''}{\vth\smb{0}{b}} = \left(-1\right)^{b+1} 8 \pi^2 \sqrt{q} + \dots \qquad \quad \eta^{12} = q^{\frac{1}{2}}+ \dots
\end{aline}
the divergent contribution arises from the $a=k=h_1=h_2=H_3 = 0 $ sector
%\begin{aline}
%    A = & \frac{1}{2^{7}} \sum_{indices}  (-1)^{b+l+b H_1+H_1 G_1+l H_2+H_2 G_2 }  
%    \sum_{p^1 _{L}, p^1 _{R} \in \Gamma^1}
%    \sum_{p^2 _{L}, p^2 _{R} \in \Gamma^2}
%    \;  \int d^2 \tau \, \frac{1}{\tau_2 ^2 \; \eta^{12}\bar{\eta}^{12}} \; \vth\smb{a}{b}^4 \vthb\smb{k}{l}^4
%     \\[0.3cm]
%    & \times \tau_2 \left( -2 \pi \left( k^1 _{L} \cdot p^1 _{L} \right)^2  + \frac{1}{2 \pi} \frac{\vth\smb{0}{b}''}{\vth\smb{0}{b}} \right) \left( -2 \pi \left( k^2 _{R} \cdot p^2 _{R} \right)^2  + \frac{1}{2 \pi} \frac{\vthb\smb{0}{l}''}{\vthb\smb{0}{l}}  \right)  \\[0.3cm]
%    &\times 
%    q^{\frac{1}{2}\left(p^1_{L}\right) ^2} \bar{q}^{\frac{1}{2}\left(p^1_{R}\right) ^2} \left( -1 \right)^{G_1 m_{1,2}} \;
%    q^{\frac{1}{2}\left(p^2_{L}\right) ^2} \bar{q}^{\frac{1}{2}\left(p^2_{R}\right) ^2} \left( -1 \right)^{G_2 m_{2,2}} 
%\end{aline}
\begin{aline}
    A_{divergent}& = \frac{1}{2^{7}} \sum_{indices}  (-1)^{b+l+b H_1+H_1 G_1+l H_2+H_2 G_2 }  
    \sum_{p^1 _{L}, p^1 _{R} \in \Gamma^1}
    \sum_{p^2 _{L}, p^2 _{R} \in \Gamma^2}
    \;  \int d^2 \tau \, \frac{1}{\tau_2 ^2 \; q^{\frac{1}{2}} \bar{q}^{\frac{1}{2}}} 
     \\[0.3cm]
    & \times \tau_2 \left( -2 \pi \left( k^1 _{L} \cdot p^1 _{L} \right)^2  + \left(-1\right)^{b+1} 4 \pi q^{\frac{1}{2}}  \right) \left( -2 \pi \left( k^2 _{R} \cdot p^2 _{R} \right)^2  + \left(-1\right)^{l+1} 4 \pi \bar{q}^{\frac{1}{2}}  \right)  \\[0.3cm]
    &\times 
    q^{\frac{1}{2}\left(p^1_{L}\right) ^2} \bar{q}^{\frac{1}{2}\left(p^1_{R}\right) ^2} \left( -1 \right)^{G_1 m_{1,2}} \;
    q^{\frac{1}{2}\left(p^2_{L}\right) ^2} \bar{q}^{\frac{1}{2}\left(p^2_{R}\right) ^2} \left( -1 \right)^{G_2 m_{2,2}} 
\end{aline}
%\begin{aline}
%    A = & \frac{1}{2^{7}} \sum_{indices}  (-1)^{b+l+b H_1+H_1 G_1+l H_2+H_2 G_2 }  
%    \sum_{p^1 _{L}, p^1 _{R} \in \Gamma^1}
%    \sum_{p^2 _{L}, p^2 _{R} \in \Gamma^2}
%    \;  \int d^2 \tau \, \frac{4 \pi^2}{\tau_2 ^2 \; q^{\frac{1}{2}} \bar{q}^{\frac{1}{2}}} 
%     \\[0.3cm]
%    & \times \tau_2 \left( \left( k^1 _{L} \cdot p^1 _{L} \right)^2 \left( k^2 _{R} \cdot p^2 _{R} \right)^2 +  2 \left(-1\right)^l \left( k^1 _{L} \cdot p^1 _{L} \right)^2 \bar{q}^{\frac{1}{2}} + 2 \left(-1\right)^b \left( k^2 _{R} \cdot p^2 _{R} \right)^2 q^{\frac{1}{2}} + 4 \left(-1\right)^{b+l} q^{\frac{1}{2}} \bar{q}^{\frac{1}{2}} \right) \\[0.3cm]
%    &\times 
%    q^{\frac{1}{2}\left(p^1_{L}\right) ^2} \bar{q}^{\frac{1}{2}\left(p^1_{R}\right) ^2} \left( -1 \right)^{G_1 m_{1,2}} \;
%    q^{\frac{1}{2}\left(p^2_{L}\right) ^2} \bar{q}^{\frac{1}{2}\left(p^2_{R}\right) ^2} \left( -1 \right)^{G_2 m_{2,2}} 
%\end{aline}
The non-vanishing divergence from the $\left( k^1 _{L} \cdot p^1 _{L} \right)^2$ term arises setting $H_1 = 1$ with $p^{1} _L = \pm k^1 _L$ and $p^1 _R =  \left(0,0\right)^T$. For the $\left(-1\right)^b q^{\frac{1}{2}}$ term, the contribution arises by setting $H_1 = 0$ with $p^{1} _L = p^{1} _L = \left(0,0\right)^T$. Same applies for the Right-side considering the second torus, with other terms cancelling out upon summation over the indices. Then
\begin{aline}
    A_{divergent} & = \frac{1}{2^{7}} \sum_{b,l,G_1,G_2,G_3}  
    \;  \int d^2 \tau \, \frac{64 \pi^2}{\tau_2 } = 16 \pi^2 \int \frac{d^2 \tau}{\tau_2} 
\end{aline}
The divergence arises from massless scalars propagating in the loop with non-derivative 3-point couplings from the $\boldsymbol{0}$, $\boldsymbol{T_1}$, $\boldsymbol{T_2}$ and $\boldsymbol{T_1 +T_2}$ sectors. 
The relevant one loop Feynman diagrams for the `double charged' fields' mass in $\eqref{double_charged_1_2}$ in the IR limit are as in Figure \ref{fig:Feynman_1}

and for the scalars in $\eqref{double_charged_2_1}$ as in Figure \ref{fig:Feynman_2}

\begin{figure}[H]
    \centering
    \includegraphics[width=1.\linewidth]{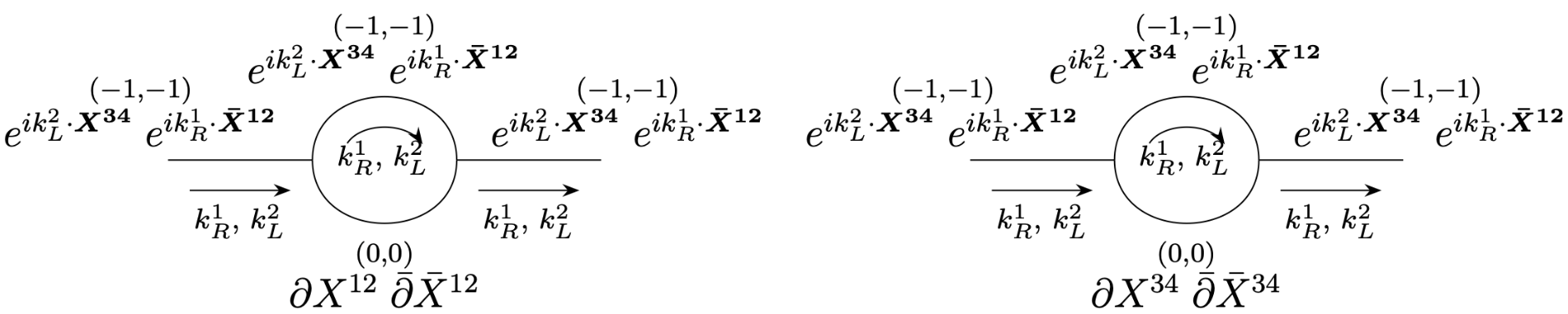}
    \captionsetup{justification=centering}
    \caption{IR divergent One loop Feynman diagrams for $\eqref{double_charged_2_1}$}
    \label{fig:Feynman_2}
\end{figure}

The finite part obtained %In order to give a numerical value one can use the minimal regulator, adopted 
by removing by hand the divergent piece, leads to positive contributions for all scalars;
%although modular invariance is lost. In this case the integral will be finite and the final regulated masses for the scalars turn out to be positive. 
for the scalars in $\eqref{double_charged_1_2}$
\begin{aline}\label{mfT1}
    m^2 _{T_L ^{\left(1\right)}T_R ^{\left(2\right)}}\large{\mid}_\text{finite} & = %0.06084 %00615739236 
%    \text{ }M_s ^{2} g_s ^2\\
    m^2 _{T_L ^{\left(3\right)}T_R ^{\left(2\right)}}\large{\mid}_\text{finite} & = 0.06084 %00615739236 
    \text{ }M_s ^{2} g_s ^2
\end{aline}
and for the scalars in $\eqref{double_charged_2_1}$
\begin{aline}\label{mfT2}
    m^2 _{T_L ^{\left(2\right)}T_R ^{\left(1\right)}}\large{\mid}_\text{finite} & = %0.04081 %068443981663 
%    \text{ }M_s ^{2} g_s ^2\\
    m^2 _{T_L ^{\left(2\right)}T_R ^{\left(3\right)}}\large{\mid}_\text{finite} & = 0.04081 %068443981663 
    \text{ }M_s ^{2} g_s ^2
\end{aline}

\subsection{Twisted scalars}
\label{twisted}

Massless twisted scalars arise from three $\mathbb{Z}_2$-twisted sectors where one plane remains untwisted. For instance, those arising from the three untwisted planes are given, in the three columns respectively, by
\begin{center}
\begin{tabular}{lcclccl} 
    $\bm{b_1}+\bm{S}+\bm{\bar{S}}$ & & & $\bm{b_2}+\bm{S}+\bm{\bar{S}}$ & & & $\bm{b_3}+\bm{S}+\bm{\bar{S}}$ \\
    $\bm{b_1}+\bm{S}+\bm{\bar{S}}+\bm{T_2}$ & & & $\bm{b_2}+\bm{S}+\bm{\bar{S}}+\bm{T_1}$ & & & $\bm{b_3}+\bm{S}+\bm{\bar{S}}+\bm{T_1}$ \\
    $\bm{b_1}+\bm{S}+\bm{\bar{S}}+\bm{T_3}$ & & & $\bm{b_2}+\bm{S}+\bm{\bar{S}}+\bm{T_3}$ & & & $\bm{b_3}+\bm{S}+\bm{\bar{S}}+\bm{T_2}$ \\
    $\bm{b_1}+\bm{S}+\bm{\bar{S}}+\bm{T_2}+\bm{T_3}$ & & & $\bm{b_2}+\bm{S}+\bm{\bar{S}}+\bm{T_1}+\bm{T_3}$ & & & $\bm{b_3}+\bm{S}+\bm{\bar{S}}+\bm{T_1}+\bm{T_2}$ \\
\end{tabular}
\end{center}

Bosonizing the complex fermions
\begin{equation}
    \chi^{IJ} \left(z\right) \sim e^{+ i \phi\left(z\right)}, \quad \quad \chi^{IJ*} \left(z\right) \sim e^{- i \phi \left(z\right)}
\end{equation}
the $SO\left(2\right)$ spin operators are given by \cite{Spin_1}
\begin{equation}
    S_{\chi^{IJ}}^+ \left(z\right)  \sim  e^{+ \frac{i \phi \left(z\right)}{2}},  \quad \quad  S_{\chi^{IJ}}^- \left(z\right) \sim  e^{- \frac{i \phi \left(z\right)}{2}} 
\end{equation}

The complex conjugate is given by exchanging $S^{\pm} \rightarrow S^{\mp}$. Also notice that the spin field is already invariant under the orbifold action. Each twisted plane thus gives a total of 16 complex massless scalars. \\

The spin fields have the following OPE 
\begin{align}
    & \chi^{IJ} \left(z\right)\; S^- \left(w\right) \sim \frac{1}{\left(z-w\right)^{1/2}} \; S^+ \\
    & \chi^{IJ*} \left(z\right)\; S^+ \left(w\right) \sim \frac{1}{\left(z-w\right)^{1/2}} \; S^- \\
    & \chi^{I} \left(z\right)\; S^{\pm} \left(w\right) \sim \frac{1}{\left(z-w\right)^{1/2}} \; S^{\pm} \\
    & S^+ \left(z\right) \; S^- \left(w\right) \sim \frac{1}{\left(z-w\right)^{1/4}}
\end{align}
which is used to find the vertex operators in the $0$ picture. The full list of massless twisted states for the three planes is given in Appendix \ref{appendix:spectrum}. \\

We will examine the first of these states, given in the $-1$ picture by 
\begin{equation}
S_{\chi^{34}}^+ S_{\chi^{56}}^- S_{y^{34}}^{\pm} S_{y^{56}}^{\pm} 
    \;
    \bar{S}_{\bar{\chi}^{34}}^+ \bar{S}_{\bar{\chi}^{56}}^- \bar{S}_{\bar{y}^{34}}^{\pm} \bar{S}_{\bar{y}^{56}}^{\pm}
\end{equation}
and in the $0$ picture by
\begin{equation}
    \frac{1}{4\sqrt{2}} \; S_{\chi^{34}}^+ S_{\chi^{56}}^- S_{y^{34}}^{\pm} S_{y^{56}}^{\pm} 
     \;
    \bar{S}_{\bar{\chi}^{34}}^+ \bar{S}_{\bar{\chi}^{56}}^- \bar{S}_{\bar{y}^{34}}^{\pm} \bar{S}_{\bar{y}^{56}}^{\pm}
    \left[
    \left( \omega^{3} + \omega^{4} \right) \left( \bar{\omega}^{3} + \bar{\omega}^{4} \right)
    +
    \left( \omega^{5} + \omega^{6} \right)\left( \bar{\omega}^{5} + \bar{\omega}^{6} \right)
    \right]    
\end{equation}

To compute the 2-point function for the  scalar we contract the spin fields using \eqref{correlator_twisted} 

\begin{aline}
    \frac{1}{2} \int & \frac{d^2 z}{\pi^2}  \;
    \langle S^+_{\chi^{34}} S^-_{\chi^{34}} \rangle
    \langle S^-_{\chi^{56}} S^+_{\chi^{56}} \rangle
    \langle S^-_{y^{34}} S^+_{y^{34}} \rangle
    \langle S^-_{y^{56}} S^+_{y^{56}} \rangle 
    \;
    \langle \bar{S}^+_{\bar{\chi}^{34}} \bar{S}^-_{\bar{\chi}^{34}} \rangle
    \langle \bar{S}^-_{\bar{\chi}^{56}} \bar{S}^+_{\bar{\chi}^{56}} \rangle
    \langle \bar{S}^-_{\bar{y}^{34}} \bar{S}^+_{\bar{y}^{34}} \rangle
    \langle \bar{S} ^-_{\bar{y}^{56}} \bar{S} ^+ _{\bar{y} ^{56}} \rangle
    \\
    & 
    \left[
    \left( \langle \omega^I \omega^J \rangle \;G_{IJ} \right)
    \left( \langle \bar{\omega}^I \bar{\omega}^J \rangle \;G_{IJ}\right)
    +
    \left( \langle \omega^M \omega^N \;G_{MN} \right)
    \left( \langle \bar{\omega}^M \bar{\omega}^N \rangle \;G_{MN}\right)
    \right]
    \\
    = & \; \frac{1}{2} \int \frac{d^2 z}{\pi^2}  \;  
    \frac{\vth_1^{'2} \left(0 \right)}{\vth_1 ^2 \left(z \right)}  
    \frac{\vthb_1^{'2} \left(0 \right)}{\vthb_1 ^2 \left(\bar{z} \right)}
    \left(
    \frac{\vth \smb{\zeta_2}{\delta_2}_{\omega^{34}} \left( z \right)}
    {\vth \smb{\zeta_2}{\delta_2}_{\omega^{34}} \left( 0 \right)}
    \;
    \frac{\vthb \smb{\zeta_2}{\delta_2}_{\bar{\omega}^{34}} \left( \bar{z} \right)}
    {\vthb \smb{\zeta_2}{\delta_2}_{\bar{\omega}^{34}} \left( 0 \right)}
    +
    \frac{\vth \smb{\zeta_3}{\delta_3}_{\omega^{56}} \left( z \right)}
    {\vth \smb{\zeta_3}{\delta_3}_{\omega^{56}} \left( 0 \right)}
    \;
    \frac{\vthb \smb{\zeta_3}{\delta_3}_{\bar{\omega}^{56}} \left( \bar{z} \right)}
    {\vthb \smb{\zeta_3}{\delta_3}_{\bar{\omega}^{56}} \left( 0 \right)}
    \right)\\
    & \times  
    \frac{\vth \smb{a+h_1}{b+g_1}_{\chi^{34}} \left( \frac{z}{2} \right)} 
    {\vth \smb{a+h_1}{b+g_1}_{\chi^{34}} \left( 0 \right)}
    \frac{\vth \smb{a-h_1-h_2}{b-g_1-g_2}_{\chi^{56}} \left( - \frac{z}{2} \right)}
    {\vth \smb{a-h_1-h_2}{b-g_1-g_2}_{\chi^{56}} \left( 0 \right)}
    \frac{\vth \smb{\zeta_2 +h_1}{\delta_2 +g_1}_{y^{34}} \left( - \frac{z}{2} \right)}
    {\vth \smb{\zeta_2 +h_1}{\delta_2 +g_1}_{y^{34}} \left( 0 \right)}
    \frac{\vth \smb{\zeta_3 +h_1+h_2}{\delta_3 +g_1+g_2}_{y^{56}} \left( - \frac{z}{2} \right)}
    {\vth \smb{\zeta_3 +h_1+h_2}{\delta_3 +g_1+g_2}_{y^{56}} \left( 0 \right)} \\
    & \times  
    \frac{\vthb \smb{k+h_1}{l+g_1}_{\bar{\chi}^{34}} \left( \frac{\bar{z}}{2} \right)} 
    {\vthb \smb{k+h_1}{l+g_1}_{\bar{\chi}^{34}} \left( 0 \right)}
    \frac{\vthb \smb{k-h_1-h_2}{l-g_1-g_2}_{\bar{\chi}^{56}} \left( - \frac{\bar{z}}{2} \right)}
    {\vthb \smb{k-h_1-h_2}{l-g_1-g_2}_{\bar{\chi}^{56}} \left( 0 \right)}
    \frac{\vthb \smb{\zeta_2 +h_1}{\delta_2 +g_1}_{\bar{y}^{34}} \left( - \frac{\bar{z}}{2} \right)}
    {\vthb \smb{\zeta_2 +h_1}{\delta_2 +g_1}_{\bar{y}^{34}} \left( 0 \right)}
    \frac{\vthb \smb{\zeta_3 +h_1+h_2}{\delta_3 +g_1+g_2}_{\bar{y}^{56}} \left( - \frac{\bar{z}}{2} \right)}
    {\vthb \smb{\zeta_3 +h_1+h_2}{\delta_3 +g_1+g_2}_{\bar{y}^{56}} \left( 0 \right)}  
\end{aline}
with $I,J = 3,4$, $M,N = 5,6$.

The integrand can be checked to be modular covariant and double periodic under $z \rightarrow z+1$ by redefining  $g_1 \rightarrow g_1 + 1$ and under $z \rightarrow z+\tau$ by redefining  $h_1 \rightarrow h_1 + 1$. Note that the $z$-independent denominators cancel by multiplying the 2-point function with the partition function. For this amplitude the summation over spin structures include odd ones, essential for periodicity. \\

Also for this integral, having a double pole at $z=0$ and double periodic over the torus we can rewrite it in terms of the Weierstrass function. By defining 
\begin{aline}  
    A_{tw} = & \frac{1}{2} \int d^2 z  \;  
    \left(
    \frac{\vth \smb{\zeta_2}{\delta_2}_{\omega^{34}} \left( z \right)}
    {\vth \smb{\zeta_2}{\delta_2}_{\omega^{34}} \left( 0 \right)}
    \;
    \frac{\vthb \smb{\zeta_2}{\delta_2}_{\bar{\omega}^{34}} \left( \bar{z} \right)}
    {\vthb \smb{\zeta_2}{\delta_2}_{\bar{\omega}^{34}} \left( 0 \right)}
    +
    \frac{\vth \smb{\zeta_3}{\delta_3}_{\omega^{56}} \left( z \right)}
    {\vth \smb{\zeta_3}{\delta_3}_{\omega^{56}} \left( 0 \right)}
    \;
    \frac{\vthb \smb{\zeta_3}{\delta_3}_{\bar{\omega}^{56}} \left( \bar{z} \right)}
    {\vthb \smb{\zeta_3}{\delta_3}_{\bar{\omega}^{56}} \left( 0 \right)}
    \right)\\
    & \times  
    \frac{\vth \smb{a+h_1}{b+g_1}_{\chi^{34}} \left( \frac{z}{2} \right)} 
    {\vth \smb{a+h_1}{b+g_1}_{\chi^{34}} \left( 0 \right)}
    \frac{\vth \smb{a-h_1-h_2}{b-g_1-g_2}_{\chi^{56}} \left( - \frac{z}{2} \right)}
    {\vth \smb{a-h_1-h_2}{b-g_1-g_2}_{\chi^{56}} \left( 0 \right)}
    \frac{\vth \smb{\zeta_2 +h_1}{\delta_2 +g_1}_{y^{34}} \left( - \frac{z}{2} \right)}
    {\vth \smb{\zeta_2 +h_1}{\delta_2 +g_1}_{y^{34}} \left( 0 \right)}
    \frac{\vth \smb{\zeta_3 +h_1+h_2}{\delta_3 +g_1+g_2}_{y^{56}} \left( - \frac{z}{2} \right)}
    {\vth \smb{\zeta_3 +h_1+h_2}{\delta_3 +g_1+g_2}_{y^{56}} \left( 0 \right)} \\
    & \times  
    \frac{\vthb \smb{k+h_1}{l+g_1}_{\bar{\chi}^{34}} \left( \frac{\bar{z}}{2} \right)} 
    {\vthb \smb{k+h_1}{l+g_1}_{\bar{\chi}^{34}} \left( 0 \right)}
    \frac{\vthb \smb{k-h_1-h_2}{l-g_1-g_2}_{\bar{\chi}^{56}} \left( - \frac{\bar{z}}{2} \right)}
    {\vthb \smb{k-h_1-h_2}{l-g_1-g_2}_{\bar{\chi}^{56}} \left( 0 \right)}
    \frac{\vthb \smb{\zeta_2 +h_1}{\delta_2 +g_1}_{\bar{y}^{34}} \left( - \frac{\bar{z}}{2} \right)}
    {\vthb \smb{\zeta_2 +h_1}{\delta_2 +g_1}_{\bar{y}^{34}} \left( 0 \right)}
    \frac{\vthb \smb{\zeta_3 +h_1+h_2}{\delta_3 +g_1+g_2}_{\bar{y}^{56}} \left( - \frac{\bar{z}}{2} \right)}
    {\vthb \smb{\zeta_3 +h_1+h_2}{\delta_3 +g_1+g_2}_{\bar{y}^{56}} \left( 0 \right)}  
\end{aline}

the 2-point function is given by
\begin{equation}
    \left\langle  \tau_2 \left( \frac{1}{2 \pi}\partial^2 _z + \frac{1}{\tau_2}   \right)  \left( \frac{1}{2 \pi} \bar{\partial}^2 _{\bar{z}} + \frac{1}{\tau_2} \right) A_{tw}|_{z,\bar{z} = 0}  \right\rangle 
\end{equation}
Also in this case the presence of the $\frac{1}{\tau_2}$ term signals a tadpole contribution, which is removed. The expressions for the scalars from the $\bm{b_1}+\bm{S}+\bm{\bar{S}} $ and $\bm{b_3}+\bm{S}+\bm{\bar{S}}$ sectors are IR divergent, and by removing the divergent piece one finds the following numerical one-loop masses for the different sectors
\begin{aline}
\label{eq:twisted}
    & \left\{ m^2 _{\bm{b_1}+\bm{S}+\bm{\bar{S}}} = m^2 _{\bm{b_1}+\bm{S}+\bm{\bar{S}}+\bm{T_2}} = m^2 _{\bm{b_1}+\bm{S}+\bm{\bar{S}}+\bm{T_3}} = m^2 _{\bm{b_1}+\bm{S}+\bm{\bar{S}}+\bm{T_2}+\bm{T_3}}\right\} \large{\mid}_\text{finite}
    = 0.031 %07839830930184 
    \text{ }M_s ^{2} g_s ^2\\ 
    & \left\{ m^2 _{\bm{b_3}+\bm{S}+\bm{\bar{S}}} = m^2 _{\bm{b_3}+\bm{S}+\bm{\bar{S}}+\bm{T_1}} = m^2 _{\bm{b_3}+\bm{S}+\bm{\bar{S}}+\bm{T_2}} = m^2 _{\bm{b_3}+\bm{S}+\bm{\bar{S}}+\bm{T_1}+\bm{T_2}}\right\} \large{\mid}_\text{finite}
    = 0.031 %07839830930184 
    \text{ }M_s ^{2} g_s ^2\\ 
\end{aline}

By fermionising the double charged scalars 
\begin{aline}
\frac{1}{2} \; &  k_L ^2 \cdot \boldsymbol{\chi^{34}} \left( e^{i k_L ^2 \cdot \boldsymbol{X^{34}}} - e^{-i k_L ^2 \cdot \boldsymbol{X^{34}}} \right)  \; k_R ^3 \cdot \boldsymbol{\bar{\chi}^{56}} \left( e^{i k_R ^3 \cdot \boldsymbol{\bar{X}^{56}}} - e^{-i k_R ^3 \cdot \boldsymbol{\bar{X}^{56}}} \right) \\
& \sim
\chi^{I} y^{I} \;
\bar{\chi}^{J} \bar{y}^{J}, \quad I = 3,4, \quad J = 5,6
\end{aline}
the relevant one loop Feynman diagrams for the $\bm{b_1}+\bm{S}+\bm{\bar{S}} $ twisted fields' mass the IR limit are given as follows

\begin{figure}[H]
    \centering
    \includegraphics[width=0.6\linewidth]{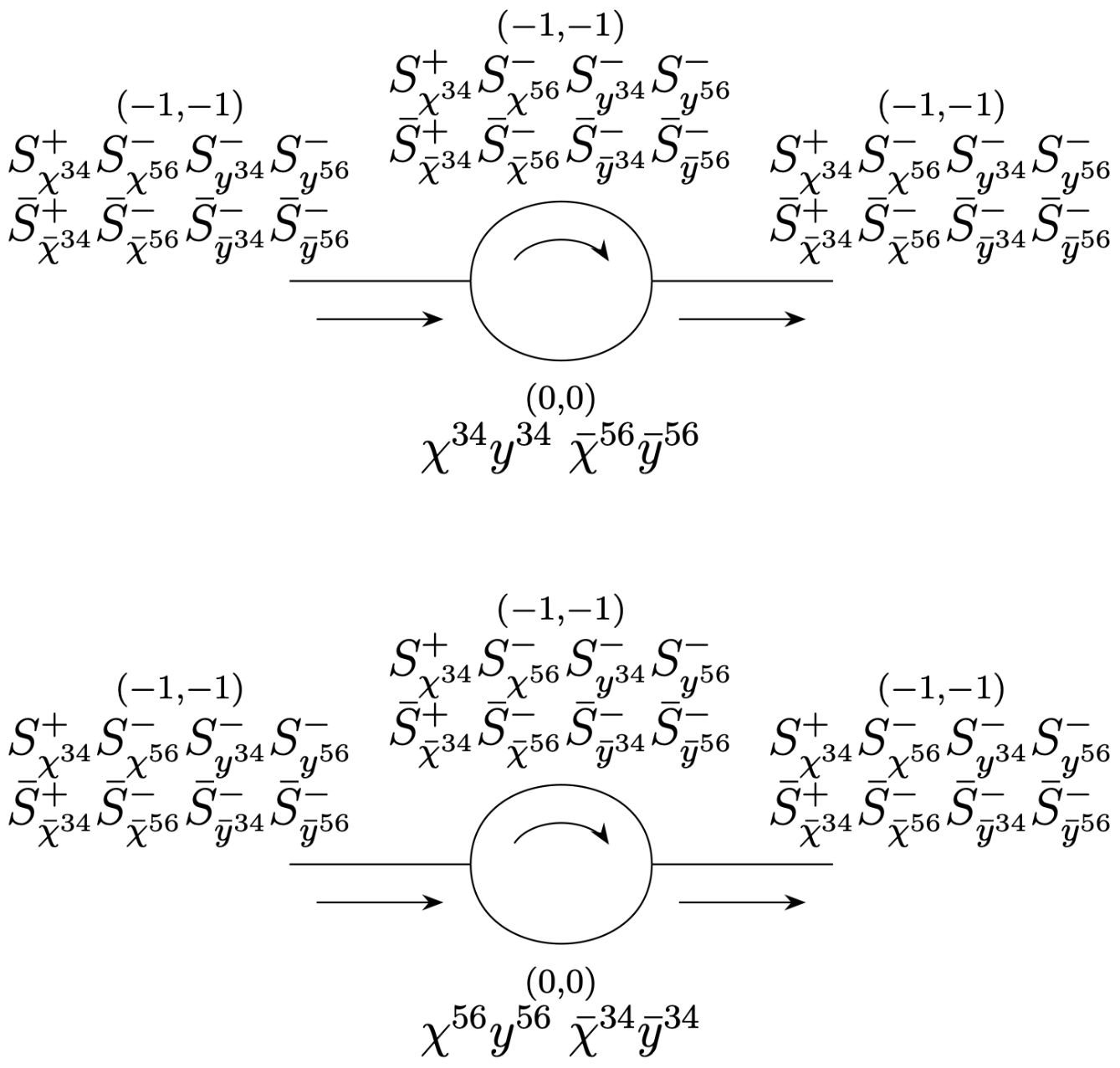}
    \captionsetup{justification=centering}
    \caption{IR divergent One loop Feynman diagrams for $\eqref{eq:twisted}$}
    \label{fig:Feynman_2}
\end{figure}

Note that the spin field brings a $\frac{1}{2}$ fermionic charge, while the double charged scalars does not have a fermionic charge being real.  \\
Note also that the same tree-level 3-scalar coupling twisted-twisted-untwisted enters in the evaluation of the double charged scalar masses but it does not bring divergences, since its contribution is cancelled by an additional contribution of propagating twisted scalars with opposite relative charges. \\

The masses for the scalars from the $\bm{b_2} + \bm{S} + \bm{\bar{S}}$ sector are instead  finite, since there is no scalar with Left internal momentum from the first torus and Right internal momentum from the third torus, or vice-versa, which would give the divergent tree-level coupling with the twisted states. The finite one-loop masses are then given by 
\begin{aline} 
    & m^2 _{\bm{b_2}+\bm{S}+\bm{\bar{S}}} = m^2 _{\bm{b_2}+\bm{S}+\bm{\bar{S}}+\bm{T_1}} = m^2 _{\bm{b_2}+\bm{S}+\bm{\bar{S}}+\bm{T_3}} = m^2 _{\bm{b_2}+\bm{S}+\bm{\bar{S}}+\bm{T_1}+\bm{T_3}} = 0.03116207447899192 \text{ }M_s ^{2} g_s ^2\\ 
\end{aline}

%Twisted scalar from  $\bm{b_1}$ and $\bm{b_1} + \Sv$
%\begin{equation}
%    e^{\frac{i \psi^\mu}{2} + \frac{i \chi^{12}}{2} + \frac{i y^{34}}{2} + \frac{iy^{56} }{2}} \; e^{\frac{i \bar{\psi}^\mu}{2} + \frac{i \bar{\chi}^{12}}{2} + \frac{i \bar{y}^{34}}{2} + \frac{i \bar{y}^{56} }{2}} 
%\end{equation}
%\begin{equation}
%    e^{\frac{i}{2} \left( \psi^\mu + \chi^{12} + y^{34} + y^{56} \right)} \; e^{\frac{i}{2} \left( \bar{\psi}^\mu + \bar{\chi}^{12} + \bar{y}^{34} + \bar{y}^{56} \right)}
%\end{equation}

We focus now to the divergent term for both, the double charged scalars as well as the twisted scalars. The divergent piece
\begin{equation}
    I_\text{div}=\alpha  \int_{1} ^{\infty} d \tau_2  \; \tau_2 ^{-1}  
\end{equation}
with $\alpha$ a positive constant given by the $q$-expansion of the corresponding amplitude, can be regularised 
by considering a small non-zero external momentum $p^2$. Using the correspondence of the modular parameter $\tau_2$ with the one entering in the Schwinger representation of the internal propagators, we get the regularised form of $I_\text{div}$ (in string units)
\begin{equation}
    I_\text{reg} = \alpha \int_{1} ^{\infty} d \tau_2  \; \tau_2 ^{-1}  e^{- \tau_2 | p^2 |}
\end{equation}
or by changing variable $\tau_2 | p^2 | \rightarrow \tau_2 $
\begin{equation}
   I_\text{reg} = \alpha \int_{| p^2 |} ^{\infty} d \tau_2  \; \tau_2 ^{-1}  e^{- \tau_2} = 
    \alpha 
    \; \Gamma \left( 0, | p^2 | \right)
\end{equation}
with $\Gamma$ the incomplete Gamma function. Using the expansion
\begin{equation}
    \Gamma \left(0, x \right) = - \gamma - \ln x + \sum_{n=1} ^{\infty} \frac{\left(-1\right)^n x^n}{n \cdot n!} \sim - \gamma - \ln x
\end{equation}
with $\gamma$ the Euler-Mascheroni constant, one finds that the one loop contribution becomes in the limit $| p^2 |\to 0$:
\begin{equation} 
  I_1= I_\text{fin}  - \alpha \gamma - \alpha \ln | p^2 | = - \alpha \ln \frac{|p^2|}{\mu^2}
\end{equation}
with $\mu ^2 = e^{- \gamma+I_{fin}/\alpha}$, where $I_\text{fin}$ stands for the finite contribution computed in equations \eqref{mfT1}, \eqref{mfT2}, \eqref{eq:twisted}. \\

The one loop mass is then given by the zeroes of the self-energy comprising the one-loop and tree-level contribution of the 2-point function:
\begin{equation}
    I = I_\text{tree}+I_1 = p^2 - \alpha \ln \frac{|p^2|}{\mu^2} = - m^2 - \alpha \ln \frac{m^2}{\mu^2}
\end{equation}
We find
\begin{equation}
    \alpha \ln \frac{m^2}{\mu^2} + m^2 = 0 
    \quad \rightarrow \quad 
    \frac{m^2}{\mu^2} = e^{\frac{-m^2}{ \alpha}} 
    \quad \rightarrow \quad 
    \frac{m^2}{\alpha} \;  e^{\frac{m^2}{\alpha}} = \frac{\mu^2}{\alpha}
\end{equation}
For the double charged scalars $\eqref{double_charged_1_2}$ the coefficient $\alpha = 16 \pi^2 g_s^2$, for the double charged scalars $\eqref{double_charged_2_1}$ the coefficient $\alpha = 8 \pi^2 g_s^2$, for the twisted scalars $\alpha = \frac{3}{4} \pi^2 g_s^2$. In all cases then $\frac{\mu^2}{\alpha} > 0$ and the unique solution is given by the Lambert function $W_0$:
\begin{equation}
    \frac{m^2}{\alpha} = W_0 \left( \frac{\mu^2}{\alpha} \right)
    \quad \rightarrow \quad 
    m^2 = \alpha  W_0 \left( \frac{\mu^2}{\alpha} \right) > 0
\end{equation}

%\begin{figure}[H]
%    \centering
%    \subfigure[]{\includegraphics[width=0.49\linewidth]{Lambert_1a.png}} 
%    \subfigure[]{\includegraphics[width=0.49
%    \linewidth]{Lambert_1b.png}} 
%    \captionsetup{justification=centering}
%    \caption{Plot of the functions $- \alpha \ln \frac{m^2}{\mu^2}$ and $m^2$ as a function of $m^2$ (left), with the zoomed plot (right) with $\alpha = 16 \pi^2$.}       
%    \label{fig:Lambert_1}
%\end{figure}

%\begin{figure}[H]
%    \centering
%    \subfigure[]{\includegraphics[width=0.49\linewidth]{Lambert_2a.png}} 
%    \subfigure[]{\includegraphics[width=0.49\linewidth]{Lambert_2b.png}} 
%    \captionsetup{justification=centering}
%    \caption{Plot of the functions $\frac{m^2}{\mu^2} e^{\frac{m^2}{\alpha}} -1$ as a function of $m^2$ (left), with the zoomed plot (right)  with $\alpha = 16 \pi^2$.}        \label{fig:Lambert_2}
%\end{figure}
The expansion of the Lambert function  $W_0 \left(\frac{1}{\alpha} \right)$ for $\alpha \rightarrow 0$ is
\begin{equation}
    W_0 \left(\frac{1}{\alpha} \right) = \ln \frac{1}{\alpha}  -  \ln \ln \frac{1}{\alpha} + \dots
\end{equation}
such that $\alpha W_0 \left( \frac{\mu^2}{\alpha} \right) \rightarrow 0 $ as $\alpha \rightarrow 0$ in the weak string coupling limit.
\begin{figure}[H]
    \centering
    \subfigure[]{\includegraphics[width=0.61\linewidth]{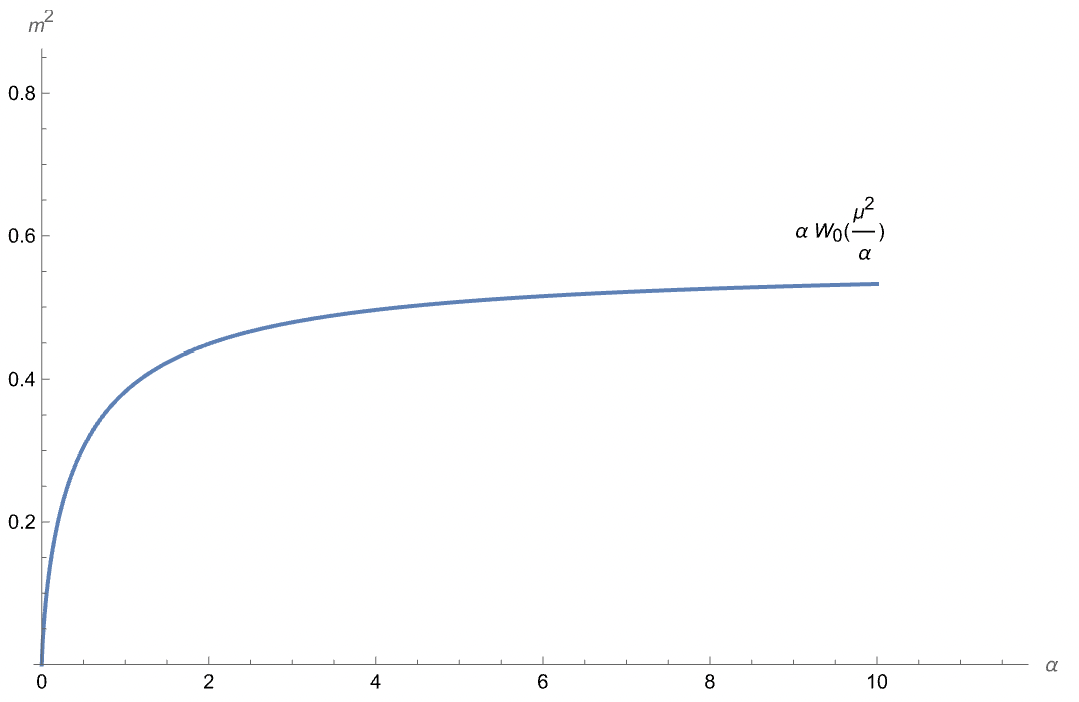}}  
    \captionsetup{justification=centering}
    \caption{Plot of the squared mass as a function of $\alpha$.}        \label{fig:Lambert_2}
\end{figure}

\section{Conclusions}

%\section{Dilaton stabilisation}

Interest in non-supersymmetric string vacua arose from the early days of string phenomenology in the mid-eighties 
\cite{dh, AGMV, kltclas, gv, itoyama} 
and continued in earnest in more recent years
\cite{bglr, adm, aafs, spwsp, fmpPS, fmpFSU5}. 
While the $SO(16)\times SO(16)$ heterotic-string in ten dimensions is
tachyon-free, the non-supersymmetric string vacua compactified to
four dimensions generically are connected to points in the moduli
space that give rise to physical tachyons, which indicates that these 
non-supersymmetric string vacua are not stable. The main issue
concerning the non-supersymmetric string vacua is their stability. In
general, it is possible to project the physical tachyons using the
Generalised GSO projections at special points in the moduli space. 
However, given that the moduli fields in these vacua are generically not fixed, 
entails that the vacua are connected to unstable points in the moduli space. 

In this paper, we found a specific four dimensional configuration in Type II closed string theory which is tachyon-free at tree-level. Notably, this property seems to be related to the way in which the Scherk-Schwarz supersymmetry breaking mechanism is implemented, suggesting that this mechanism plays a crucial role in determining whether a configuration remains free of tachyons at tree level. 
Another way to address this problem is to use asymmetric boundary conditions to 
project out the moduli fields \cite{moduli}. It is possible to find 
string vacua in which all the moduli of the internal compactified 
dimensions are projected out. In these cases, it can be argued that 
the internal space is frozen and the vacua are "stable"
\cite{Angelantonj:2006ut, stable, Baykara:2024vss, Basaad:2024lno, Angelantonj:2024jtu, Aldazabal:2025zht}.

There are several caveats to these arguments for the potential stability
of string vacua with broken supersymmetry. 
The first concerns stability beyond tree-level, at higher loop--orders, which we also examined in this paper. 
The stability claim is typically made at leading order 
in the perturbative string expansion. However, given that the leading order parameters are not protected by supersymmetric non-renormalisation arguments in the vacua without supersymmetry, loop--order corrections to massless scalar 
states may give rise to tachyonic instabilities. We demonstrated that this does
not happen at one-loop in the specific vacuum that we analysed in detail in the
current paper. 
This is the first demonstration of this kind and provides a proof of concept that a stable vacuum can be realised in a perturbatively controlled way. 
 
The second caveat concerns the issue of dilaton stabilisation. 
In Section~2.4, we found that the minimum of the scalar potential at the self-dual point has negative energy. This does not imply that the vacuum is AdS because a cosmological constant in the string frame becomes a runaway dilaton potential in the Einstein frame which vanishes from below at zero string coupling. A way to stabilise the dilaton  
%is by adding a tree-level potential which is equivalent of going `slightly' off criticality. The resulting potential in the Einstein frame reads:
%\begin{aline}\label{dilatonpot}
%V(\phi) = {1\over 2}\delta\hat{c}\, e^{\phi} + \Lambda\, e^{2\phi}\,,
%\end{aline}
%where $\phi$ is the canonically normalised dilaton, whose vacuum expectation value determines the string coupling $g_s$,$e^{\langle\phi\rangle}=g_s^2$, and $\delta\hat{c}$ is the central charge deficit, so that the total central charge is $c={3\over 2}(10-\delta\hat{c})$. The above potential has a minimum at
%\begin{aline}\label{dilatonpotmin}
%\langle e^{\phi}\rangle=-{1\over 4}{\delta\hat{c}\over\Lambda}\quad;\quad V_\text{min}=-{1\over 16}{(\delta\hat{c})^2\over\Lambda}
%\end{aline}
on the type IIB side is by turning on 3-form fluxes along the universal holomorphic $(3,0)$ cycles of $T^6$ from the NS-NS (Neuveu-Schwarz) and R-R (Ramond) sectors and their magnetic duals. The price to pay is the generation of a positive 3-brane charge that can be canceled by considering an appropriate orientifold of this model. This can also stabilise the twisted moduli at the orbifold point~\cite{Antoniadis:2024hqd}. It would be very interesting to do such an analysis in order to conclude on the existence of a perturbatively stable
non-supersymmetric string vacuum on AdS but it goes beyond the scope of this work.
We further note that 
an alternative perturbative method of dilaton stabilisation was proposed
in the context of `super no-scale’ models \cite{ivano}.

Finally, it ought to be remarked that while string theory provides a mundane
extension of the point-particle quantum field theories, which is consistent
with perturbative quantum gravity, our understanding of it is still
very rudimentary. In the current paper and in most of the contemporary
literature on non-supersymmetric string vacua, the discussion revolves 
about tachyon-free configurations. Ultimately, an improved understanding of 
the tachyonic vacua, as in \cite{Kaidi}, may prove to be important for 
understanding the string dynamics in the early universe. 

\section*{Acknowledgments}
AEF and ARDA would like to thank the LPTHE
Sorbonne Universit\'e for hospitality and support. AEF acknowledges the HEP Research Unit of Faculty of Science in Chulalongkorn University for support and hospitality.
The work of ARDA is supported in part by EPSRC grant EP/T517975/1.
I.A. is supported by the Second Century Fund (C2F), Chulalongkorn University.
This work was supported by a Royal Society Exchange grant IES/R1/221199, ``Phenomenological studies of string vacua".

\appendix

\section{Theta Functions}
\label{appendix:Theta Functions}
We give the definition of the functions used for calculations.\\
The Jacobi Theta function is defined as follows
\begin{equation}
    \vth\smb{a}{b} \left(z|\tau\right) = \sum_{n \in \mathbb{Z}} q^{\frac{1}{2} \left(  n- \frac{a}{2}\right)^2} e^{2 \pi i \left( z - \frac{b}{2} \right)  \left( n- \frac{a}{2} \right)}
\end{equation}
with $q = e^{2 \pi i \tau}$. The Dedekind $\eta$ function is defined as 
\begin{equation}
    \eta = q^{\frac{1}{24}} \prod_{n=1} ^\infty \left(1-q^n\right)
\end{equation}
Under modular transformations
\begin{aline}
    & \vth\smb{a}{b} \left(z|\tau+1\right) = e^{- \frac{i \pi }{4} a \left(a-2\right)} \; \vth\smb{a}{a+b-1} \left(z|\tau\right) \\
    & \vth\smb{a}{b} \left(\frac{z}{\tau}|-\frac{1}{\tau}\right) = \sqrt{-i\tau} \;e^{\frac{i \pi}{2}ab+i \pi \frac{z^2}{\tau}}\; \vth\smb{b}{-a} \left(z|\tau\right) \\
    & \eta\left(\tau+1\right) = e^{\frac{i \pi}{12}} \; \eta\left(\tau\right) \\
    & \eta\left(- \frac{1}{\tau}\right) = \sqrt{-i \tau} \; \eta\left(\tau\right) \\
\end{aline}
Under periodic transformation along the torus
\begin{aline}
    \vth\smb{a}{b} \left(z + z_1 + z_2 \tau |\tau\right) = e^{-i \pi z_2 \left(2z - b\right) - 2 \pi i z_1 z_2} \; q^{- z_2 ^2} \; \vth\smb{a- 2 z_2}{b-2 z_1} \left(z|\tau\right) 
\end{aline}
The Weierstrass function is defined as
\begin{equation}
    \wp \left(z \right) = - \frac{\pi^2 }{3} E_2 - \partial_z ^2 \log \vth_1 \left(z\right) = \frac{1}{z^2} + \mathcal{O}\left(z^2\right)
\end{equation}
with the Eisenstein series $E_2$ 
\begin{equation}
    E_2 = \frac{12}{i \pi} \partial_\tau \log \eta = - \frac{1}{\pi^2} \frac{\vth_1 ^{'''}}{\vth_1 ^{'}}
\end{equation}
and following modular transformations
\begin{aline}
    & \wp \left(z | \tau+1\right) = \wp \left(z|\tau \right) \\
    & \wp \left(\frac{z}{\tau} | - \frac{1}{\tau}\right) = \tau^2 \; \wp \left(z|\tau \right) 
\end{aline}
and periodicity transformations
\begin{aline}
    & \wp \left(z +1\right) = \wp \left(z +\tau\right) = \wp \left(z \right) 
\end{aline}
The multidimensional Poisson resummation formula is 
\begin{equation}
    \label{Poisson}
    \sum_{m_i \in \mathbb{Z}} e^{- \pi m_i m_j A_{ij}+ \pi B_i m_i} = \frac{1}{\sqrt{\det A}} \sum_{m_i \in \mathbb{Z}} e^{-\pi \left( m_i + \frac{i B_i}{2} \right) A^{-1} _{ij} \left( m_j +  \frac{i B_j}{2} \right)}
\end{equation}

Below are some useful world-sheet integrals used for the computation of one-loop amplitudes
\begin{equation}
    \int d^2z \; \partial^2 \log \frac{\vth_1 \left(z\right)}{\vth_1 ' \left(0\right)} = - \pi
    \label{integral_1}
\end{equation}
\begin{equation}
    \int d^2 z \;  \partial^2 \log \frac{\vth_1 \left(z\right)}{\vth_1 ' \left(0\right)}  \; \bar{\partial}^2 \log \frac{\vthb_1\left(\bar{z}\right)}{\vthb_1 '\left(0\right)}  = \frac{\pi^2}{\tau_2}
    \label{integral_2}
\end{equation}
\begin{equation}
    \int d^2z \; \wp = \frac{\tau_2}{3} \frac{\vth_1 '''}{\vth_1 '} + \pi 
\end{equation}
\begin{equation}
    \int d^2z \; \wp \; \bar{\wp} = \frac{\pi^2}{\tau_2 } + \frac{\pi}{3} \frac{\vth_1 '''}{\vth_1 '} + \frac{\pi}{3} \frac{\vthb_1 '''}{\vthb_1 '} + \frac{\tau_2}{9 } \frac{\vth_1 '''}{\vth_1 '} \frac{\vthb_1 '''}{\vthb_1 '}
\end{equation}

\section{Supersymmetric Massless spectrum}

With the basis vectors definded by $\{\one,\Sv,\bar{\Sv},\bv{1},\bv{2} \}$, this model corresponds to a $\mathcal{N} = 2$ symmetric $\mathbb{Z}_2 \times \mathbb{Z}_2$ symmetric orbifold in which six gravitinos are projected out. The spectrum can be read from the GSO phases. In particular the massless untwisted states are given by the following sectors

\begin{center}
\begin{tabular}{ l l }
    $
    \begin{aligned}[t]
    & \mathbf{NS} \text{ sector} \\
    & \psi^\mu \otimes \bar{\psi}^\nu \; \text{ graviton} \\
    & \psi^\mu \otimes \bar{\psi}^{\nu *} \; \text{dilaton pseudoscalar} \\
    & \begin{rcases*}
        \chi^{1,2} \otimes \bar{\chi}^{1,2} \;\\
        \chi^{3,4} \otimes \bar{\chi}^{3,4} \;\\
        \chi^{5,6} \otimes \bar{\chi}^{5,6} \;\\
    \end{rcases*} \parbox{8em}{ \raggedright complex structure moduli $U^1$,$U^2$,$U^3$} \\
    & \begin{rcases*}
        \chi^{1,2} \otimes \bar{\chi}^{1,2 *} \;\\
        \chi^{3,4} \otimes \bar{\chi}^{3,4 *} \;\\
        \chi^{5,6} \otimes \bar{\chi}^{5,6 *} \;\\
    \end{rcases*}  \parbox{8em}{ \raggedright K\"ahler moduli $T^1$,$T^2$,$T^3$} 
    \end{aligned}$
&
    $
    \begin{aligned}[t]
    &\mathbf{S+\bar{S}} \text{ sector} \\
    &\begin{rcases*}
        \ket{++++} \otimes \ket{++++}  \;\\
        \ket{++++} \otimes \ket{----} \;\\
        \ket{++--} \otimes \ket{++--} \;\\
        \ket{++--} \otimes \ket{--++} \;\\
    \end{rcases*} \parbox{7em}{ \raggedright four Gauge vector bosons} \\
    &\begin{rcases*}
        \ket{+-+-} \otimes \ket{+-+-}  \;\\
        \ket{+-+-} \otimes \ket{-+-+} \;\\
        \ket{+--+} \otimes \ket{+--+} \;\\
        \ket{+--+} \otimes \ket{-++-} \;\\
    \end{rcases*} \parbox{7em}{ \raggedright four complex scalars} \\
    \end{aligned}$
\end{tabular}
\end{center}

\begin{center}
\begin{tabular}{ l l }
    $
    \begin{aligned}[t]
    &\mathbf{S} \text{ sector} \\
    &\ket{+--+} \otimes \bar{\psi}^\mu \; \parbox{7em}{ \raggedright gravitino} \\
    &\ket{+--+} \otimes \bar{\psi}^{\mu*} \; \parbox{7em}{ \raggedright dilatino} \\
    &\begin{rcases*}
        \ket{+-+-} \otimes \bar{\chi}^{1,2} \;\\
        \ket{++--} \otimes \bar{\chi}^{3,4} \;\\
        \ket{++++} \otimes \bar{\chi}^{5,6} \;\\
        \ket{+-+-} \otimes \bar{\chi}^{1,2*} \;\\
        \ket{++--} \otimes \bar{\chi}^{3,4*} \;\\
        \ket{++++} \otimes \bar{\chi}^{5,6*} \;\\
    \end{rcases*} \parbox{7em}{ \raggedright six spin-1/2 fermions} \\ 
    \end{aligned}
    $
&
    $
    \begin{aligned}[t]
    &\mathbf{\bar{S}} \text{ sector} \\
    &\psi^\mu \otimes  \ket{+--+}  \; \parbox{7em}{ \raggedright gravitino} \\
    &\psi^{\mu*} \otimes  \ket{+--+}\; \parbox{7em}{ \raggedright dilatino} \\
    &\begin{rcases*}
        \chi^{1,2} \otimes \ket{+-+-}  \;\\
        \chi^{3,4} \otimes \ket{++--} \;\\
        \chi^{5,6} \otimes \ket{++++} \;\\
        \chi^{1,2*} \otimes \ket{+-+-}  \;\\
        \chi^{3,4*} \otimes \ket{++--} \;\\
        \chi^{5,6*} \otimes \ket{++++} \;\\
    \end{rcases*} \parbox{7em}{ \raggedright  six spin-$1/2$ fermions} \\
    \end{aligned}
    $\\
\end{tabular}
\end{center}

\noindent
It is understood all states come into pair with its complex conjugate, where $\ket{\pm \pm \pm \pm}^* = \ket{\mp \mp \mp \mp}$. \\

\noindent
The spectrum is also given in the $-1/2$ bosonic vertex operator representation. Introducing the $SU(2)$ R-parity currents $e^{\pm \frac{i}{2}H_i}$,$e^{\pm \frac{i}{2}\bar{H}_i}$, $i=0,1,2,3$ with
\begin{equation}
    \partial H_0 = \psi^1\psi^2, \;\;\; \partial H_1 = \chi^1\chi^2, \;\;\; \partial H_2 = \chi^3\chi^4, \;\;\; \partial H_3 = \chi^5\chi^6  
\end{equation}
\begin{equation}
    \partial \bar{H}_0 = \bar{\psi}^1\bar{\psi}^2, \;\;\; \partial \bar{H}_1 = \bar{\chi}^1\bar{\chi}^2, \;\;\; \partial \bar{H}_2 = \bar{\chi}^3\bar{\chi}^4, \;\;\; \partial \bar{H}_3 = \bar{\chi}^5\bar{\chi}^6  
\end{equation}
the states are represented as follows\\

\begin{center}
\hspace{-1cm}
\begin{tabular}{ p{6cm} p{0.5cm} p{6cm} }
    $
    \begin{aligned}[t]
    & \mathbf{NS} \text{ sector} \\
    & e^{\pm i\left( H_0 +\bar{H}_0 \right)} \; \text{ \raggedright graviton} \\
    & e^{\pm i\left( H_0 -\bar{H}_0 \right)} \; \text{ \raggedright dilaton pseudoscalar} \\
    & e^{\pm i\left( H_i +\bar{H}_i \right)} \; \text{ \raggedright structure moduli $U^i$} \\
    & e^{\pm i\left( H_i -\bar{H}_i \right)} \; \text{ \raggedright K\"ahler moduli $T^i$} \\
    \end{aligned}
    $
& &
    $
    \begin{aligned}[t]
    & \mathbf{S+\bar{S}} \text{ sector} \\
    & \begin{rcases*}
        e^{\pm\frac{i}{2}\left(H_0 + H_1 + H_2 + H_3 + \bar{H}_0 + \bar{H}_1 + \bar{H}_2 + \bar{H}_3 \right)}  \;\\  
        e^{\pm\frac{i}{2}\left(H_0 + H_1 - H_2 - H_3 + \bar{H}_0 + \bar{H}_1 - \bar{H}_2 - \bar{H}_3 \right)} \;\\
        e^{\pm\frac{i}{2}\left(H_0 - H_1 + H_2 - H_3 + \bar{H}_0 - \bar{H}_1 + \bar{H}_2 - \bar{H}_3 \right)}  \;\\
        e^{\pm\frac{i}{2}\left(H_0 - H_1 - H_2 + H_3 + \bar{H}_0 - \bar{H}_1 - \bar{H}_2 + \bar{H}_3 \right)} \;\\
    \end{rcases*} \parbox{6em}{ \raggedright 4 Gauge  bosons} \\
    & \begin{rcases*}
        e^{\pm\frac{i}{2}\left(H_0 + H_1 - H_2 - H_3 - \bar{H}_0 - \bar{H}_1 + \bar{H}_2 + \bar{H}_3 \right)} \;\\
        e^{\pm\frac{i}{2}\left(H_0 + H_1 + H_2 + H_3 - \bar{H}_0 - \bar{H}_1 - \bar{H}_2 - \bar{H}_3 \right)} \;\\
        e^{\pm\frac{i}{2}\left(H_0 - H_1 + H_2 - H_3 - \bar{H}_0 + \bar{H}_1 - \bar{H}_2 + \bar{H}_3 \right)} \;\\
        e^{\pm\frac{i}{2}\left(H_0 - H_1 - H_2 + H_3 - \bar{H}_0 + \bar{H}_1 + \bar{H}_2 - \bar{H}_3 \right)} \;\\
    \end{rcases*} \parbox{7em}{ \raggedright 4 complex scalars} \\
    \end{aligned}
    $
\end{tabular}
\end{center}

\begin{center}
\hspace{-1cm}
\begin{tabular}{ p{6cm} p{0.5cm} p{6cm} }
    $
    \begin{aligned}[t]
    & \mathbf{S} \text{ sector} \\
    & e^{\pm\frac{i}{2}\left( 2\bar{H}_0 + H_0 - H_1 - H_2 + H_3 \right)} \; \parbox{7em}{ \raggedright  gravitino} \\
    & e^{\pm\frac{i}{2}\left( - 2\bar{H}_0 + H_0 - H_1 - H_2 + H_3 \right)} \; \parbox{7em}{ \raggedright  dilatino} \\
    & \begin{rcases*}
        e^{\pm i \bar{H}_1}e^{\pm\frac{i}{2}\left(H_0 - H_1 + H_2 - H_3 \right)}  \;\\
        e^{\pm i \bar{H}_2}e^{\pm\frac{i}{2}\left( H_0 + H_1 - H_2 - H_3 \right)} \;\\
        e^{\pm i \bar{H}_3}e^{\pm\frac{i}{2}\left(H_0 + H_1 + H_2 + H_3 \right)} \;\\
    \end{rcases*}  \parbox{7em}{ \raggedright  6 spin-$1/2$ fermions} \\
    \end{aligned}
    $
& &
    $
    \begin{aligned}[t]
    & \mathbf{\bar{S}} \text{ sector} \\
    & e^{\pm\frac{i}{2}\left( 2H_0 + \bar{H}_0 - \bar{H}_1 - \bar{H}_2 + \bar{H}_3 \right)} \; \parbox{7em}{ \raggedright gravitino} \\
    & e^{\pm\frac{i}{2}\left(- 2H_0 + \bar{H}_0 - \bar{H}_1 - \bar{H}_2 + \bar{H}_3 \right)} \; \parbox{7em}{ \raggedright dilatino} \\
    & \begin{rcases*}
        e^{\pm i H_1}e^{\pm\frac{i}{2}\left(\bar{H}_0 - \bar{H}_1 + \bar{H}_2 - \bar{H}_3 \right)}  \;\\
        e^{\pm i H_2}e^{\pm\frac{i}{2}\left(\bar{H}_0 + \bar{H}_1 - \bar{H}_2 - \bar{H}_3 \right)} \;\\
        e^{\pm i H_3}e^{\pm\frac{i}{2}\left(\bar{H}_0 + \bar{H}_1 + \bar{H}_2 + \bar{H}_3 \right)} \;\\
    \end{rcases*}  \parbox{7em}{ \raggedright 6 spin-$1/2$ fermions} \\
    \end{aligned}
    $\\
\end{tabular}
\end{center}

From the representation of the gravitons and gravitinos we can read the generators of the $\mathcal{N} = 2$ supersymmetry 
\begin{equation}
    Q=e^{-\frac{i}{2}\left( H_0 + H_1 + H_2 - H_3 \right)}, \;\;\; \bar{Q}=e^{-\frac{i}{2}\left( \bar{H}_0 + \bar{H}_1 + \bar{H}_2 - \bar{H}_3 \right)} 
\end{equation}
The $J^\pm$ operators 
\begin{equation}
    J^+ = e^{\frac{i}{2}\left(H_0 + H_1 +H_2 - H_3 \right)} e^{-\frac{i}{2}\left(\bar{H}_0 + \bar{H}_1 +\bar{H}_2 -\bar{H}_3 \right)}
\end{equation}
with $J^- \equiv J^{+*}$, such that
\begin{equation}
    J^+ Q \rightarrow \bar{Q},  \;\;\; J^- \bar{Q} \rightarrow Q
\end{equation}
destroy the holomorphic (antiholomorphic) charge and create the antiholomorphic (holomorphic) charge. Thus $J^+$, $J^-$, $J^0 = J^+ J^-$ are the generators of the $SU(2)$ algebra that rotates the two supercharges.\\

\noindent
The states fit into the $\mathcal{N}=2$ representation as follows
\begin{itemize}

    \item[] Gravity multiplet \vspace{0.3cm} \\ 
  $e^{\pm i\left( H_0 +\bar{H}_0 \right)}$ \; graviton \\
  $\begin{rcases*}
        e^{\pm\frac{i}{2}\left( 2\bar{H}_0 + H_0 - H_1 - H_2 +H_3 \right)} \;\\
        e^{\pm\frac{i}{2}\left( 2H_0 + \bar{H}_0 - \bar{H}_1 - \bar{H}_2 +\bar{H}_3 \right)} \;\\
    \end{rcases*}$ two gravitinos \\
  $e^{\pm\frac{i}{2}\left(H_0 - H_1 - H_2 + H_3 + \bar{H}_0 - \bar{H}_1 - \bar{H}_2 + \bar{H}_3 \right)}$ \; Vector  
    
    \item[] Vector multiplet \vspace{0.3cm} \\ 
    $\begin{rcases*}
        e^{\pm\frac{i}{2}\left(H_0 + H_1 + H_2 + H_3 + \bar{H}_0 + \bar{H}_1 + \bar{H}_2 + \bar{H}_3 \right)}  \;\\  
        e^{\pm\frac{i}{2}\left(H_0 + H_1 - H_2 - H_3 + \bar{H}_0 + \bar{H}_1 - \bar{H}_2 - \bar{H}_3 \right)} \;\\
        e^{\pm\frac{i}{2}\left(H_0 - H_1 + H_2 - H_3 + \bar{H}_0 - \bar{H}_1 + \bar{H}_2 - \bar{H}_3 \right)}  \;\\
    \end{rcases*}$ three Gauge vector bosons \\
    $\begin{rcases*}
        e^{\pm\frac{i}{2}\left(2\bar{H}_1+H_0 - H_1 + H_2 - H_3 \right)}  \;\\
        e^{\pm\frac{i}{2}\left( 2 \bar{H}_2 + H_0 + H_1 - H_2 - H_3 \right)} \;\\
        e^{\pm\frac{i}{2}\left(2 \bar{H}_3 + H_0 + H_1 + H_2 + H_3 \right)} \; \\
        e^{\pm\frac{i}{2}\left(2 H_1 +\bar{H}_0 - \bar{H}_1 + \bar{H}_2 - \bar{H}_3 \right)}  \;\\
        e^{\pm\frac{i}{2}\left(2 H_2 + \bar{H}_0 + \bar{H}_1 - \bar{H}_2 - \bar{H}_3 \right)} \;\\
        e^{\pm\frac{i}{2}\left(2 H_3 + \bar{H}_0 + \bar{H}_1 + \bar{H}_2 + \bar{H}_3 \right)} \;\\
    \end{rcases*}$ six spin-$1/2$ fermions \\
    $e^{\pm i\left( H_i +\bar{H}_i \right)}$ \; $i=1,2,3$, complex structure moduli $U^1$,$U^2$,$U^3$ 
    
    \item[] Hypermultiplet \vspace{0.3cm} \\
    $\begin{rcases*}
        e^{\pm\frac{i}{2}\left( - 2\bar{H}_0 + H_0 - H_1 - H_2 + H_3 \right)} \; \text{ dilatino} \;\\
        e^{\pm\frac{i}{2}\left(- 2H_0 + \bar{H}_0 - \bar{H}_1 - \bar{H}_2 + \bar{H}_3 \right)}  \; \text{ dilatino} \;\\
        e^{\pm\frac{i}{2}\left(-2\bar{H}_1+H_0 - H_1 + H_2 - H_3 \right)}  \;\\
        e^{\pm\frac{i}{2}\left( -2 \bar{H}_2 + H_0 + H_1 - H_2 - H_3 \right)} \;\\
        e^{\pm\frac{i}{2}\left(-2 \bar{H}_3 + H_0 + H_1 + H_2 + H_3 \right)} \; \\
        e^{\pm\frac{i}{2}\left(-2 H_1 +\bar{H}_0 - \bar{H}_1 + \bar{H}_2 - \bar{H}_3 \right)}  \;\\
        e^{\pm\frac{i}{2}\left(-2 H_2 + \bar{H}_0 + \bar{H}_1 - \bar{H}_2 - \bar{H}_3 \right)} \;\\
        e^{\pm\frac{i}{2}\left(-2 H_3 + \bar{H}_0 + \bar{H}_1 + \bar{H}_2 + \bar{H}_3 \right)} \;\\
    \end{rcases*}$ eight spin-$1/2$ fermions \\

    $\begin{rcases*}
        e^{\pm i\left( H_0 -\bar{H}_0 \right)} \; \text{ dilaton pseudoscalar} \;\\
        e^{\pm i\left( H_i -\bar{H}_i \right)} \; \text{ $i=1,2,3$, K\"ahler moduli $T^1,T^2,T^3$} \;\\
        e^{\pm\frac{i}{2}\left(H_0 + H_1 - H_2 - H_3 - \bar{H}_0 - \bar{H}_1 + \bar{H}_2 + \bar{H}_3 \right)} \;\\
        e^{\pm\frac{i}{2}\left(H_0 + H_1 + H_2 + H_3 - \bar{H}_0 - \bar{H}_1 - \bar{H}_2 - \bar{H}_3 \right)} \;\\
        e^{\pm\frac{i}{2}\left(H_0 - H_1 + H_2 - H_3 - \bar{H}_0 + \bar{H}_1 - \bar{H}_2 + \bar{H}_3 \right)} \;\\
        e^{\pm\frac{i}{2}\left(H_0 - H_1 - H_2 + H_3 - \bar{H}_0 + \bar{H}_1 + \bar{H}_2 - \bar{H}_3 \right)} \;\\
    \end{rcases*}$ eight complex scalars \\
\end{itemize}

The twisted states will also contribute with massless states from the following sectors
\begin{center}
\begin{tabular}{ p{4.5cm} p{4.5cm} p{4.5cm} }
    $
    \begin{aligned}[t]
    & \mathbf{b_1} \text{ sector} \\
    & \mathbf{b_1}+\mathbf{T_2} \text{ sector} \\
    & \mathbf{b_1}+\mathbf{T_3} \text{ sector} \\
    & \mathbf{b_1}+\mathbf{T_2}+\mathbf{T_3} \text{ sector} \\
    \end{aligned}
    $
& 
    $
    \begin{aligned}[t]
    & \mathbf{b_2} \text{ sector} \\
    & \mathbf{b_2}+\mathbf{T_1} \text{ sector} \\
    & \mathbf{b_2}+\mathbf{T_3} \text{ sector} \\
    & \mathbf{b_2}+\mathbf{T_1}+\mathbf{T_3} \text{ sector} \\
    \end{aligned}
    $
&
    $
    \begin{aligned}[t]
    & \mathbf{b_3} \text{ sector} \\
    & \mathbf{b_3}+\mathbf{T_1} \text{ sector} \\
    & \mathbf{b_3}+\mathbf{T_2} \text{ sector} \\
    & \mathbf{b_3}+\mathbf{T_1}+\mathbf{T_2} \text{ sector} \\
    \end{aligned}
    $ \\
\end{tabular}
\end{center}
along with the sectors with the addition of the $\mathbf{S}$, $\mathbf{\bar{S}}$ and $\mathbf{S}+ \mathbf{\bar{S}}$ basis vectors. Each of these sectors contribute with four complex massless states which corresponds to the 16 hypermultiplets.

\section{Non-Supersymmetric Massless spectrum}
\label{appendix:spectrum}
Sector-by-sector partition function contribution, for non-Susy and Susy (supersymmetry) restored in the large volume limit $T_2 ^{\left(i\right)} \rightarrow \infty$
\begin{center}
\scriptsize
\begin{longtable}{c| c |c } 
Sector & Non-Susy & Susy $\left(\times  \;  T_2 ^{\left(1\right)}  T_2 ^{\left(2\right)}  T_2 ^{\left(3\right)} \tau_2 ^{-3}\right) $\\ [3ex] 
$  \mathbf {0} $ & $16 + 448\bar {q} + 768 q^{1/2}\bar {q}^{1/2} + 
                    448 q + 24832 q\bar {q} $ & $2 + 32\bar {q} + 
                  32 q + 512 q\bar {q} $ \\ [0.6 ex] $\Sv$ & $ - 
               1024 q^{1/2}\bar {q}^{1/2} - 16384 q\bar {q} $ & $ - 
            2 - 32\bar {q} - 32 q - 
            512 q\bar {q} $ \\ [0.6 ex] $\bar {\Sv} $ & $ - 
         128 q^{1/4}\bar {q}^{1/4} - 
         4096 q^{3/4}\bar {q}^{3/4} $ & $ - 2 - 32\bar {q} - 
      32 q - 
      512 q\bar {q} $ \\ [0.6 ex] $\bar {\Sv} + \Sv$ & $8192q^{3/
        4}\bar {q}^{3/4} $ & $2 + 32\bar {q} + 32 q + 
 512 q\bar {q} $ \\ [2.4 ex]
$ \T {2} $ & $8 + 352\bar {q} + 8 q^{1/2}\bar {q}^{-1/2} + 
                    640 q^{1/2}\bar {q}^{1/2} + 480 q %+ 
                    %23168 q\bar {q} 
                    $ & $\sim \left( q \bar{q} \right)^{T^{\left(2\right)} _2}$ \\ [
                   0.6 ex] $\Sv + \T {2} $ & $ - 
              512 q^{1/2}\bar {q}^{1/2} - 512 q - 
              22528 q\bar {q} $ & $\sim - \left( q \bar{q} \right)^{T^{\left(2\right)} _2}$ \\ [
             0.6 ex] $\bar {\Sv} + \T {2} $ & $ - 
        128 q^{1/4}\bar {q}^{1/4} - 
        4096 q^{3/4}\bar {q}^{3/4} $ & $\sim -\left( q \bar{q} \right)^{T^{\left(2\right)} _2}$ \\ [
       0.6 ex] $\bar {\Sv} + \Sv + \T {2} $ & $8192q^{3/4}\bar {q}^{3/
      4} $ & $\sim \left( q \bar{q} \right)^{T^{\left(2\right)} _2}$ \\ [2.4 ex]
$ \T {1} $ & $8 + 480\bar {q} + 8 q^{-1/2}\bar {q}^{1/2} + 
                    640 q^{1/2}\bar {q}^{1/2} + 352 q %+ 
                    %23168 q\bar {q} 
                    $ & $\sim \left( q \bar{q} \right)^{T^{\left(1\right)} _2}$ \\ [
                   0.6 ex] $\Sv + \T {1} $ & $ - 
              1024 q^{1/2}\bar {q}^{1/2} - 16384 q\bar {q} $ & $\sim - \left( q \bar{q} \right)^{T^{\left(1\right)} _2}$ \\ [
             0.6 ex] $\bar {\Sv} + \T {1} $ & $ - 
        64 q^{-1/4}\bar {q}^{3/4} - 64 q^{1/4}\bar {q}^{1/4} - 
        3968 q^{3/4}\bar {q}^{3/4} $ & $\sim - \left( q \bar{q} \right)^{T^{\left(1\right)} _2}$ \\ [
       0.6 ex] $\bar {\Sv} + \Sv + \T {1} $ & $8192q^{3/4}\bar {q}^{3/
      4} $ & $\sim \left( q \bar{q} \right)^{T^{\left(1\right)} _2}$ \\ [2.4 ex]
$ \T {1} + \T {2} $ & $8 
%+ 416\bar {q} 
+ 4 q^{-1/2}\bar {q}^{1/2} + 
                    4 q^{1/2}\bar {q}^{-1/2} + 
                    512 q^{1/2}\bar {q}^{1/2} 
                    %+ 416 q + 
                    %21120 q\bar {q} 
                    $ & $\sim \left( q \bar{q} \right)^{T^{\left(1\right)} _2+T^{\left(2\right)} _2}$ \\ [
                   0.6 ex] $\Sv + \T {1} + \T {2} $ & $ - 
              512 q^{1/2}\bar {q}^{1/2} - 512 q - 
              22528 q\bar {q} $ & $\sim -\left( q \bar{q} \right)^{T^{\left(1\right)} _2+T^{\left(2\right)} _2}$ \\ [
             0.6 ex] $\bar {\Sv} + \T {1} + \T {2} $ & $ - 
        64 q^{-1/4}\bar {q}^{3/4} - 64 q^{1/4}\bar {q}^{1/4} - 
        3968 q^{3/4}\bar {q}^{3/4} $ & $\sim -\left( q \bar{q} \right)^{T^{\left(1\right)} _2+T^{\left(2\right)} _2}$ \\ [
       0.6 ex] $\bar {\Sv} + \Sv + \T {1} + \T {2} $ & $8192q^{3/
       4}\bar {q}^{3/4} $ & $\sim \left( q \bar{q} \right)^{T^{\left(1\right)} _2+T^{\left(2\right)} _2}$ \\ [2.4 ex]
$ \bv {2} $ & $64q^{1/4}\bar {q}^{1/4} + 
                    4096 q^{3/4}\bar {q}^{3/4} $ & $1 + 40\bar {q} + 
                   64 q^{1/2}\bar {q}^{1/2} + 40 q + 
                   1600 q\bar {q} $ \\ [
                    0.6 ex] $\bv {2} + \Sv$ & $ - 
                64 q^{1/4}\bar {q}^{1/4} - 
                4096 q^{3/4}\bar {q}^{3/4} $ & $ - 1 - 40\bar {q} - 
             64 q^{1/2}\bar {q}^{1/2} - 40 q - 
             1600 q\bar {q} $ \\ [
               0.6 ex] $\bar {\Sv} + \bv {2} $ & $ - 8 - 
          352\bar {q} - 640 q^{1/2}\bar {q}^{1/2} - 352 q - 
          23680 q\bar {q} $ & $ - 1 - 40\bar {q} - 
       64 q^{1/2}\bar {q}^{1/2} - 40 q - 
       1600 q\bar {q} $ \\ [
         0.6 ex] $\bar {\Sv} + \bv {2} + \Sv$ & $8 + 352\bar {q} + 
    640 q^{1/2}\bar {q}^{1/2} + 352 q + 23680 q\bar {q} $ & $1 + 
 40\bar {q} + 64 q^{1/2}\bar {q}^{1/2} + 40 q + 
 1600 q\bar {q} $ \\ [2.4 ex]
$ \bv {2} + \T {1} $ & $64q^{1/4}\bar {q}^{1/4} + 
                    4096 q^{3/4}\bar {q}^{3/4} $ & $1 + 40\bar {q} + 
                   64 q^{1/2}\bar {q}^{1/2} + 40 q + 
                   1600 q\bar {q} $ \\ [
                    0.6 ex] $\bv {2} + \Sv + \T {1} $ & $ - 
                64 q^{1/4}\bar {q}^{1/4} - 
                4096 q^{3/4}\bar {q}^{3/4} $ & $ - 1 - 40\bar {q} - 
             64 q^{1/2}\bar {q}^{1/2} - 40 q - 
             1600 q\bar {q} $ \\ [
               0.6 ex] $\bar {\Sv} + \bv {2} + \T {1} $ & $ - 8 - 
          352\bar {q} - 640 q^{1/2}\bar {q}^{1/2} - 352 q - 
          23680 q\bar {q} $ & $ - 1 - 40\bar {q} - 
       64 q^{1/2}\bar {q}^{1/2} - 40 q - 
       1600 q\bar {q} $ \\ [
         0.6 ex] $\bar {\Sv} + \bv {2} + \Sv + \T {1} $ & $8 + 
    352\bar {q} + 640 q^{1/2}\bar {q}^{1/2} + 352 q + 
    23680 q\bar {q} $ & $1 + 40\bar {q} + 
 64 q^{1/2}\bar {q}^{1/2} + 40 q + 1600 q\bar {q} $ \\ [2.4 ex]
 $ \bv {1} $ & $64q^{1/4}\bar {q}^{1/4} + 
                    4096 q^{3/4}\bar {q}^{3/4} $ & $1 + 40\bar {q} + 
                   64 q^{1/2}\bar {q}^{1/2} + 40 q + 
                   1600 q\bar {q} $ \\ [
                    0.6 ex] $\bv {1} + \Sv$ & $ - 8 - 352\bar {q} - 
                640 q^{1/2}\bar {q}^{1/2} - 352 q - 
                23680 q\bar {q} $ & $ - 1 - 40\bar {q} - 
             64 q^{1/2}\bar {q}^{1/2} - 40 q - 
             1600 q\bar {q} $ \\ [
               0.6 ex] $\bar {\Sv} + \bv {1} $ & $ - 
          64 q^{1/4}\bar {q}^{1/4} - 
          4096 q^{3/4}\bar {q}^{3/4} $ & $ - 1 - 40\bar {q} - 
       64 q^{1/2}\bar {q}^{1/2} - 40 q - 
       1600 q\bar {q} $ \\ [
         0.6 ex] $\bar {\Sv} + \bv {1} + \Sv$ & $8 + 352\bar {q} + 
    640 q^{1/2}\bar {q}^{1/2} + 352 q + 23680 q\bar {q} $ & $1 + 
 40\bar {q} + 64 q^{1/2}\bar {q}^{1/2} + 40 q + 
 1600 q\bar {q} $ \\ [2.4 ex]
$ \bv {1} + \T {2} $ & $64q^{1/4}\bar {q}^{1/4} + 
                    4096 q^{3/4}\bar {q}^{3/4} $ & $1 + 40\bar {q} + 
                   64 q^{1/2}\bar {q}^{1/2} + 40 q + 
                   1600 q\bar {q} $ \\ [
                    0.6 ex] $\bv {1} + \Sv + \T {2} $ & $ - 8 - 
                352\bar {q} - 640 q^{1/2}\bar {q}^{1/2} - 352 q - 
                23680 q\bar {q} $ & $ - 1 - 40\bar {q} - 
             64 q^{1/2}\bar {q}^{1/2} - 40 q - 
             1600 q\bar {q} $ \\ [
               0.6 ex] $\bar {\Sv} + \bv {1} + \T {2} $ & $ - 
          64 q^{1/4}\bar {q}^{1/4} - 
          4096 q^{3/4}\bar {q}^{3/4} $ & $ - 1 - 40\bar {q} - 
       64 q^{1/2}\bar {q}^{1/2} - 40 q - 
       1600 q\bar {q} $ \\ [
         0.6 ex] $\bar {\Sv} + \bv {1} + \Sv + \T {2} $ & $8 + 
    352\bar {q} + 640 q^{1/2}\bar {q}^{1/2} + 352 q + 
    23680 q\bar {q} $ & $1 + 40\bar {q} + 
 64 q^{1/2}\bar {q}^{1/2} + 40 q + 1600 q\bar {q} $ \\ [2.4 ex]
$ \bv {1} + \bv {2} $ & $8 + 352\bar {q} + 
                    640 q^{1/2}\bar {q}^{1/2} + 352 q + 
                    23680 q\bar {q} $ & $1 + 40\bar {q} + 
                   64 q^{1/2}\bar {q}^{1/2} + 40 q + 
                   1600 q\bar {q} $ \\ [
                    0.6 ex] $\bv {1} + \bv {2} + \Sv$ & $ - 
                64 q^{1/4}\bar {q}^{1/4} - 
                4096 q^{3/4}\bar {q}^{3/4} $ & $ - 1 - 40\bar {q} - 
             64 q^{1/2}\bar {q}^{1/2} - 40 q - 
             1600 q\bar {q} $ \\ [
               0.6 ex] $\bar {\Sv} + \bv {1} + \bv {2} $ & $ - 8 - 
          352\bar {q} - 640 q^{1/2}\bar {q}^{1/2} - 352 q - 
          23680 q\bar {q} $ & $ - 1 - 40\bar {q} - 
       64 q^{1/2}\bar {q}^{1/2} - 40 q - 
       1600 q\bar {q} $ \\ [
         0.6 ex] $\bar {\Sv} + \bv {1} + \bv {2} + \Sv$ & $64q^{1/
         4}\bar {q}^{1/4} + 4096 q^{3/4}\bar {q}^{3/4} $ & $1 + 
 40\bar {q} + 64 q^{1/2}\bar {q}^{1/2} + 40 q + 
 1600 q\bar {q} $ \\ [2.4 ex]
$ \bv {1} + \bv {2} + \T {2} $ & $8 + 352\bar {q} + 
                    640 q^{1/2}\bar {q}^{1/2} + 352 q + 
                    23680 q\bar {q} $ & $1 + 40\bar {q} + 
                   64 q^{1/2}\bar {q}^{1/2} + 40 q + 
                   1600 q\bar {q} $ \\ [
                    0.6 ex] $\bv {1} + \bv {2} + \Sv + \T {2} $ & $ - 
                64 q^{1/4}\bar {q}^{1/4} - 
                4096 q^{3/4}\bar {q}^{3/4} $ & $ - 1 - 40\bar {q} - 
             64 q^{1/2}\bar {q}^{1/2} - 40 q - 
             1600 q\bar {q} $ \\ [
               0.6 ex] $\bar {\Sv} + \bv {1} + \bv {2} + \T {2} $ & $ \
- 8 - 352\bar {q} - 640 q^{1/2}\bar {q}^{1/2} - 352 q - 
          23680 q\bar {q} $ & $ - 1 - 40\bar {q} - 
       64 q^{1/2}\bar {q}^{1/2} - 40 q - 
       1600 q\bar {q} $ \\ [
         0.6 ex] $\bar {\Sv} + \bv {1} + \bv {2} + \Sv + \T {2} $ & \
$64q^{1/4}\bar {q}^{1/4} + 4096 q^{3/4}\bar {q}^{3/4} $ & $1 + 
 40 \bar {q} + 64 q^{1/2}\bar {q}^{1/2} + 40 q + 
 1600 q\bar {q} $ \\ [2.4 ex]
$ \bv {1} + \bv {2} + \T {1} $ & $8 + 352\bar {q} + 
                    640 q^{1/2}\bar {q}^{1/2} + 352 q + 
                    23680 q\bar {q} $ & $1 + 40\bar {q} + 
                   64 q^{1/2}\bar {q}^{1/2} + 40 q + 
                   1600 q\bar {q} $ \\ [
                    0.6 ex] $\bv {1} + \bv {2} + \Sv + \T {1} $ & $ - 
                64 q^{1/4}\bar {q}^{1/4} - 
                4096 q^{3/4}\bar {q}^{3/4} $ & $ - 1 - 40\bar {q} - 
             64 q^{1/2}\bar {q}^{1/2} - 40 q - 
             1600 q\bar {q} $ \\ [
               0.6 ex] $\bar {\Sv} + \bv {1} + \bv {2} + \T {1} $ & $ \
- 8 - 352\bar {q} - 640 q^{1/2}\bar {q}^{1/2} - 352 q - 
          23680 q\bar {q} $ & $ - 1 - 40\bar {q} - 
       64 q^{1/2}\bar {q}^{1/2} - 40 q - 
       1600 q\bar {q} $ \\ [
         0.6 ex] $\bar {\Sv} + \bv {1} + \bv {2} + \Sv + \T {1} $ & \
$64q^{1/4}\bar {q}^{1/4} + 4096 q^{3/4}\bar {q}^{3/4} $ & $1 + 
 40\bar {q} + 64 q^{1/2}\bar {q}^{1/2} + 40 q + 
 1600 q\bar {q} $ \\ [2.4 ex]
$ \bv {1} + \bv {2} + \T {1} + \T {2} $ & $8 + 352\bar {q} + 
                    640 q^{1/2}\bar {q}^{1/2} + 352 q + 
                    23680 q\bar {q} $ & $1 + 40\bar {q} + 
                   64 q^{1/2}\bar {q}^{1/2} + 40 q + 
                   1600 q\bar {q} $ \\ [
                    0.6 ex] $\bv {1} + \bv {2} + \Sv + \T {1} + \T 
{2} $ & $ - 64 q^{1/4}\bar {q}^{1/4} - 
                4096 q^{3/4}\bar {q}^{3/4} $ & $ - 1 - 40\bar {q} - 
             64 q^{1/2}\bar {q}^{1/2} - 40 q - 
             1600 q\bar {q} $ \\ [
               0.6 ex] $\bar {\Sv} + \bv {1} + \bv {2} + \T {1} + \T 
{2} $ & $ - 8 - 352\bar {q} - 640 q^{1/2}\bar {q}^{1/2} - 352 q - 
          23680 q\bar {q} $ & $ - 1 - 40\bar {q} - 
       64 q^{1/2}\bar {q}^{1/2} - 40 q - 
       1600 q\bar {q} $ \\ [
         0.6 ex] $\bar {\Sv} + \bv {1} + \bv {2} + \Sv + \T {1} + \T 
{2} $ & $64q^{1/4}\bar {q}^{1/4} + 
    4096 q^{3/4}\bar {q}^{3/4} $ & $1 + 40\bar {q} + 
 64 q^{1/2}\bar {q}^{1/2} + 40 q + 1600 q\bar {q} $ \\ [2.4 ex]
$ \bv {2} + \one + \T {2} $ & $8 + 352\bar {q} + 
                    640 q^{1/2}\bar {q}^{1/2} + 352 q + 
                    23680 q\bar {q} $ & $1 + 40\bar {q} + 
                   64 q^{1/2}\bar {q}^{1/2} + 40 q + 
                   1600 q\bar {q} $ \\ [
                    0.6 ex] $\bv {2} + \one + \Sv + \T {2} $ & $ - 
                8 - 352\bar {q} - 640 q^{1/2}\bar {q}^{1/2} - 352 q - 
                23680 q\bar {q} $ & $ - 1 - 40\bar {q} - 
             64 q^{1/2}\bar {q}^{1/2} - 40 q - 
             1600 q\bar {q} $ \\ [
               0.6 ex] $\bar {\Sv} + \bv {2} + \one + \T {2} $ & $ - 
          64 q^{1/4}\bar {q}^{1/4} - 
          4096 q^{3/4}\bar {q}^{3/4} $ & $ - 1 - 40\bar {q} - 
       64 q^{1/2}\bar {q}^{1/2} - 40 q - 
       1600 q\bar {q} $ \\ [
         0.6 ex] $\bar {\Sv} + \bv {2} + \one + \Sv + \T {2} $ & \
$64q^{1/4}\bar {q}^{1/4} + 4096 q^{3/4}\bar {q}^{3/4} $ & $1 + 
 40\bar {q} + 64 q^{1/2}\bar {q}^{1/2} + 40 q + 
 1600 q\bar {q} $ \\ [2.4 ex]
$ \bv {2} + \one + \T {1} + \T {2} $ & $8 + 352\bar {q} + 
                    640 q^{1/2}\bar {q}^{1/2} + 352 q + 
                    23680 q\bar {q} $ & $1 + 40\bar {q} + 
                   64 q^{1/2}\bar {q}^{1/2} + 40 q + 
                   1600 q\bar {q} $ \\ [
                    0.6 ex] $\bv {2} + \one + \Sv + \T {1} + \T {2} $ \
& $ - 8 - 352\bar {q} - 640 q^{1/2}\bar {q}^{1/2} - 352 q - 
                23680 q\bar {q} $ & $ - 1 - 40\bar {q} - 
             64 q^{1/2}\bar {q}^{1/2} - 40 q - 
             1600 q\bar {q} $ \\ [
               0.6 ex] $\bar {\Sv} + \bv {2} + \one + \T {1} + \T {2} \
$ & $ - 64 q^{1/4}\bar {q}^{1/4} - 4096 q^{3/4}\bar {q}^{3/4} $ & $ - 
       1 - 40\bar {q} - 64 q^{1/2}\bar {q}^{1/2} - 40 q - 
       1600 q\bar {q} $ \\ [
         0.6 ex] $\bar {\Sv} + \bv {2} + \one + \Sv + \T {1} + \T {2} \
$ & $64q^{1/4}\bar {q}^{1/4} + 4096 q^{3/4}\bar {q}^{3/4} $ & $1 + 
 40 \bar {q} + 64 q^{1/2}\bar {q}^{1/2} + 40 q + 
 1600 q\bar {q} $ \\ [2.4 ex]
$ \bv {1} + \one + \T {1} $ & $8 + 352\bar {q} + 
                    640 q^{1/2}\bar {q}^{1/2} + 352 q + 
                    23680 q\bar {q} $ & $1 + 40\bar {q} + 
                   64 q^{1/2}\bar {q}^{1/2} + 40 q + 
                   1600 q\bar {q} $ \\ [
                    0.6 ex] $\bv {1} + \one + \Sv + \T {1} $ & $ - 
                64 q^{1/4}\bar {q}^{1/4} - 
                4096 q^{3/4}\bar {q}^{3/4} $ & $ - 1 - 40\bar {q} - 
             64 q^{1/2}\bar {q}^{1/2} - 40 q - 
             1600 q\bar {q} $ \\ [
               0.6 ex] $\bar {\Sv} + \bv {1} + \one + \T {1} $ & $ - 
          8 - 352\bar {q} - 640 q^{1/2}\bar {q}^{1/2} - 352 q - 
          23680 q\bar {q} $ & $ - 1 - 40\bar {q} - 
       64 q^{1/2}\bar {q}^{1/2} - 40 q - 
       1600 q\bar {q} $ \\ [
         0.6 ex] $\bar {\Sv} + \bv {1} + \one + \Sv + \T {1} $ & \
$64q^{1/4}\bar {q}^{1/4} + 4096 q^{3/4}\bar {q}^{3/4} $ & $1 + 
 40 \bar {q} + 64 q^{1/2}\bar {q}^{1/2} + 40 q + 
 1600 q\bar {q} $ \\ [2.4 ex]
$ \bv {1} + \one + \T {1} + \T {2} $ & $8 + 352\bar {q} + 
                    640 q^{1/2}\bar {q}^{1/2} + 352 q + 
                    23680 q\bar {q} $ & $1 + 40\bar {q} + 
                   64 q^{1/2}\bar {q}^{1/2} + 40 q + 
                   1600 q\bar {q} $ \\ [
                    0.6 ex] $\bv {1} + \one + \Sv + \T {1} + \T {2} $ \
& $ - 64 q^{1/4}\bar {q}^{1/4} - 4096 q^{3/4}\bar {q}^{3/4} $ & $ - 
             1 - 40\bar {q} - 64 q^{1/2}\bar {q}^{1/2} - 40 q - 
             1600 q\bar {q} $ \\ [
               0.6 ex] $\bar {\Sv} + \bv {1} + \one + \T {1} + \T {2} \
$ & $ - 8 - 352\bar {q} - 640 q^{1/2}\bar {q}^{1/2} - 352 q - 
          23680 q\bar {q} $ & $ - 1 - 40\bar {q} - 
       64 q^{1/2}\bar {q}^{1/2} - 40 q - 
       1600 q\bar {q} $ \\ [
         0.6 ex] $\bar {\Sv} + \bv {1} + \one + \Sv + \T {1} + \T {2} \
$ & $64q^{1/4}\bar {q}^{1/4} + 4096 q^{3/4}\bar {q}^{3/4} $ & $1 + 
 40\bar {q} + 64 q^{1/2}\bar {q}^{1/2} + 40 q + 
 1600 q\bar {q} $ \\ [2.4 ex]
$ \bar {\Sv} + \one + \Sv + \T {1} $ & $8 
%+ 416\bar {q} 
+ 4 q^{-1/2}\bar {q}^{1/2} + 
                    4 q^{1/2}\bar {q}^{-1/2} + 
                    512 q^{1/2}\bar {q}^{1/2} 
                    %+ 416 q  
                    %+21120 q\bar {q} 
                    $ & $\sim \left( q \bar{q} \right)^{T^{\left(2\right)} _2+T^{\left(3\right)} _2}$ \\ [
                   0.6 ex] $\bar {\Sv} + \one + \T {1} $ & $ - 
              512 q^{1/2}\bar {q}^{1/2} - 512 q - 
              22528 q\bar {q} $ & $\sim -\left( q \bar{q} \right)^{T^{\left(1\right)} _2+T^{\left(2\right)} _2}$ \\ [
             0.6 ex] $\one + \Sv + \T {1} $ & $ - 
        64 q^{-1/4}\bar {q}^{3/4} - 64 q^{1/4}\bar {q}^{1/4} - 
        3968 q^{3/4}\bar {q}^{3/4} $ & $\sim -\left( q \bar{q} \right)^{T^{\left(1\right)} _2+T^{\left(2\right)} _2}$ \\ [
       0.6 ex] $\one + \T {1} $ & $8192q^{3/4}\bar {q}^{3/
      4} $ & $\sim \left( q \bar{q} \right)^{T^{\left(1\right)} _2+T^{\left(2\right)} _2}$ \\ [2.4 ex]
$ \bar {\Sv} + \one + \Sv + \T {1} + \T {2} $ & $8 
%+ 480\bar {q} 
+ 
                    8 q^{-1/2}\bar {q}^{1/2} + 
                    640 q^{1/2}\bar {q}^{1/2} 
                    %+ 352 q 
                    %+ 23168 q\bar {q} 
                    $ & $\sim \left( q \bar{q} \right)^{T^{\left(3\right)} _2}$ \\ [
                   0.6 ex] $\bar {\Sv} + \one + \T {1} + \T {2} $ & $ \
- 1024 q^{1/2}\bar {q}^{1/2} - 16384 q\bar {q} $ & $\sim -\left( q \bar{q} \right)^{T^{\left(3\right)} _2}$ \\ [
             0.6 ex] $\one + \Sv + \T {1} + \T {2} $ & $ - 
        64 q^{-1/4}\bar {q}^{3/4} - 64 q^{1/4}\bar {q}^{1/4} - 
        3968 q^{3/4}\bar {q}^{3/4} $ & $\sim -\left( q \bar{q} \right)^{T^{\left(3\right)} _2}$ \\ [
       0.6 ex] $\one + \T {1} + \T {2} $ & $8192q^{3/4}\bar {q}^{3/
      4} $ & $ \sim \left( q \bar{q} \right)^{T^{\left(3\right)} _2} $ \\
\label{tab:Spectrum}
\end{longtable}
% \captionof{table}{Sector-by-sector spectrum for massless states, for non-Susy and Susy restored theories.}
\end{center}

The twisted state of section \ref{twisted} in the $-1$ picture are given as follows

\allowdisplaybreaks
\begin{aline}
& S_{\chi^{34}}^+ S_{\chi^{56}}^- S_{y^{34}}^{\pm} S_{y^{56}}^{\pm} 
    \;
    \bar{S}_{\bar{\chi}^{34}}^+ \bar{S}_{\bar{\chi}^{56}}^- \bar{S}_{\bar{y}^{34}}^{\pm} \bar{S}_{\bar{y}^{56}}^{\pm}
     \\
& S_{\chi^{34}}^+ S_{\chi^{56}}^- S_{\omega^{34}}^{\pm} S_{y^{56}}^{\pm} 
     \; 
    \bar{S}_{\bar{\chi}^{34}}^+ \bar{S}_{\bar{\chi}^{56}}^- \bar{S}_{\bar{\omega}^{34}}^{\pm} \bar{S}_{\bar{y}^{56}}^{\pm}
     \\ 
& S_{\chi^{34}}^+ S_{\chi^{56}}^- S_{y^{34}}^{\pm} S_{\omega^{56}}^{\pm} 
     \; 
    \bar{S}_{\bar{\chi}^{34}}^+ \bar{S}_{\bar{\chi}^{56}}^- \bar{S}_{\bar{y}^{34}}^{\pm} \bar{S}_{\bar{\omega}^{56}}^{\pm}
     \\
& S_{\chi^{34}}^+ S_{\chi^{56}}^- S_{\omega^{34}}^{\pm} S_{\omega^{56}}^{\pm} 
    \; 
    \bar{S}_{\bar{\chi}^{34}}^+ \bar{S}_{\bar{\chi}^{56}}^- \bar{S}_{\bar{\omega}^{34}}^{\pm} \bar{S}_{\bar{\omega}^{56}}^{\pm}
     \\[10pt]
& S_{\chi^{12}}^+ S_{\chi^{56}}^- S_{y^{12}}^{\pm} S_{y^{56}}^{\pm} 
     \; 
    \bar{S}_{\bar{\chi}^{12}}^+ \bar{S}_{\bar{\chi}^{56}}^- \bar{S}_{\bar{y}^{12}}^{\pm} \bar{S}_{\bar{y}^{56}}^{\pm}
     \\
& S_{\chi^{12}}^+ S_{\chi^{56}}^- S_{\omega^{12}}^{\pm} S_{y^{56}}^{\pm} 
     \; 
    \bar{S}_{\bar{\chi}^{12}}^+ \bar{S}_{\bar{\chi}^{56}}^- \bar{S}_{\bar{\omega}^{12}}^{\pm} \bar{S}_{\bar{y}^{56}}^{\pm}
     \\
& S_{\chi^{12}}^+ S_{\chi^{56}}^- S_{y^{12}}^{\pm} S_{\omega^{56}}^{\pm} 
     \; 
    \bar{S}_{\bar{\chi}^{12}}^+ \bar{S}_{\bar{\chi}^{56}}^- \bar{S}_{\bar{y}^{12}}^{\pm} \bar{S}_{\bar{\omega}^{56}}^{\pm}
     \\
& S_{\chi^{12}}^+ S_{\chi^{56}}^- S_{\omega^{12}}^{\pm} S_{\omega^{56}}^{\pm} 
     \; 
    \bar{S}_{\bar{\chi}^{12}}^+ \bar{S}_{\bar{\chi}^{56}}^- \bar{S}_{\bar{\omega}^{12}}^{\pm} \bar{S}_{\bar{\omega}^{56}}^{\pm}
     \\[10pt]
& S_{\chi^{12}}^+ S_{\chi^{34}}^- S_{y^{12}}^{\pm} S_{y^{34}}^{\pm} 
     \; 
    \bar{S}_{\bar{\chi}^{12}}^+ \bar{S}_{\bar{\chi}^{34}}^- \bar{S}_{\bar{y}^{12}}^{\pm} \bar{S}_{\bar{y}^{34}}^{\pm}
     \\
& S_{\chi^{12}}^+ S_{\chi^{34}}^- S_{\omega^{12}}^{\pm} S_{y^{34}}^{\pm} 
     \; 
    \bar{S}_{\bar{\chi}^{12}}^+ \bar{S}_{\bar{\chi}^{34}}^- \bar{S}_{\bar{\omega}^{12}}^{\pm} \bar{S}_{\bar{y}^{34}}^{\pm}
     \\
& S_{\chi^{12}}^+ S_{\chi^{34}}^- S_{y^{12}}^{\pm} S_{\omega^{34}}^{\pm} 
     \; 
    \bar{S}_{\bar{\chi}^{12}}^+ \bar{S}_{\bar{\chi}^{34}}^- \bar{S}_{\bar{y}^{12}}^{\pm} \bar{S}_{\bar{\omega}^{34}}^{\pm}
     \\
& S_{\chi^{12}}^+ S_{\chi^{34}}^- S_{\omega^{12}}^{\pm} S_{\omega^{34}}^{\pm} 
     \; 
    \bar{S}_{\bar{\chi}^{12}}^+ \bar{S}_{\bar{\chi}^{34}}^- \bar{S}_{\bar{\omega}^{12}}^{\pm} \bar{S}_{\bar{\omega}^{34}}^{\pm} \\
\end{aline}
\\

and in the $0$ picture they read

\vspace{0em}

\begin{align}
\frac{1}{4\sqrt{2}} &\; S_{\chi^{34}}^+ S_{\chi^{56}}^- S_{y^{34}}^{\pm} S_{y^{56}}^{\pm} 
     \;
    \bar{S}_{\bar{\chi}^{34}}^+ \bar{S}_{\bar{\chi}^{56}}^- \bar{S}_{\bar{y}^{34}}^{\pm} \bar{S}_{\bar{y}^{56}}^{\pm}
    \left[
    \left( \omega^{3} + \omega^{4} \right) \left( \bar{\omega}^{3} + \bar{\omega}^{4} \right)
    +
    \left( \omega^{5} + \omega^{6} \right)\left( \bar{\omega}^{5} + \bar{\omega}^{6} \right)
    \right] \nonumber \\
\frac{1}{4\sqrt{2}} &\; S_{\chi^{34}}^+ S_{\chi^{56}}^- S_{\omega^{34}}^{\pm} S_{y^{56}}^{\pm} 
     \; 
    \bar{S}_{\bar{\chi}^{34}}^+ \bar{S}_{\bar{\chi}^{56}}^- \bar{S}_{\bar{\omega}^{34}}^{\pm} \bar{S}_{\bar{y}^{56}}^{\pm}
    \left[\left( y^{3} + y^{4}  \right)
    \left( \bar{y}^{3} + \bar{y}^{4} \right) 
    +
    \left( \omega^{5} + \omega^{6} \right)
    \left(  \bar{\omega}^{5} + \bar{\omega}^{6} \right)
    \right]
    \nonumber \\ 
\frac{1}{4\sqrt{2}} &\; S_{\chi^{34}}^+ S_{\chi^{56}}^- S_{y^{34}}^{\pm} S_{\omega^{56}}^{\pm} 
     \; 
    \bar{S}_{\bar{\chi}^{34}}^+ \bar{S}_{\bar{\chi}^{56}}^- \bar{S}_{\bar{y}^{34}}^{\pm} \bar{S}_{\bar{\omega}^{56}}^{\pm}
    \left[\left( \omega^{3} + \omega^{4} \right)
    \left( \bar{\omega}^{3} + \bar{\omega}^{4} \right) 
    +
    \left( y^{5} + y^{6} \right)
    \left(  \bar{y}^{5} + \bar{y}^{6} \right)
    \right]
    \nonumber \\
\frac{1}{4\sqrt{2}} &\; S_{\chi^{34}}^+ S_{\chi^{56}}^- S_{\omega^{34}}^{\pm} S_{\omega^{56}}^{\pm} 
    \; 
    \bar{S}_{\bar{\chi}^{34}}^+ \bar{S}_{\bar{\chi}^{56}}^- \bar{S}_{\bar{\omega}^{34}}^{\pm} \bar{S}_{\bar{\omega}^{56}}^{\pm}
    \left[\left( y^{3} + y^{4} \right)
    \left( \bar{y}^{3} + \bar{y}^{4} \right) 
    +
    \left( y^{5} + y^{6} \right)
    \left( \bar{y}^{5} +\bar{y}^{6} \right)
    \right]
    \nonumber \\[10pt]
\frac{1}{4\sqrt{2}} &\; S_{\chi^{12}}^+ S_{\chi^{56}}^- S_{y^{12}}^{\pm} S_{y^{56}}^{\pm} 
     \; 
    \bar{S}_{\bar{\chi}^{12}}^+ \bar{S}_{\bar{\chi}^{56}}^- \bar{S}_{\bar{y}^{12}}^{\pm} \bar{S}_{\bar{y}^{56}}^{\pm}
    \left[
    \left( \omega^{1} + \omega^{2} \right)\left( \bar{\omega}^{1} + \bar{\omega}^{2}\right)
    +
    \left( \omega^{5} + \omega^{6} \right)\left( \bar{\omega}^{5} + \bar{\omega}^{6} \right)
    \right] \nonumber \\
\frac{1}{4\sqrt{2}} & \; S_{\chi^{12}}^+ S_{\chi^{56}}^- S_{\omega^{12}}^{\pm} S_{y^{56}}^{\pm} 
     \; 
    \bar{S}_{\bar{\chi}^{12}}^+ \bar{S}_{\bar{\chi}^{56}}^- \bar{S}_{\bar{\omega}^{12}}^{\pm} \bar{S}_{\bar{y}^{56}}^{\pm}
    \left[\left( y^{1} + y^{2}  \right)
    \left( \bar{y}^{1} + \bar{y}^{2} \right) 
    +
    \left( \omega^{5} + \omega^{6} \right)
    \left( \bar{\omega}^{5} + \bar{\omega}^{6} \right)
    \right]
    \nonumber \\
\frac{1}{4\sqrt{2}} &\; S_{\chi^{12}}^+ S_{\chi^{56}}^- S_{y^{12}}^{\pm} S_{\omega^{56}}^{\pm} 
     \; 
    \bar{S}_{\bar{\chi}^{12}}^+ \bar{S}_{\bar{\chi}^{56}}^- \bar{S}_{\bar{y}^{12}}^{\pm} \bar{S}_{\bar{\omega}^{56}}^{\pm}
    \left[\left( \omega^{1} + \omega^{2} \right)
    \left( \bar{\omega}^{1} + \bar{\omega}^{2} \right) 
    +
    \left( y^{5} + y^{6} \right)
    \left( \bar{y}^{5} + \bar{y}^{6} \right)
    \right]
    \nonumber \\
\frac{1}{4\sqrt{2}} &\; S_{\chi^{12}}^+ S_{\chi^{56}}^- S_{\omega^{12}}^{\pm} S_{\omega^{56}}^{\pm} 
     \; 
    \bar{S}_{\bar{\chi}^{12}}^+ \bar{S}_{\bar{\chi}^{56}}^- \bar{S}_{\bar{\omega}^{12}}^{\pm} \bar{S}_{\bar{\omega}^{56}}^{\pm}
    \left[\left( y^{1} + y^{2} \right)
    \left( \bar{y}^{1} + \bar{y}^{2} \right) 
    +
    \left( y^{5} + y^{6} \right)
    \left( \bar{y}^{5} + \bar{y}^{6} \right)
    \right]
    \nonumber \\[10pt]
\frac{1}{4\sqrt{2}} &\; S_{\chi^{12}}^+ S_{\chi^{34}}^- S_{y^{12}}^{\pm} S_{y^{34}}^{\pm} 
     \; 
    \bar{S}_{\bar{\chi}^{12}}^+ \bar{S}_{\bar{\chi}^{34}}^- \bar{S}_{\bar{y}^{12}}^{\pm} \bar{S}_{\bar{y}^{34}}^{\pm}
    \left[
    \left( \omega^{1} + \omega^{2} \right)\left( \bar{\omega}^{1} + \bar{\omega}^{2} \right)
    +
    \left(\omega^{3} + \omega^{4} \right)\left( \bar{\omega}^{3} + \bar{\omega}^{4} \right)
    \right]
    \nonumber \\
\frac{1}{4\sqrt{2}} &\; S_{\chi^{12}}^+ S_{\chi^{34}}^- S_{\omega^{12}}^{\pm} S_{y^{34}}^{\pm} 
    \; 
    \bar{S}_{\bar{\chi}^{12}}^+ \bar{S}_{\bar{\chi}^{34}}^- \bar{S}_{\bar{\omega}^{12}}^{\pm} \bar{S}_{\bar{y}^{34}}^{\pm}
    \left[\left( y^{1} + y^{2} \right)
    \left( \bar{y}^{1} + \bar{y}^{2} \right) 
    +
    \left( \omega^{3} + \omega^{4} \right)
    \left( \bar{\omega}^{3} + \bar{\omega}^{4} \right)
    \right]
    \nonumber \\
\frac{1}{4\sqrt{2}} &\; S_{\chi^{12}}^+ S_{\chi^{34}}^- S_{y^{12}}^{\pm} S_{\omega^{34}}^{\pm} 
     \; 
    \bar{S}_{\bar{\chi}^{12}}^+ \bar{S}_{\bar{\chi}^{34}}^- \bar{S}_{\bar{y}^{12}}^{\pm} \bar{S}_{\bar{\omega}^{34}}^{\pm}
    \left[\left( \omega^{1} + \omega^{2} \right)
    \left( \bar{\omega}^{1} + \bar{\omega}^{2} \right) 
    +
    \left( y^{3} + y^{4} \right)
    \left( \bar{y}^{3} + \bar{y}^{4} \right)
    \right]
    \nonumber \\
\frac{1}{4\sqrt{2}} &\; S_{\chi^{12}}^+ S_{\chi^{34}}^- S_{\omega^{12}}^{\pm} S_{\omega^{34}}^{\pm} 
     \; 
    \bar{S}_{\bar{\chi}^{12}}^+ \bar{S}_{\bar{\chi}^{34}}^- \bar{S}_{\bar{\omega}^{12}}^{\pm} \bar{S}_{\bar{\omega}^{34}}^{\pm}
    \left[\left( y^{1} + y^{2} \right)
    \left( \bar{y}^{1} + \bar{y}^{2} \right) 
    +
    \left(y^{3} + y^{4} \right)
    \left(\bar{y}^{3} + \bar{y}^{4} \right)
    \right]
\end{align}

\section{One-loop correlators}
\label{appendix:correlators}

The following one-loop correlators \cite{Verlinde_1, Verlinde_2, Verlinde_3, Dixon_1, CFT_1, CFT_2, CFT_3, CFT_4, CFT_5}
have been used for the evaluation of one-loop amplitudes. \\
Fermion correlator on the torus for even spin structure
\begin{equation}
    \langle \psi^I \left(z \right) \psi^J \left( w\right) \rangle \smb{a}{b} = G^{IJ} \frac{\vth_1 ' \left(0\right)}{\vth_1 \left(z-w\right)} \; \frac{\vth\smb{a}{b} \left(z-w \right) }{\vth\smb{a}{b} \left(0 \right) } 
\end{equation}
Torus correlator for non-compact bosons
\begin{aline}
    \langle X^\mu \left(z, \bar{z} \right) X^\nu \left(w, \bar{w}\right) \rangle = - \eta^{\mu \nu} \log \left| \frac{\vth_1 \left(z-w\right)}{\vth_1 ' \left(0\right)} \right|^2 + 2 \pi \;\eta^{\mu \nu} \frac{\text{Im} \left(z-w\right)^2}{\tau_2}
\end{aline}
Correlator for chiral compact boson  
\begin{equation}
    \langle e^{i k \cdot X\left(z\right)} e^{-ik \cdot X\left(w\right)} \rangle = e^{k^I k^M G_{IJ} G_{MN} \langle X^J \left(z\right) X^N \left(w\right) \rangle} = \left(  \frac{\vth_1 \left(z-w\right)}{\vth_1 ' \left(0\right)} \right)^{-k^2} e^{2 \pi i \; k \cdot p_L \; \left(z-w \right)} 
\end{equation}
\begin{equation}
    \langle \partial_z X^I  \left(z\right) \partial_w X^J \left(w\right) \rangle = G^{IJ} \partial_z \partial_w  \log \left(  \frac{\vth_1 \left(z-w\right)}{\vth_1 ' \left(0\right)} \right)^{-1} + \left(2 \pi p_L ^I \right)\left(2 \pi p_L ^J \right)
\end{equation}
with the symmetrized product $\cdot$ defined as
\begin{equation}
    \alpha \cdot \beta = \alpha^I G_{IJ} \beta^J, \qquad \alpha^2 = \alpha^I G_{IJ} \alpha^J
\end{equation}
The chiral anti-chiral correlator for the compact boson is 
\begin{equation}
    \langle \partial_z X^I  \left(z\right) \partial_{\bar{z}} X^J \left(\bar{z}\right) \rangle = p_L ^I p_R ^J
\end{equation}

It can be checked that performing the correlator for holomorphic + antiholomorphic part
\begin{aline}
    \sum_{\left(p^I_L, p^I_R\right) \in \Gamma}&  \frac{q^{\frac{1}{2}p _L ^2}\; \bar{q}^{\frac{1}{2} p _R ^2}}{\eta^2 \bar{\eta}^2} \langle  e^{i k _L \cdot X_L\left(z\right)+i k _R \cdot X_R\left(\bar{z}\right)}  e^{-i k _L \cdot X_L\left(0\right)-i k _R \cdot X_R\left(0\right)} \rangle  \\
    & = \sum_{\left(p^I_L, p^I_R\right) \in \Gamma} \frac{q^{\frac{1}{2}p _L ^2}\; \bar{q}^{\frac{1}{2} p _R ^2}}{\eta^2 \bar{\eta}^2} \langle e^{i k _L \cdot X_L\left(z\right)}e^{-i  k _L \cdot X_L\left(0\right)} \rangle \langle e^{i k _R \cdot X_R\left(\bar{z}\right)}e^{-i k _R \cdot X_R\left(0\right)} \rangle \\
    & = \sum_{\left(p^I_L, p^I_R\right) \in \Gamma} \frac{q^{\frac{1}{2}p _L ^2}\; \bar{q}^{\frac{1}{2} p _R ^2}}{\eta^2 \bar{\eta}^2}
    \left( \frac{\vth_1 \left(z\right)}{\vth_1 ' \left(0\right)} \right)^{-k_L ^2}
    \left( \frac{\vthb_1 \left(\bar{z}\right)}{\vthb_1 ' \left(0\right)} \right)^{-k_R ^2}
    e^{2 \pi i \left(k_L \cdot p_L  z - k_R \cdot p_R \bar{z} \right)} \\
    % & \stackrel{k^I_L = k^I_R \equiv k^I}{=} \sum_{\left(p^I_L, p^I_R\right) \in \Gamma} \frac{q^{\frac{1}{2}p _L ^2}\; \bar{q}^{\frac{1}{2} p _R ^2}}{\eta^2 \bar{\eta}^2}
    % \left|   E\left(z \right) \right|^{-2 k ^2}
    % e^{2 \pi i k \cdot \left(p_L  z - p_R \bar{z} \right)} \\
    % & \stackrel{p^I_L = p^I_R \equiv p^I}{=} \sum_{p^I \in \Gamma} \frac{q^{\frac{1}{2}p ^2}\; \bar{q}^{\frac{1}{2} p ^2}}{\eta^2 \bar{\eta}^2}
    % \left|  E\left(z \right) \right|^{-2 k ^2}
    % e^{2 \pi i k \cdot p \; \left(z - \bar{z} \right)} \\
\end{aline}
setting $k_L ^I = k_R ^I = k/\sqrt{2}$ and $p_L ^I = p_R ^I = p/\sqrt{2}$, switching the sum to integration
\begin{equation}
    \sum_{p^I \in \Gamma} \rightarrow  \text{Vol}_{\Gamma} \int dp^1 dp^2  \sqrt{ \det G}
\end{equation}
with $\text{Vol}_{\Gamma}  = \tau_2$, and performing the Gaussian integral
\begin{aline}
    & = \frac{\tau_2 \sqrt{\det G}}{\eta^2 \bar{\eta}^2} \left|  \frac{\vth_1 \left(z\right)}{\vth_1 ' \left(0\right)} \right|^{-k^2} \int dp^1 dp^2 e^{-\pi \tau_2  p^2 - 2 \pi  k \cdot p \operatorname{Im} z } = \frac{1}{\eta^2 \bar{\eta}^2} \left| \frac{\vth_1 \left(z\right)}{\vth_1 ' \left(0\right)} \right|^{-k^2}  e^{ \frac{\pi k^2}{\tau_2} \operatorname{Im} z^2} 
\end{aline}
% Using $\langle e^A e^B \rangle = e^{\frac{AB}{2}}$, 
with $k^2 = 1$, one gets the torus correlator of non-compact bosons
\begin{aline}
    \langle X^I \left( z, \bar{z}\right) X^J \left(0\right) \rangle = G^{IJ} \left[ - \log \left| \frac{\vth_1 \left(z\right)}{\vth_1 ' \left(0\right)} \right|^2 + \frac{2 \pi}{\tau_2} \operatorname{Im}z^2 \right]
\end{aline}
with indices $I,J \rightarrow \mu, \nu$. \\
Correlator for spin fields 
\begin{equation}
    \langle \; S^\pm _{\phi} \left( z \right) S^\mp _{\phi} \left(0\right)\; \rangle = \frac{\vartheta_{\phi} \left( \pm \frac{z}{2}\right)}{\vartheta_{\phi} \left(0 \right)} \frac{\vartheta^{\prime 1/4} _{1}     \left(0 \right)}{\vartheta_1 ^{1/4} \left( z\right)}
    \label{correlator_twisted}
\end{equation}
\
The following identity for the square of the fermion correlator has been used in the paper
\begin{equation}
    \langle \psi \left(z \right) \psi \left( 0\right) \rangle ^ 2 \smb{a}{b} = - \partial^2 _z \log \vth_1 \left(z \right) + \partial^2 _z \log \vth \smb{a}{b} \left(0 \right) 
    \label{square_Szeko}
\end{equation}

\subsection{Torus Periodicity}

Using the formulae in Appendix \ref{appendix:Theta Functions}, the fermion propagator on the torus transforms as
        \begin{aline}
            \langle \psi^I\left(z +1\right) \psi^J \left( 0\right) \rangle \smb{a}{b} &  = e^{i \pi a}  \langle \psi^I\left(z\right) \psi^J \left( 0\right) \rangle \smb{a}{b}
        \label{periodicity_fermion_1}
        \end{aline}
        \begin{aline}
            \langle \psi^I\left(z +\tau\right) \psi^J \left( 0\right) \rangle \smb{a}{b} & = - e^{i \pi b}  \langle \psi^I\left(z\right) \psi^J \left( 0\right) \rangle \smb{a}{b}
        \label{periodicity_fermion_2}
        \end{aline}
The propagator on the torus for the non-compact bosons is inviariant. 
The propagator for the (holomorphic) compact boson
    \begin{aline}
        \langle e^{i k \cdot X\left(z + 1\right)} e^{-ik \cdot X\left(0\right)} \rangle =  \left(  \frac{\vth_1 \left(z+1\right)}{\vth_1 ' \left(0\right)} \right)^{-k^2} e^{2 \pi i \; k \cdot p_L \left(z+1 \right)} =\left(  \frac{\vth_1 \left(z\right)}{\vth_1 ' \left(0\right)} \right)^{-k^2} e^{2 \pi i \; k \cdot p_L  z } \; e^{2 \pi i \; k \cdot p_L}
        \label{periodicity_compact_boson_1}
    \end{aline}
    with $k \cdot p_L = - \frac{H_1}{2} - \frac{m_2}{2} + \text{integers}$, so not integer in general, and
    \begin{aline}
        \sum_{p_L, p_R \in \Gamma} & 
        \frac{q^{\frac{1}{2}p_L ^2} \bar{q}^{\frac{1}{2}p_R ^2}}{\eta^2\bar{\eta}^2} \;  \langle e^{i k \cdot X\left(z + \tau\right)} e^{-ik \cdot X\left(0\right)} \rangle = \sum_{p_L, p_R \in \Gamma}  
        \frac{q^{\frac{1}{2}p_L ^2} \bar{q}^{\frac{1}{2}p_R ^2}}{\eta^2\bar{\eta}^2} \;  \left(  \frac{\vth_1 \left(z+\tau\right)}{\vth_1 ' \left(0\right)} \right)^{-k^2} e^{2 \pi i \; k \cdot p_L \; \left(z+\tau \right)} \\
        & = \sum_{p_L, p_R \in \Gamma}  
        \frac{q^{\frac{1}{2}p_L ^2} \bar{q}^{\frac{1}{2}p_R ^2}}{\eta^2\bar{\eta}^2} \;  \left(  - \frac{\vth_1 \left(z\right)}{\vth_1 ' \left(0\right)} \right)^{-k^2} e^{i \pi \tau k^2} e^{2 \pi i z k^2} e^{2 \pi i \; k \cdot p_L \; \tau } e^{2 \pi i \; k \cdot p_L \; z }  \\
        & = \sum_{p_L, p_R \in \Gamma}  
        \frac{q^{\frac{1}{2}\left(p_L + k \right) ^2} \bar{q}^{\frac{1}{2}p_R ^2}}{\eta^2\bar{\eta}^2} \;  \left(   - \frac{\vth_1 \left(z\right)}{\vth_1 ' \left(0\right)} \right)^{-k^2}  e^{2 \pi i \; k \cdot \left(p_L+k\right) \; z }  
        \label{periodicity_compact_boson_2}
    \end{aline}
    with Left momentum shifted $p_L \rightarrow p_L +k$. Finally the correlator for the compact boson $\langle \partial_z X^I \left(z\right) \partial_w X^J \left(w\right) \rangle$ is trivially periodic under $z\rightarrow z+1$, $z\rightarrow z+\tau$. \\

The product of a fermion and the exponential of a non-compact boson propagator, as for the single and double charged scalars, gives a periodic expression. 
Under $z\rightarrow z+1$ the fermion correlator gets the phase $(-1)^{a}$, while the exponential of compact boson gets the phase $e^{2 \pi i k \cdot p_L} = \left(-1\right)^{H_i - m_2}$, with the bosons in the $\left(i\right)$-th torus and with $k \cdot p_L = \frac{H_i}{2} - \frac{m_2}{2} + $integers. The total additional phase $\left(-1\right)^{a+H_i-m_2}$ can be reabsorbed by performing $G_i \rightarrow G_i + 1$ in the sum. \\
Under $z\rightarrow z+\tau$ the fermion propagator gets the phase $\left(-1\right)^{b}$, while the exponential of compact boson gets a shift in the Left momentum $p_L \rightarrow p_L + k$. The momentum shift in the lattice can be achieved by performing $H_i \rightarrow H_i +1$, $m_2 \rightarrow m_2 -1$, such that $p_L \rightarrow p_L +k$, $p_R \rightarrow p_R$ and getting the additional phase $\left(-1\right)^{b}$. \\

For the spin correlator we have the following periodicities
\begin{equation}
    \langle \; S^\pm _{\phi} \left(z +1\right) S^\mp _{\phi} \left(0\right)\; \rangle = 
    \frac{\vth \smb{a}{b}_{\phi} \left(\pm \frac{z}{2} \pm \frac{1}{2}\right)}{\vartheta \smb{a}{b}_{\phi} \left(0 \right)} \frac{\vartheta^{\prime 1/4} _{1}     \left(0 \right)}{\vartheta_1 ^{1/4} \left(z\right)} = 
    \frac{\vth \smb{a}{b \mp 1}_{\phi} \left(\pm \frac{z}{2} \right)}{\vartheta \smb{a}{b}_{\phi} \left(0 \right)} \frac{\vartheta^{\prime 1/4} _{1}     \left(0 \right)}{\vartheta_1 ^{1/4} \left(z\right)} 
\end{equation}
\begin{equation}
    \langle \; S^\pm _{\phi} \left(z + \tau\right) S^\mp _{\phi} \left(0\right)\; \rangle = 
    \frac{\vth \smb{a}{b}_{\phi} \left(\pm \frac{z}{2} \pm \frac{\tau}{2}\right)}{\vartheta \smb{a}{b}_{\phi} \left(0 \right)} \frac{\vartheta^{\prime 1/4} _{1}     \left(0 \right)}{\vartheta_1 ^{1/4} \left(z\right)} =  e^{- \frac{i \pi \tau}{4}} e^{-\frac{i \pi}{2} \left(z \mp n  \right)}
    \frac{\vth \smb{a \mp 1}{b }_{\phi} \left(\pm \frac{z}{2} \right)}{\vartheta \smb{a}{b}_{\phi} \left(0 \right)} \frac{\vartheta^{\prime 1/4} _{1}     \left(0 \right)}{\vartheta_1 ^{1/4} \left(z\right)} 
\end{equation}
which are invariant only upon proper combination with other spin correlators.

\subsection{Modular Invariance}
Using the modular transformations in Appendix \ref{appendix:Theta Functions} and the following identities
\begin{aline}
    & \frac{\vth_1 \left(z|\tau+1\right)}{\vth_1 ' \left(0|\tau+1\right)} =  \frac{\vth_1 \left(z\right)}{\vth_1 ' \left(0\right)}
    , \quad \quad \frac{\vth_1 \left(\frac{z}{\tau}| -\frac{1}{\tau}\right)}{\vth_1 ' \left(0| -\frac{1}{\tau}\right)} = \frac{e^{i \pi \frac{z^2}{\tau}}}{\tau} \; \frac{\vth_1 \left(z|\tau\right)}{\vth_1 ' \left(0|\tau\right)} \\
    & \frac{\Gamma^{(i)}_{2,2}\smbar{H_i}{G_i}{h_i}{g_i}(T^{(i)},U^{(i)}) \left(\tau+1\right)}{\eta^2 \left(\tau+1\right) \bar{\eta}^2 \left(\bar{\tau}+1\right)} = \frac{\Gamma^{(i)}_{2,2}\smbar{H_i}{H_i + G_i}{h_i}{g_i}(T^{(i)},U^{(i)}) \left(\tau\right)}{\eta^2 \left(\tau\right) \bar{\eta}^2 \left(\bar{\tau}\right)} \\
    & \frac{\Gamma^{(i)}_{2,2}\smbar{H_i}{G_i}{h_i}{g_i}(T^{(i)},U^{(i)}) \left(- \frac{1}{\tau}\right)}{\eta^2 \left(-\frac{1}{\tau}\right) \bar{\eta}^2 \left(-\frac{1}{\bar{\tau}}\right)} = \frac{\Gamma^{(i)}_{2,2}\smbar{G_i}{-H_i}{h_i}{g_i}(T^{(i)},U^{(i)}) \left(\tau\right)}{\eta^2 \left(\tau\right) \bar{\eta}^2 \left(\bar{\tau}\right)}
\end{aline}
the fermion correlator on the torus transforms as
\begin{aline}
        & 
        \langle \psi^I\left(z \right) \psi^J \left( 0\right) \rangle \smb{a}{b} \left( \tau+1 \right) =  
        G^{IJ}\frac{\vth_1 ' \left(0| \tau+1\right)}{\vth_1  \left(z| \tau+1\right)} \; \frac{\vth\smb{a}{b} \left(z| \tau+1\right) }{\vth\smb{a}{b} \left(0 | \tau+1\right) } =  
        \langle \psi^I\left(z \right) \psi^J \left( 0\right) \rangle \smb{a}{a+b-1} \left(  \tau \right) \\
        & 
        \langle \psi^I\left(\frac{z}{\tau} \right) \psi^J \left( 0 \right) \rangle \smb{a}{b} \left( -\frac{1}{\tau} \right) =
        G^{IJ}\frac{\vth_1 ' \left(0| -\frac{1}{\tau}\right)}{\vth_1  \left(\frac{z}{\tau}| -\frac{1}{\tau}\right)} \; \frac{\vth\smb{a}{b} \left(\frac{z}{\tau}| -\frac{1}{\tau} \right) }{\vth\smb{a}{b} \left(0 |  -\frac{1}{\tau} \right) }
        =
        \tau \; \langle \psi^I\left(z \right) \psi^J \left( 0\right) \rangle \smb{b}{-a} \left( \tau \right)
   \end{aline}
The propagator on the torus for non-compact bosons is invariant, while for the compact bosons
\begin{aline}
    \frac{\Gamma^{(i)}_{2,2}\smbar{H_i}{G_i}{h_i}{g_i}(T^{(i)},U^{(i)}) \left(\tau+1\right)}{\eta^2 \left(\tau+1\right) \bar{\eta}^2 \left(\bar{\tau}+1\right)} \; \langle e^{i k \cdot X\left(z\right)} e^{-ik \cdot X\left(0\right)} \rangle \left(\tau+1\right) = \\
    \frac{\Gamma^{(i)}_{2,2}\smbar{H_i}{H_i+G_i}{h_i}{g_i}(T^{(i)},U^{(i)}) \left(\tau\right)}{\eta^2 \left(\tau\right) \bar{\eta}^2 \left(\bar{\tau}\right)} \; \langle e^{i k \cdot X\left(z\right)} e^{-ik \cdot X\left(0\right)} \rangle \left(\tau\right)
\end{aline}
\begin{aline}
    \frac{\Gamma^{(i)}_{2,2}\smbar{H_i}{G_i}{h_i}{g_i}(T^{(i)},U^{(i)}) \left(- \frac{1}{\tau}\right)}{\eta^2 \left(- \frac{1}{\tau}\right) \bar{\eta}^2 \left(- \frac{1}{\bar{\tau}}\right)} \; \langle e^{i k \cdot X\left(z\right)} e^{-ik \cdot X\left(0\right)} \rangle \left(- \frac{1}{\tau}\right) = \\
    \tau^{k^2} \; \frac{\Gamma^{(i)}_{2,2}\smbar{H_i}{H_i}{h_i}{g_i}(T^{(i)},U^{(i)}) \left(\tau\right)}{\eta^2 \left(\tau\right) \bar{\eta}^2 \left(\bar{\tau}\right)} \; \langle e^{i k \cdot X\left(z\right)} e^{-ik \cdot X\left(0\right)} \rangle \left(\tau\right)
\end{aline}
and
\begin{aline}
    \frac{\Gamma^{(i)}_{2,2}\smbar{H_i}{G_i}{h_i}{g_i}(T^{(i)},U^{(i)}) \left(\tau+1\right)}{\eta^2 \left(\tau+1\right) \bar{\eta}^2 \left(\bar{\tau}+1\right)} \; \langle \partial_{z} X^I \left(z\right) \partial_{w} X^J \left(w\right) \rangle \left(\tau+1\right) = \\
    \frac{\Gamma^{(i)}_{2,2}\smbar{H_i}{G_i}{h_i}{g_i}(T^{(i)},U^{(i)}) \left(\tau\right)}{\eta^2 \left(\tau\right) \bar{\eta}^2 \left(\bar{\tau}\right)} \; \langle \partial_{z} X^I \left(z\right) \partial_{w} X^J \left(w\right) \rangle \left(\tau\right)
\end{aline}
\begin{aline}
    \frac{\Gamma^{(i)}_{2,2}\smbar{H_i}{G_i}{h_i}{g_i}(T^{(i)},U^{(i)}) \left(- \frac{1}{\tau}\right)}{\eta^2 \left(- \frac{1}{\tau}\right) \bar{\eta}^2 \left(- \frac{1}{\bar{\tau}}\right)} \; \langle \partial_{\frac{z}{\tau}} X^I \left(\frac{z}{\tau}\right) \partial_{\frac{w}{\tau}} X^J \left(\frac{w}{\tau}\right) \rangle \left(-\frac{1}{\tau}\right) = \\
    \tau^2 \; \frac{\Gamma^{(i)}_{2,2}\smbar{H_i}{G_i}{h_i}{g_i}(T^{(i)},U^{(i)}) \left(\tau\right)}{\eta^2 \left(\tau\right) \bar{\eta}^2 \left(\bar{\tau}\right)} \; \langle \partial_{z} X^I \left(z\right) \partial_{w} X^J \left(w\right) \rangle \left(\tau\right)
\end{aline}
having used the Poisson resummation formula \eqref{Poisson}. \\
The spin correlator, similarly to the fermion correlator, transforms as follows
\begin{equation}
    \langle \; S^\pm  \left(z \right) S^\mp \left(0\right)\; \rangle \smb{a}{b} \left(\tau+1\right) = \langle \; S^\pm  \left(z \right) S^\mp \left(0\right)\; \rangle \smb{a}{a+b-1} \left(\tau\right) 
\end{equation}
\begin{equation}
    \langle \; S^\pm  \left(z \right) S^\mp \left(0\right)\; \rangle \smb{a}{b} \left(- \frac{1}{\tau}\right) = \tau^{\frac{1}{4}} \; \langle \; S^\pm  \left(z \right) S^\mp \left(0\right)\; \rangle \smb{b}{-a} \left(\tau\right)
\end{equation}

\vfill\eject

%\newpage
%-----------------------------------------------%	BIBLIOGRAPHY
%----------------------------------------------
% \printbibliography[heading=bibintoc]
%-----------------------------------------------

\end{document}